\documentclass[article, aps, 12pt, floatfix, longbibliography, a4paper]{revtex4-2}

\usepackage{graphicx}
\usepackage{epstopdf}

\usepackage{color}
\usepackage[normalem]{ulem}

\usepackage[utf8]{inputenc}
\usepackage{xcolor}
\usepackage{physics}
\usepackage{multirow}

\begin{document}

\title[Non-coplanar helimagnetism in the layered van-der-Waals metal DyTe3]{Non-coplanar helimagnetism\\ in the layered van-der-Waals metal DyTe$_3$}

\author{Shun Akatsuka$^{1}$}
\thanks{These two authors contributed equally to this work}
\author{Sebastian Esser$^{1,*}$}\email{esser@g.ecc.u-tokyo.ac.jp}
\author{Shun Okumura$^{1}$}
\author{Ryota Yambe$^{1}$}
\author{Rinsuke Yamada$^{1}$}
\author{Moritz M. Hirschmann$^{2}$}
\author{Seno Aji$^{3}$}
\thanks{Present address: Department of Physics, Faculty of Mathematics and Natural Sciences, Universitas Indonesia, Depok 16424, Indonesia.}
\author{Jonathan S. White$^{4}$}
\author{Shang Gao$^{5}$}
\author{Yoshichika Onuki$^{2}$}
\author{Taka-hisa Arima$^{2,6}$} 
\author{Taro Nakajima$^{3,6}$}
\author{Max Hirschberger$^{1,2,7}$}\email{hirschberger@ap.t.u-tokyo.ac.jp}

\affiliation{$^{1}$Department of Applied Physics, The University of Tokyo, Bunkyo-ku, Tokyo 113-8656, Japan}
\affiliation{$^{2}$RIKEN Center for Emergent Matter Science (CEMS), Wako, Saitama 351-0198, Japan}
\affiliation{$^{3}$The Institute for Solid State Physics, The University of Tokyo, Kashiwa 277-8581, Japan}
\affiliation{$^{4}$Laboratory for Neutron Scattering and Imaging (LNS), Paul Scherrer Institute (PSI), 5232, Villigen, Switzerland}
\affiliation{$^{5}$Department of Physics, University of Science and Technology of China, Hefei 230026, China}
\affiliation{$^{6}$Department of Advanced Materials Science, The University of Tokyo, Kashiwa 277-8561, Japan}
\affiliation{$^{7}$Quantum-Phase Electronics Center (QPEC), The University of Tokyo, Bunkyo-ku, Tokyo 113-8656, Japan}
\maketitle
\newpage


\begin{center}
\Large{Abstract}
\end{center}
\textbf{Magnetic materials with highly anisotropic chemical bonding can be exfoliated to realize ultrathin sheets or interfaces with highly controllable optical or spintronics responses, while also promising novel cross-correlation phenomena between electric polarization and the magnetic texture. The vast majority of these van-der-Waals magnets are collinear ferro-, ferri-, or antiferromagnets, with a particular scarcity of lattice-incommensurate helimagnets of defined left- or right-handed rotation sense, or helicity. Here we use polarized neutron scattering to reveal cycloidal, or conical, magnetic structures in DyTe$_3$, with coupled commensurate and incommensurate order parameters, where covalently bonded double-slabs of dysprosium square nets are separated by highly metallic tellurium layers. Based on this ground state and its evolution in a magnetic field as probed by small-angle neutron scattering (SANS), we establish a one-dimensional spin model with off-diagonal on-site terms, spatially modulated by the unconventional charge order in DyTe$_3$. The CDW-driven term couples to antiferromagnetism, or to the net magnetization in applied magnetic field, and creates a complex magnetic phase diagram indicative of competing interactions in an easily cleavable helimagnet. Our work paves the way for twistronics research, where helimagnetic layers can be combined to form complex spin textures on-demand, using the vast family of rare earth chalcogenides and beyond. }
\newpage

%%==================================%%
%%          MAIN TEXT               %%
%%==================================%%

\begin{center}
\Large{Main text}
\end{center}
Magnetism in layered materials, held together by weak van-der-Waals interactions, is an active field of research spurred on by the discovery of magnetic ordering in monolayer sheets of ferromagnets and antiferromagnets~\cite{Huang2017, Gong2017, Burch2018, Gong2019}. At the frontier of this field, helimagnetic layered systems, where magnetic order has a fixed, left- or right-handed rotation sense, have been predicted to host complex spin textures~\cite{Amoroso2020, Shimizu2021} and to serve as controllable multiferroics platforms, where magnetic order is readily tuned by electric fields or currents~\cite{Jiang2020, Ohe2021, Masuda2022, Wang2022_NiI2}. However, most layered van-der-Waals magnets are commensurate ferro-, antiferro-, or ferrimagnets~\cite{Burch2018, Gong2019}; the rare helimagnets provided to us by nature are often modulated along the stacking direction, with relatively simple spin arrangement in individual layers (Table \ref{ETab1}). 

\begin{figure}[h]%
\centering
\includegraphics[width=0.9\textwidth]{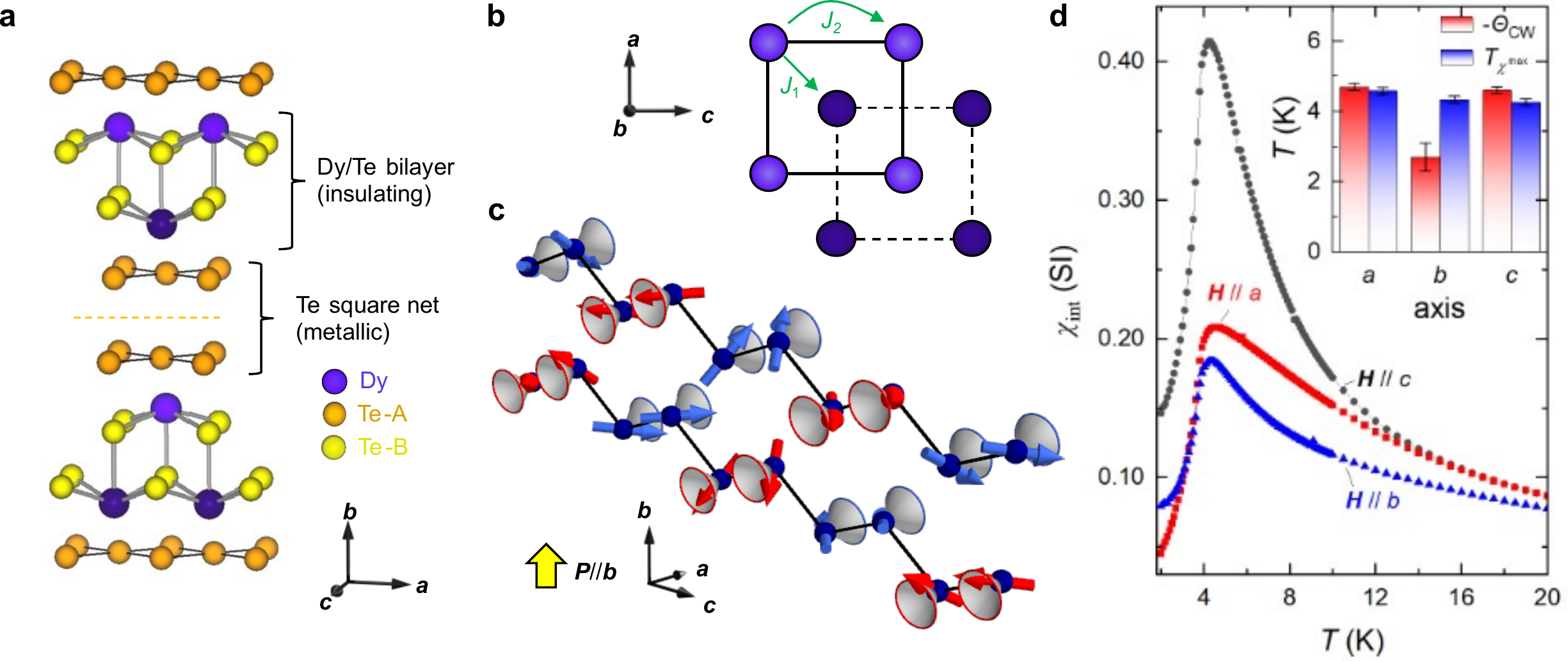}
\caption{\textbf{Conical helimagnetism in the layered square-lattice antiferromagnet DyTe$_\mathbf{3}$.} \textbf{a}, Crystallographic unit cell with covalently bonded DyTe bilayers and metallic Te bilayers, with natural cleaving plane (dashed).  \textbf{b}, Magnetic exchange interactions in a single DyTe double-square net bilayer. \textbf{c}, Zigzag chain illustration of double-square net structure in DyTe$_3$. The interactions $J_1$, $J_2$ connect nearest and next-nearest neighbours in the zigzag chain model, respectively; but the inter-atomic distance in the full crystallographic structure is shorter for $J_2$, and its antiferromagnetic coupling strength is dominant. Conical, noncoplanar helimagnetism is resolved in the zero-field ground state by neutron scattering. The cone direction, parallel to the $a$-axis, alternates both between pairs of magnetic sites in a zigzag chain, and between stacked zigzag chains. This texture causes polarization along the $b$-axis, i.e. perpendicular to square net bilayers (yellow arrow). Note: The full magnetic unit cell extends two times further along the chain direction (Fig. \ref{EfigMstructure}). \textbf{d}, Weak anisotropy of the magnetic susceptibility $\chi$ in DyTe$_3$. The softest direction is $\mathbf{H}\parallel c$, consistent with the modulation direction of the magnetic order in panel c. The inset shows Curie-Weiss temperatures and temperatures $T_\chi$ of maximal $\chi$, measured in a small magnetic fields along three crystallographic directions.}\label{fig1}
\end{figure}

In the quest for helimagnetism in layered structures with weak van-der-Waals bonds, we focus on rare earth tritellurides $R$Te$_3$ ($R$: rare earth element). These materials form a highly active arena of research regarding the interplay of correlations and topological electronic states~\cite{Schmitt2011, Kogar2020, Dolgirev2020, Gonzalez2022, Wang2022_RTe3, Gonzalez2022}. Their structure, which can be exfoliated down to the thickness of a few monolayers~\cite{Lei2020, Chen2019}, is composed of tellurium Te$_2$ double-layers and covalently bonded $R$Te slabs, with characteristic square net motifs in both (Fig. \ref{fig1}\,\textbf{a})~\cite{Malliakas2006}. Tellurium $5p$ electrons are highly localized in Te$_2$ square net bilayers ($ac$ plane), in which they form highly dispersive bands with elevated Fermi velocity~\cite{Laverock2005, Brouet2008, Chikina2022, Lei2020}. This quasi two-dimensional electronic structure amplifies correlation phenomena, such as the formation of charge-density wave (CDW) order~\cite{Ru2008} and superconductivity~\cite{Zocco2011, Zocco2015}, while also hosting protected band degeneracies~\cite{Lei2020, Sarkar2022}. In view of intense research efforts on $R$Te$_3$, it is remarkable that their magnetism has never been discussed in detail; in particular, no full refinement of magnetic structures is available~\cite{Iyeiri2003, Pfuner2011,Yang2020,Guo2021, Volkova2022, Chillal2020}.

Here, we report on helimagnetic, cone-type orders of DyTe$_3$ using polarized elastic neutron scattering. We reveal the magnetic texture in real space, probe its evolution with temperature and magnetic field, and reveal its relationship to charge-density wave formation. In DyTe$_3$, dysprosium moments are arranged in square net bilayers, where each ion has neighbours within its own layer, and within the respective other layer (Fig. \ref{fig1}\,\textbf{b}). As all zero-field magnetic orders of DyTe$_3$ are uniform along the crystallographic $a$-axis, it is reasonable to understand each square net bilayer as an effective zigzag chain of magnetic rare earth ions and to define magnetic interactions $J_1$ and $J_2$ in terms of nearest- and next-nearest neighbours on the zigzag chain, respectively. On such chains, our experiment shows that pairs of ions have cones pointing along the same direction, followed by a flip of the cone axis (Fig. \ref{fig1}\,\textbf{c}, which illustrates half a magnetic unit cell). The coupling between two DyTe bilayers, i.e. between two zigzag chains, is antiferromagnetic. Despite this complex cone arrangement, the magnetic structure defines a fixed sense of rotation, or helicity. Our theoretical spin model shows that charge-density wave order in rare earth tellurides causes local symmetry breaking, allows for off-diagonal on-site coupling terms in the Hamiltonian, and drives lattice-incommensurate magnetism when combined with antiferromagnetic interactions or with a net magnetization. We further discuss magnetocrystalline anisotropy in this layered structure, with an unconventional combination of metallic and covalent bonds. Helimagnetism of Dy rare earth moments with $4f^9$ magnetic shell emerges despite the naive expectation of strong preference for easy-axis or easy-plane anisotropy for $^{2S+1}L_J = ^{6}H_{15/2}$, with large orbital angular momentum $L = 5$. \\

\textbf{Magnetic properties of DyTe$_3$}\\
Some essential magnetic properties of DyTe$_3$ are apparent already from the magnetic susceptibility $\chi$ in Fig. \ref{fig1}\,\textbf{d}. In the high temperature regime, anisotropy in the Curie-Weiss law indicates easy-plane behaviour of magnetic moments, favouring the $ac$ plane with uniaxial anisotropy constant $K_1 =22(1)\,\mathrm{kJ}\,\mathrm{m}^{-3}$ (Methods).  At low temperatures, the strongest enhancement of $\chi$ occurs when the magnetic field $\mathbf{H}$ is along the $c$-axis, i.e. parallel to the zigzag direction defined in Fig. \ref{fig1}\,\textbf{c}. We may deduce that the magnetic moments are aligned, predominantly, along the $a$ and $b$ axes. All susceptibility curves show maxima around $T_\chi = 4.5\,$K, quite far above the onset of three-dimensional, long-range magnetic order, as shown in the following.

\begin{figure}[h]%
\centering
\includegraphics[width=0.9\textwidth]{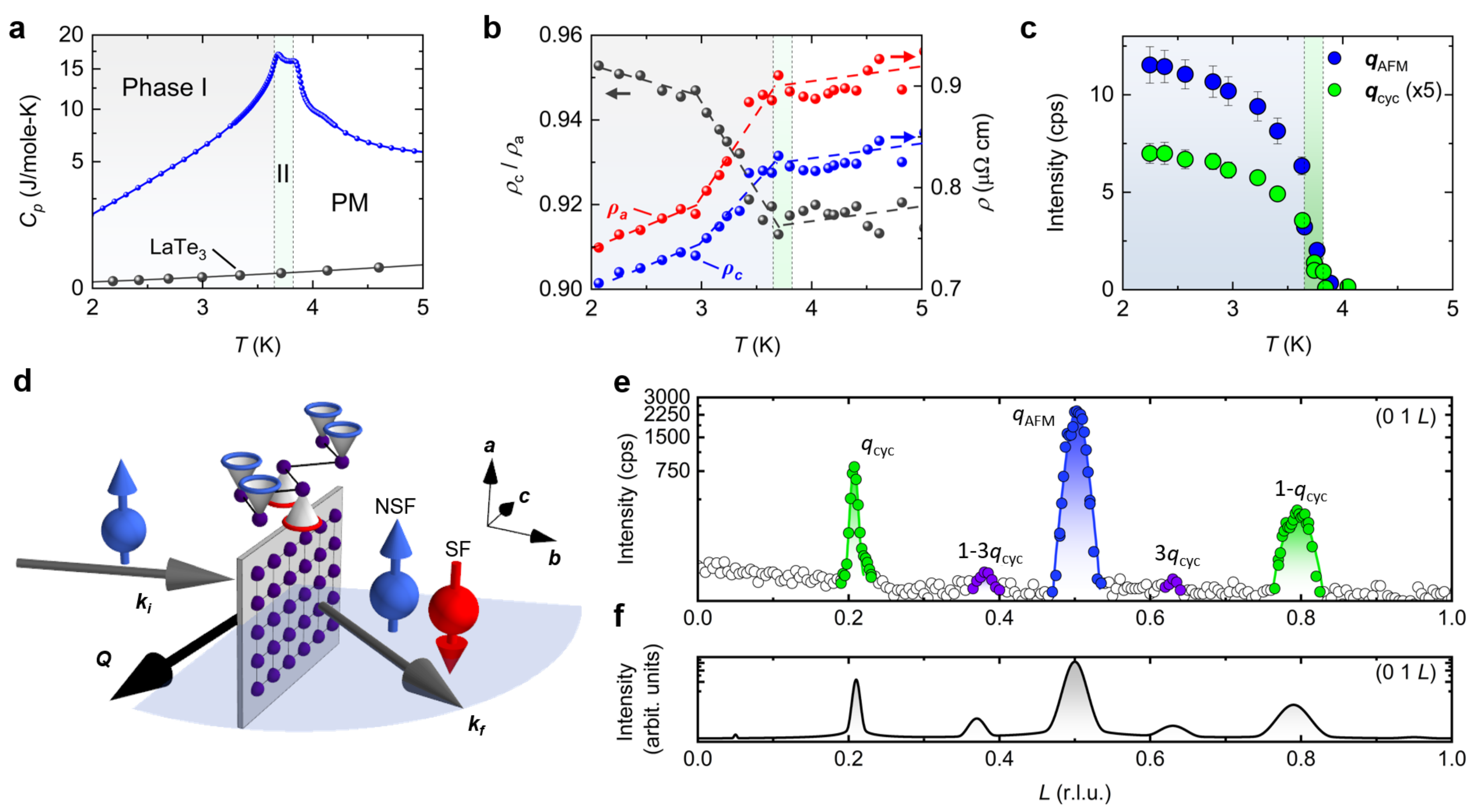}
\caption{\textbf{Bulk characterization and magnetic structure of DyTe$_3$ from elastic neutron scattering.} \textbf{a}, Specific heat showing two transitions at $T_\mathrm{N1}$ and $T_\mathrm{N2}$, with reference data from the nonmagnetic analogue LaTe$_3$~\cite{Shin2005}. \textbf{b}, Below $T_\mathrm{N1}$, where resistivity in the $ac$ basal plane drops sharply, significant anisotropy $\rho_c/\rho_a$ indicates (partial) gapping of electronic states along the $c$-axis. \textbf{c}, Temperature dependence of magnetic scattering intensity for antiferromagnetic $\mathbf{q}_\mathrm{AFM}$ and cycloidal $\mathbf{q}_{cyc}$ reflections, with Miller indices $(HKL) = (0, 1, 0.5)$ and $(0, 1, 0.207)$, respectively. Grey and green background shading mark phases I and II, respectively. \textbf{d}, Experimental geometry for polarized neutron scattering from layered DyTe$_3$ (grey plane, square net of Dy indicated). The scattering plane (blue) is spanned by wavevectors $\mathbf{k}_i$, $\mathbf{k}_f$ of incoming and outgoing neutron beams, respectively. Separating spin-flip (SF, red) and non-spin flip (NSF, blue) scattering intensities at the detector, we identify conical magnetic order (top). \textbf{e}, Linescan in momentum space, with highlights for three types of reflections including a weak higher harmonic corresponding to $3\times\mathbf{q}_{cyc}$. \textbf{f}, Simulation of magnetic intensity from magnetic structure model, corresponding to the linescan in panel \textbf{e} (Methods).}\label{fig2}
\end{figure}

We characterize the phase transition in DyTe$_3$ using thermodynamic and transport probes in Fig. \ref{fig2}\,\textbf{a},\textbf{b}. The specific heat $C(T)$ shows a two-peak anomaly, describing the transitions from the paramagnetic (PM) regime to phase II at $T_\mathrm{N2} = 3.85\,$K and to phase I at $T_\mathrm{N1} = 3.6\,$K. Below $T_\mathrm{N1}$, the resistivities in the $ac$ basal plane drop abruptly, suggesting a clear correlation between the behaviour of freely moving conduction electrons and the magnetic structure. The presence of a partial charge gap in the electronic structure, related to magnetic ordering, is inferred from an increase of the ratio of resistivities $\rho_a$ and $\rho_c$. Simulataneously, as discussed in the following, strong neutron scattering intensity appears below $T_\mathrm{N2}$ at two independent positions in reciprocal space, c.f. Fig. \ref{fig2}\,\textbf{c}. The magnetic scattering intensity rises abruptly upon cooling below $T_\mathrm{N2}$. 

To obtain this neutron data, a single-domain crystal of DyTe$_3$ is mounted on an aluminium holder and is pre-aligned by means of Laue x-ray diffraction. More quantitatively, we determine the crystallographic directions in DyTe$_3$ using the crystallographic extinction rule (Extended Section, Fig. \ref{EfigExtinctionSampleA}). Figure \ref{fig2}\,\textbf{d} describes the geometry of our neutron scattering experiment. The scattering plane that includes the incoming and outgoing neutron beams $\mathbf{k}_i$ and $\mathbf{k}_f$, is spanned by the $b$- and $c$-axes. Hence, reflections of the type $\mathbf{Q} = \mathbf{k}_f-\mathbf{k}_i$ with Miller indices $(0KL)$ can be detected, as in Fig. \ref{fig2}\,\textbf{e}, where a line scan along $(01L)$ provides sharp magnetic intensity. Three types of magnetic peaks $\mathbf{Q} = \mathbf{G}+\mathbf{q}$, with $\mathbf{G}$ a reciprocal lattice vector, are observed: 
A commensurate (C) reflection $\mathbf{q}_\mathrm{AFM} = (0,b^*,q_\mathrm{AFM})$, $q_\mathrm{AFM} = 0.5\, c^*$; an incommensurate (IC) reflection $\mathbf{q}_\mathrm{cyc}=(0,b^*,q_\mathrm{cyc})$, $q_\mathrm{cyc} = 0.207\,c^*$, where $b^* = 2\pi / b$ and $c^* = 2\pi/c$ are reciprocal lattice constants (Methods). There is also a higher harmonic ($3Q$) reflection, corresponding to three times the length of $\mathbf{q}_\mathrm{cyc}$, which describes an anharmonic distortion of the texture. As a main result of this work, we ascribe $\mathbf{q}_\mathrm{cyc}$ to a cycloidal structure in the magnetic ground state of DyTe$_3$, that results from a coupling $q_\mathrm{AFM}\pm q_\mathrm{CDW}$ between the C order and a charge-density wave (CDW) modulation $\mathbf{q}_\mathrm{CDW}$ at $(0,0,q_\mathrm{CDW})$, $q_\mathrm{CDW} = 0.29\,c^*$ in the rare earth tritelluride family.\\

\textbf{Ground state magnetic structure model}\\
We reveal the helimagnetic structure in the ground state of DyTe$_3$ using polarized neutron scattering. As shown in Fig. \ref{fig2}\,\textbf{d}, the incident neutron spins were polarized perpendicular to the scattering plane. We employ a magnetized single-crystal analyzer to select the energy and spin state of the scattered neutrons (Methods). The scattering processes in which the neutron spins are reversed (remain unchanged) is referred to as spin-flip, SF (non-spin-flip, NSF). For SF scattering, it is required that magnetic moments $\mathbf{m}$ have a component perpendicular to the spin of the incoming neutron. This means that SF and NSF scattering detect components of $\mathbf{m}$ within ($m_b$, $m_c$) and perpendicular to ($m_a$) the scattering plane, respectively. 

\begin{figure}[h]%
\centering
\includegraphics[width=0.9\textwidth]{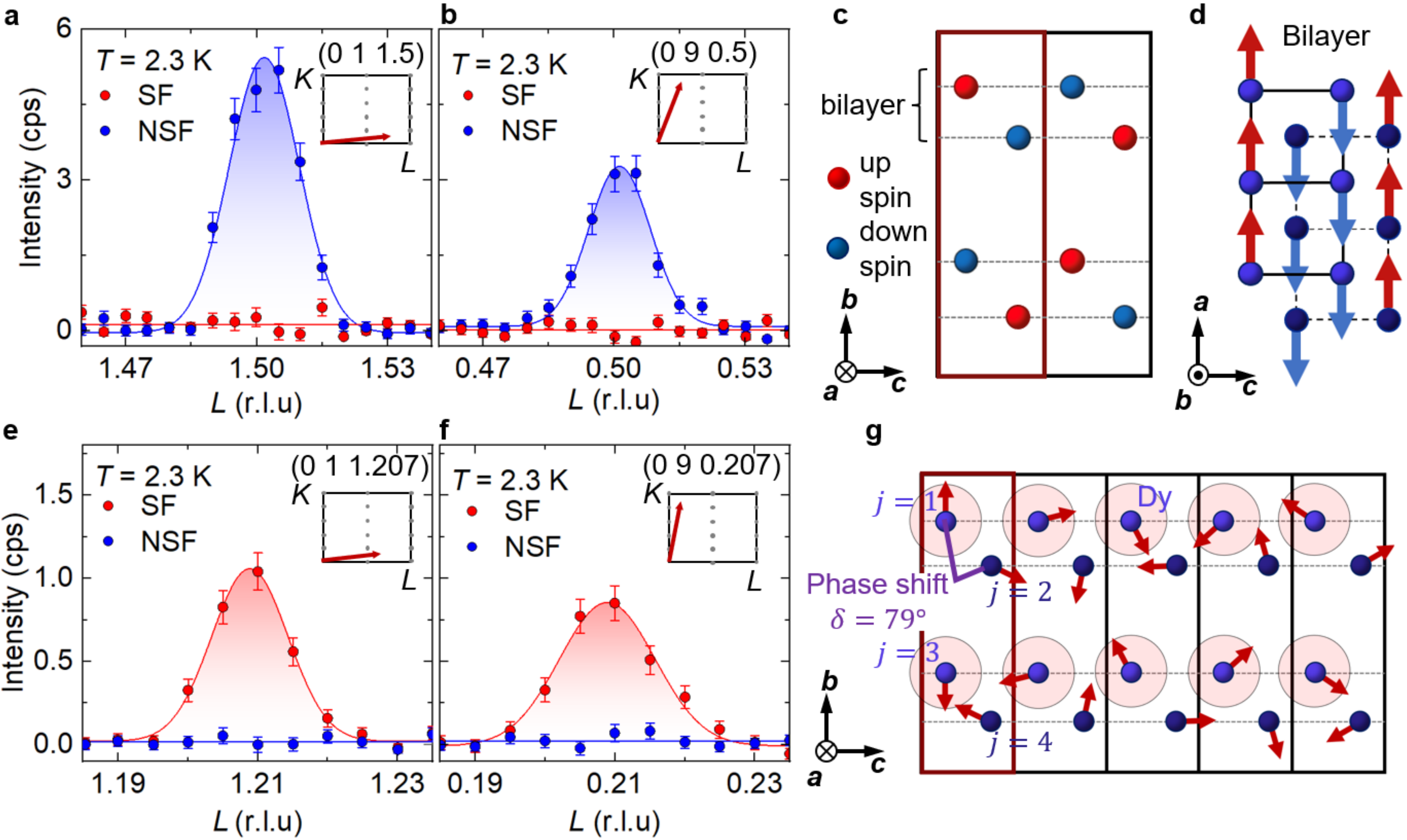}
\caption{\textbf{Two components of the cone-type order in the ground state of DyTe$_\mathbf{3}$.} In the geometry of Fig. \ref{fig2}\,\textbf{d}, non spin-flip (NSF) and spin-flip (SF) neutron scattering measure the magnetic moment along the crystallographic $a$-axis and in the $bc$ plane, respectively. \textbf{a,b}, Polarization analysis of antiferromagnetic component $\textbf{q}_\mathrm{AFM}$: absence of spin flip (SF) intensity indicates absence of $m_b$ and $m_c$, while non-spin flip (NSF) intensity reveals dominant $m_a$. \textbf{c,d}, Two views of the antiferromagnetic collinear component derived from this data, confirmed by full refinement of a number of magnetic reflections in Fig. \ref{EfigMagneticRefinement}. \textbf{e},\textbf{f}, Polarization analysis for the incommensurate cycloidal reflection $\mathbf{q}_\mathrm{cyc}$. \textbf{g}, Derived magnetic structure model for $\mathbf{q}_\mathrm{cyc}$, where $m_a$ (NSF) vanishes while $m_c$ and $m_b$ (SF) both appear at comparable magnitudes in phase I. The index $j$ labels four dysprosium atoms in four layers within the chemical unit cell (red box), where the cycloids at $j = 2, 4$ are phase-shifted with respect to cycloids at $j= 1, 3$. The numerical value of $\delta$ is determined in section \ref{Esec:magnetic_structure_analysis}. As for \textbf{a},\textbf{b},\textbf{e}, and \textbf{f}, we have subtracted background signals, and corrected the effect of imperfect beam polarization.}\label{fig3}
\end{figure}

Polarization analysis of the magnetic reflections shows that $\mathbf{q}_\mathrm{cyc}$ and $\mathbf{q}_\mathrm{AFM}$ relate to different vector components of the ordered magnetic moment (Figs. \ref{fig3}\,\textbf{a},\textbf{b} and \textbf{e},\textbf{f}). We find no hint of SF scattering at $\mathbf{q}_\mathrm{AFM}$, demonstrating collinear antiferromagnetism with magnetic moments exclusively along the $a$-direction. The incommensurate part $\mathbf{q}_\mathrm{cyc}$, in contrast, has no NSF intensity and roughly equal SF signals at various positions in reciprocal space (Fig. \ref{fig3}\,\textbf{e},\textbf{f} and insets). As neutron scattering detects the part of $\mathbf{m}$ that is orthogonal to $\mathbf{Q}$, comparison of magnetic reflections situated at nearly orthogonal directions in momentum space suggests $m_b$ and $m_c$ components are both finite in the ground state.

We determine the quantitative relationship between magnetic moments within a DyTe bilayer (within an effective zigzag chain), by comparing the observed and calculated magnetic structure factors under the constraints imposed by polarized neutron scattering, c.f. Fig. \ref{EfigMagneticRefinement}. The analysis for $\mathbf{q}_\mathrm{AFM}$ demonstrates up-up-down-down type ordering along the zigzag chain, visualized from two perspectives in Fig. \ref{fig3}\,\textbf{c},\textbf{d}. At $\mathbf{q}_\mathrm{cyc}$, the refinement yields a cycloid with a phase delay $\delta$ between the upper and lower sheets in a zigzag chain, see Fig. \ref{fig3}\,\textbf{g}. In effect, pairs of nearly parallel magnetic moments are followed by a significant rotation of the moment direction. The coupling between zigzag chains is antiferromagnetic, as imposed by the Miller index $K=1$ ($q_b = b^*$) \sout{component} in both $\mathbf{q}_\mathrm{cyc}$ and $\mathbf{q}_\mathrm{AFM}$. Superimposing the three components $m_a$, $m_b$, and $m_c$, we realize the noncoplanar, helimagnetic cone texture of Fig. \ref{fig1}\,\textbf{c} that is, to our knowledge, unique in both insulators and metals. In Extended Sections \ref{sec:ESubsec_symmetry_comm}, \ref{sec:ESubsec_symmetry_incomm}, we discuss the presence of magnetic domains in the sample and how the occurrence of higher harmonic reflections further supports our magnetic structure model.\\

\textbf{Charge density wave and magnetic order}\\
We now argue that cone-type magnetism in DyTe$_3$ is realized through (i) a spatial modulation of near-neighbour exchange interactions $J_1$, $J_2$ in presence of charge-density wave (CDW) order and (ii) unconventional single-ion anisotropy. We turn first to (i), that is the role of the CDW in stabilizing noncoplanar helimagnetism in DyTe$_3$. We use a 1D chain model to reproduce key features of the modulated magnetic order, neglecting the material's three-dimensionality (Methods). 

\begin{figure}[h]%
\centering
\includegraphics[width=0.98\textwidth]{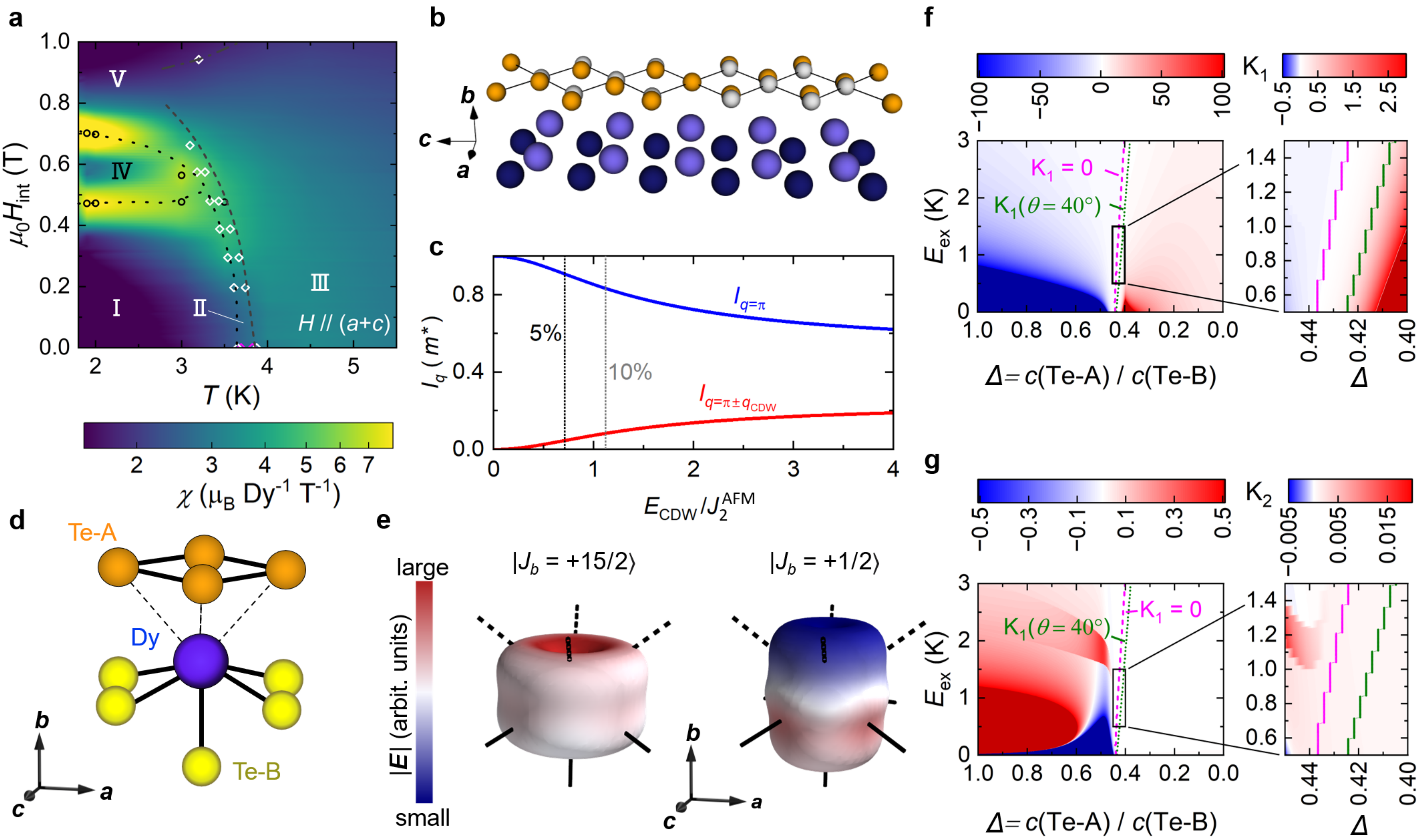}
\caption{\textbf{Modeling helimagnetism in DyTe$_3$.} \textbf{a}, Magnetic susceptibility $\chi$ for $\mathbf{H}\parallel[101]$, overlaid with phase boundaries from specific heat (white open circles) and magnetization (black circles): I (conical ground state), II, III, IV, and V. \textbf{b}, Charge density wave (CDW) on tellurium layers~\cite{Malliakas2006}, where orange (grey) spheres are distorted (undistorted) ionic positions. Coupling of Dy (violet) and CDW drives lattice-incommensurate magnetic order through spatially modulated modification of the local environment of Dy-ions. \textbf{c}, Squared moment amplitudes $I_\mathrm{AFM} = \left| S^a(q= 0.5\,c^*)\right|^2$ (blue) and $I_\mathrm{cyc} = \left| S^b(q= 0.5\,c^*\pm q_\mathrm{CDW})\right|^2+\left| S^c(q= 0.5\,c^* \pm q_\mathrm{CDW})\right|^2$ (red) as functions of $E_\mathrm{CDW}/J_2^\mathrm{AFM}$, where $E_\mathrm{CDW}=E^{ab}_\mathrm{CDW}=E^{ac}_\mathrm{CDW}$, from model calculations according to Eq. (\ref{eq:main_hamiltonian}); five and ten percent threshold for the incommensurate part indicated by black and grey dashed lines (Methods).
\textbf{d}, Local environment of Dy in DyTe$_3$, with covalent (metallic) bonds to Te-B (Te-A) depicted by solid (dashed) lines, respectively. \textbf{e}, Charge density (CD) of Dy $4f^9$ shell under the influence of crystal electric fields (CEF) from the surrounding ions, with exaggerated non-spherical part. If metallic and covalent bonds cause CEF of roughly equal strength (if metallic bonds are screened), oblate $\left|J_b = \pm 15/2\right>$ (prolate $\left|J_b = \pm 1/2\right>$) is the lowest energy CEF doublet. This favours out-of-plane (in-plane) magnetization, respectively. Colour on CD isosurfaces indicates amplitude $\left|\mathbf{E}\right|$ of the local CEF. \textbf{f},\textbf{g} Anisotropy constants $K_1$ and $K_2$ calculated for $4f^9$ multiplet in DyTe$_3$ (Methods). The abscissa describes the relative weight of CEF charges $c$ on Te-A and Te-B sites. Green lines bound an intermediate regime of weak $K_1$, where spin tilting and conical order are allowed.
}\label{fig4}
\end{figure}

In DyTe$_3$, the local environment and bond characteristics of dysprosium ions in a DyTe square net bilayer (in a zigzag chain) are spatially modulated by the CDW in the adjacent Te$_2$ sheets, c.f. Fig. \ref{fig4}\,\textbf{b}~\cite{Shin2005, Malliakas2006, Ru2008, Zocco2011, Schmitt2011, Zocco2015, Chillal2020, Dolgirev2020, Kogar2020, Gonzalez2022, Straquadine2022}. The simplest model approach is to decouple the zigzag chain, with two atoms per unit cell, into two one-dimensional chains, with one atom per unit cell. This allows for a two-parameter model, built from Ising-like exchange interactions together with a spatially modulated onsite coupling,
\begin{equation}
\label{eq:main_hamiltonian}
\mathcal{H} = \sum_{n}\left[J_2^\mathrm{AFM} S_{n}^aS_{n+1}^a - E^{ab}_\mathrm{CDW}\cos\left(q_\mathrm{CDW}z_{n}\right)S_{n}^a S_{n}^b- E^{ac}_\mathrm{CDW}\sin\left(q_\mathrm{CDW}z_{n}\right)S_{n}^a S_{n}^c\right]
\end{equation}
where $n$ counts magnetic sites, e.g., on the \textit{upper half} of the zigzag chain. The $z_{n}$ are spatial positions along the zigzag chain ($c$-axis). All the coupling constants -- $J_2^\mathrm{AFM}$, $E^{ab}_\mathrm{CDW}$, and $E^{ac}_\mathrm{CDW}$ -- are positive. The $E^{ab}_\mathrm{CDW}$ and $E^{ac}_\mathrm{CDW}$ terms are allowed by global and local mirror symmetry breaking due to the CDW, respectively. %, an oscillatory $S_n^b$ can be induced by normalizing the spin length, and we 
We may also introduce a Zeemann term to explain the behavior in a magnetic field and further inter-chain coupling to connect the two chains (Extended Section \ref{sec:ESec_Spin_Hamiltonian}).

This 1D model naturally creates different modulation period for the $a$ and $bc$ spin components and robustly reproduces two types of magnetic reflections, $q_\mathrm{AFM}/c^* = 0.5$ and $q_\mathrm{cyc}/c^* = 0.5 - 0.293 = 0.207$. In good consistency with experiment, Fig. \ref{fig4}\,\textbf{c} shows that $I_\mathrm{cyc}$ on the order of 10~\% of $I_\mathrm{AFM}$ can be induced within this model. Based on scattering techniques, we find it difficult to reveal the phase-shift between CDW and the spin cycloid, and between the antiferromagnetic and incommensurate components of the magnetic order; hence, alternative (out-of-phase) locking between cycloid and antiferromagnetic component is also possible  (Fig. \ref{EfigMstructure}). \\

\textbf{Weak magneto-crystalline anisotropy}\\
While strongly anisotropic magnetism is naively expected for dysprosium's $^6H_{15/2}$ shell, we here report a conical state with comparable $m_a$, $m_b$, $m_c$ in DyTe$_3$. Consider the local environment of a single Dy in Fig. \ref{fig4}\,\textbf{d}: Te-B ions form covalent bonds with the central Dy, while the point charges of Te-A are effectively screened by itinerant electrons in the conducting tellurium slab. We model the sequence of crystal electric field (CEF) states for the $4f^9$ shell of dysprosium as a function of the effective crystal field charge $c$ situated on Te-A and Te-B ions (Fig. \ref{EfigCEFvsCharge}). Fig. \ref{fig4}\,\textbf{e} illustrates two limiting cases: When Te-A and Te-B contribute equally to the CEF, the $4f^9$ charge cloud is compressed along the $b$-direction, with $\left|J_b=\pm 15/2\right>$ dominating the ground state wavefunction, and with effective out-of-plane magnetic anisotropy for magnetic moments. Likewise, zero contribution of Te-A, i.e. highly efficient metallic screening of CEFs, favours the prolate orbital $\left|J_b=\pm 1/2\right>$ with easy-axis anisotropy.

Adding exchange interactions $E_{ex}$ as an effective magnetic field, the CEF Hamiltonian of a point charge model is diagonalized for the orthorhombic environment of Dy (Methods). The resulting free energy density described by two parameters $K_1 \cos^2(\theta)+K_2\sin^2(\theta)\cos^2(\phi)$, where $\theta$, $\phi$ are spherical coordinates with respect to the $b$ and $c$ crystal axes, respectively. Fig. \ref{fig4}\,\textbf{f},\,\textbf{g} testify to a transition from easy-axis to easy-plane anisotropy through a sign change of $K_1$ at intermediate charge ratio (pink line)two green lines bound the regime where easy-axis (easy-plane) anisotropy is not strong enough to prevent tilting of $\mathbf{m}$ along directions intermediate between $b$-axis and the $ac$ plane. Constraining $E_{ex}$ in agreement with $T_\mathrm{N2}$ and requiring easy-plane anisotropy $K_1>0$, we identify the black box in Fig. \ref{fig4}\,\textbf{f},\,\textbf{g} to capture a parameter range well consistent with experiment. Here, the model yields $K_2>0$, meaning $m_a$ is preferred over $m_c$.\\

\textbf{Magnetic phase diagram and small-angle neutron scattering}\\
We are ready, now, to consider the evolution of magnetic order in DyTe$_3$ as a function of temperature and magnetic field. Figure \ref{fig4}\,\textbf{a} shows a contour map of the magnetic susceptibility $\chi$ (Methods), where the external magnetic field is applied along the in-plane direction $[101]$, i.e. $\mathbf{H} \parallel (a+c)$. Heating the sample above $T_\mathrm{N1} = 3.6\,$K in zero field, we observe a peak splitting of $\mathbf{q}_\mathrm{AFM}$, and a concomitant shift in $\mathbf{q}_\mathrm{cyc}$ that indicates the sustained coupling of the two ordering vectors, via the CDW, at elevated temperatures (Fig. \ref{EfigPolSampleA}). 
The sharp enhancement of $\chi_c$ in Fig. \ref{fig1}\,\textbf{a} further suggests that $m_a$, $m_b$ survive to higher temperature than $m_c$, consistent with a putative incommensurate, fan-like order in phase II, which warrants further study. 

\begin{figure}[h]%
\centering
\includegraphics[width=1.\textwidth]{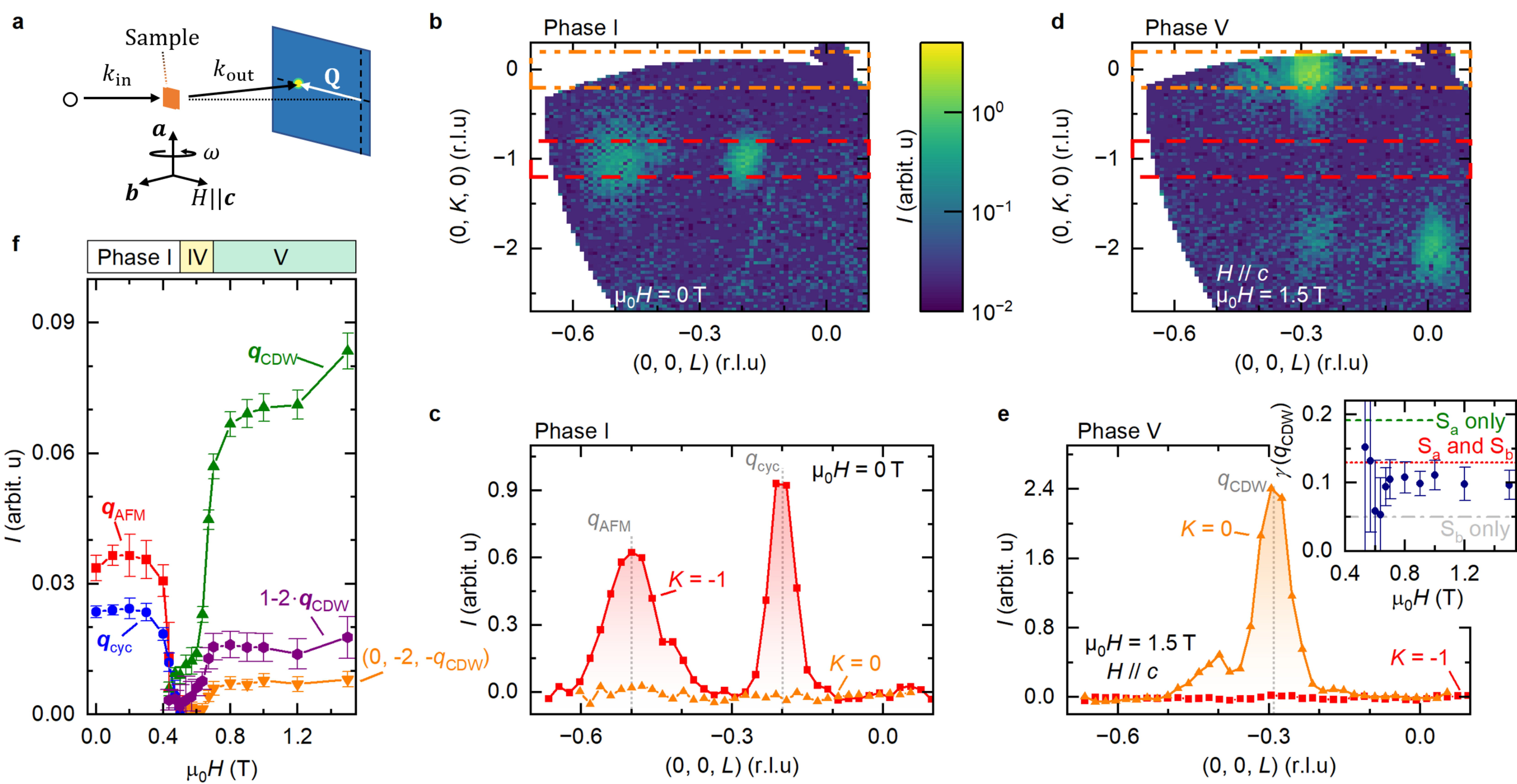}
\caption{\textbf{Coupling of magnetic order and charge-density wave in DyTe$_3$, tracked by small-angle neutron scattering (SANS) in a magnetic field along the $c$-axis at $T=2\,\mathrm{K}$.} \textbf{a}, Experimental geometry of SANS. Orange plate, blue rectangle, $\mathbf{k}_\mathrm{in}$, $\mathbf{k}_\mathrm{out}$, and $\mathbf{Q}$ are the sample, the area detector -- shifted from the center axis to catch reflections with larger $\left|\mathbf{Q}\right|$ --, incoming / outgoing neutron beam wavevector, and the momentum transfer of elastic scattering, respectively. A wedge of three-dimensional $\mathbf{Q}$ space is covered by rotating the sample in steps (angle $\omega$, Methods). \textbf{b}-\textbf{e}, Scattering intensity in $(0KL)$ cuts of momentum space (upper) and linecuts along $(00L)$ (lower), for intensity integrated along the $K$-direction within the orange and red dashed boxes. The magnetic field induces a transition to ferroic stacking, $K=0$. \textbf{f}, Integrated intensity of various reflections as a function of magnetic field, covering three magnetic phases. Inset of (e): Experimental ratio $\gamma=I(0, -2, -0.293)/ I(0, 0, -0.293)$ in Phase V, compared to simulations for fan-like orders (spin plane indicated, green and grey dashed lines) and a helimagnetic (longitudinal-cone, red dashed line) order. The experiments are consistent with a slightly distorted, conical state of the same period as the underlying charge order.}\label{fig5}
\end{figure}

To explore the regime above the critical field, confirm the coupling between $\mathbf{q}_\mathrm{AFM}$ and $\mathbf{q}_\mathrm{cyc}$ in phase I, and investigate the correlation of CDW and magnetic order, we carried out small angle neutron scattering (SANS) experiments in a magnetic field. Figure \ref{fig5}\,\textbf{a} describes the geometry of our SANS experiment and Fig. \ref{fig5}\,\textbf{b} shows the obtained zero field $(0,K,L)$ map, while Fig. \ref{fig5}\,\textbf{c} reduces the map into principal line cuts. The noncoplanar, helimagnetic cone texture with coupled $\mathbf{q}_\mathrm{AFM}$ and $\mathbf{q}_\mathrm{cyc}$ is stable up to $\mu_0H = 0.5\,\mathrm{T}$ for $\mathbf{H} \parallel c$, c.f. Fig. \ref{fig5}\,\textbf{f}. In fact, this data further supports the existence of coupled commensurate and incommensurate order parameters in phase I, and helps to exclude a domain separation scenario (Fig. \ref{EfigSANSIRatio}). A pair of phase transitions to phases IV and V is visualized in Fig. \ref{fig5}\,f and Fig. \ref{EfigSANSIRatio}. In phase V, c.f. Fig. \ref{fig5}\,\textbf{d} and \textbf{e}, which is realized when the external magnetic field exceeds $\mu_0H = 0.7\,\mathrm{T}$, strong magnetic reflections appear at momentum transfer $\mathbf{Q} = (Ha^*,Kb^*,q_\mathrm{CDW})$ with $K = \,$ even, demonstrating direct coupling between incommensurate magnetism and the CDW in absence of the antiferromagnetic order parameter. Although we cannot provide a magnetic structure model based on the available data, the comparison of intensities at Miller indices $K = 0, 2$ suggests the presence of both $m_a$ and $m_b$ spin components (inset of Fig. \ref{fig5}\,\textbf{e}).\\

\textbf{Discussion}\\
As compared to transition metal dichalcogenides, where the magnetic ion is buried inside a rather symmetric block layer~\cite{Dickinson1923, Manzeli2017}, $R$Te$_3$ harbors more complex structural features, with magnetic ions at the boundary between metallic and covalently bonded blocks. This mixed covalent / metallic environment for the magnetic ion is key to realizing the present scenario: it facilitates coupling between magnetic ions and a charge density wave (CDW) on the tellurium square net, and -- at the same time -- generates unconventional magnetocrystalline anisotropy. In fact, the present charge-transfer phenomenology is partially inspired by work on thin films of magnetic metals on insulating substrates~\cite{Pollmann2011}, on electric field control of magnetocrystalline anisotropy~\cite{Torun2015}, and on the behaviour of magnetic materials when charge transfer is induced by oxidation at the surface~\cite{Gambardella2009}. 

Symmetry breaking with cycloid / spiral magnetic order of fixed helicity is rather widely observed in zigzag chain magnets (Extended Section \ref{sec:ESubsec_symmetry_incomm}), but the present combination of antiferromagnetic and cycloidal components is unique. For example, Mn$_2$GeO$_4$ has cones arrayed on one-dimensional chains, with uniform cone direction along the chain~\cite{Honda2017}. In another metallic system, EuIn$_2$As$_2$, jumps in the rotation sense of a helimagnetic texture have recently been identified, with a short magnetic period~\cite{Riberolles2021}. In contrast, the rotation of moments in DyTe$_3$ proceeds in nearly parallel pairs, without abrupt jumps in the cycloidal component of the texture. We expect helimagnetic orders of the type observed here to be common in layered materials, and especially in rare earth tellurides and selenides. Here, rich magnetic phase diagrams have been generally observed~\cite{Lei2019, Lei2021} and could be amenable to modeling by CDW-induced terms as in Eq. (\ref{eq:main_hamiltonian}). 

This complex magnetic order, its relationship to a strain-controllable CDW~\cite{Straquadine2022}, and its (likely) rich excitation spectrum certainly warrant further research. For example, the CDW's gapped collective excitation, termed Higgs mode, shows a magnetic character in $R$Te$_3$ as observed via Raman scattering experiments~\cite{Wang2022_RTe3}, and its evolution below $T_\mathrm{N1}$ may provide insights on both the origin of magnetic order and the nature of the CDW in DyTe$_3$. Furthermore, the lowest-energy, Goldstone mode of a typical helimagnet corresponds to a spatial shift of the magnetic texture, termed phason excitation~\cite{Gruner1988}. In DyTe$_3$, the magnetic and CDW phasons~\cite{Sinchenko2012} are expected to be closely intertwined, as evident from the robust $q_\mathrm{cyc}(T)$ in phase I, and its jump -- by the same amount as $q_\mathrm{AFM}$ -- in phase II (Fig. \ref{EfigPolSampleA}). Such locking between low-energy modes may have implications for dynamic responses, further enriching the spectrum of elementary excitations in $R$Te$_3$.\\

\textbf{Concluding remarks}\\
An important open question is the stability of helimagnetism in few-layer devices of DyTe$_3$, where the cleavage plane, as well as the center of structural inversion, are situated between tellurium bilayers. As a fundamental building block of the structure, we consider a DyTe slab sandwiched by Te square nets -- that is half a unit cell in Fig. \ref{fig1}\,\textbf{a}. Being screened from top and bottom by tellurium layers, we expect no qualitative change of the local crystal field environment of Dy in the few-layer limit. However, the absence of an inversion center for odd numbers of layers, and its presence for even numbers of layers, may have a profound effect on magnetic ordering and the presence or absence of (right- or left-handed) helicity domains in the sample, considering the presence or absence of Dzyaloshinskii-Moriya interactions~\cite{Moriya1960}. 

Most appealingly, DyTe$_3$ is a potential platform for spin-Moir{\'e} engineering in solids, where complex magnetic textures can be designed by combining and twisting two or more helimagnetic sheets. Here, a plethora of noncoplanar spin textures can be engineered at will~\cite{Shimizu2021, Ghader2022}, while highly conducting tellurium square net channels may serve as a test bed for of emergent electromagnetism in a tightly controlled setting~\cite{Volovik1987, Tokura2018}.

%\bibliography{sn-bibliography}

%%%%%%%%%%%%%%%%%%%%%%%%%%%%%%%%%%%%%%%%%%%%%%%%%%%%%%%%%%%%%%%%%
%%%%%%%%%%%%%%%%%%%%%%%%%%%%%%%%%%%%%%%%%%%%%%%%%%%%%%%%%%%%%%%%%
%%                   Bibliography START
%%%%%%%%%%%%%%%%%%%%%%%%%%%%%%%%%%%%%%%%%%%%%%%%%%%%%%%%%%%%%%%%%
%%%%%%%%%%%%%%%%%%%%%%%%%%%%%%%%%%%%%%%%%%%%%%%%%%%%%%%%%%%%%%%%%


%%%%%%%%%%%%%%%%%%%%%%%%%%%%%%%%%%%%%%%%%%%%%%%%%%%%%%%%%%%%%%%%%
%%%%%%%%%%%%%%%%%%%%%%%%%%%%%%%%%%%%%%%%%%%%%%%%%%%%%%%%%%%%%%%%%
%%                   Bibliography END
%%%%%%%%%%%%%%%%%%%%%%%%%%%%%%%%%%%%%%%%%%%%%%%%%%%%%%%%%%%%%%%%%
%%%%%%%%%%%%%%%%%%%%%%%%%%%%%%%%%%%%%%%%%%%%%%%%%%%%%%%%%%%%%%%%%

\newpage
%%%%%%%%%%%%%%%%%%%%%%%%%%%%%%%%%%%%%%%%%%%%%%%%%%%%%%%%%%%%%%%%%%%%%%%
%%%%%%%%%%%%%%%%%%%%%%%%%%%%%%%%%%%%%%%%%%%%%%%%%%%%%%%%%%%%%%%%%%%%%%%
%%				METHODS
%%%%%%%%%%%%%%%%%%%%%%%%%%%%%%%%%%%%%%%%%%%%%%%%%%%%%%%%%%%%%%%%%%%%%%%
%%%%%%%%%%%%%%%%%%%%%%%%%%%%%%%%%%%%%%%%%%%%%%%%%%%%%%%%%%%%%%%%%%%%%%%
\begin{center}
\Large{Methods}
\end{center}

\textbf{Sample preparation and characterization}\\
Single crystals are grown from tellurium self-flux following the recipe in Ref.~\cite{Lei2021}: 
We set elemental Dy and Te at a ratio of $1:21.65$ in an alumina crucible, which in turn is sealed in a quartz tube in high vacuum. The raw materials are heated to $450\,^\circ\mathrm{C}$ for $36\,$hours and then to $780\,^\circ\mathrm{C}$ in $96\,$hours, where the melt remained for $48\,$hours, followed by cooling to $450\,^\circ\mathrm{C}$ at a rate of  $1.375\,^\circ\mathrm{C}/\mathrm{hour}$. 
The final product is centrifuged after renewed heating to $500^\circ$ Celsius, so that plate-shaped single crystals of typical dimensions $5\times5\times1\,\mathrm{mm}^3$ are obtained. The face of each plate is perpendicular to the $b$-axis of DyTe$_3$'s orthorhombic unit cell, and facet edges tend to be parallel to either $a$ or $c$. The existence of impurity phases above $1\,\%$ volume fraction is ruled out by single-crystal x-ray diffraction on cleaved surfaces in a Rigaku SmartLab X-ray powder diffractometer. The experiment yields lattice constants of $a = 4.27(2)\,\text{\AA}$, $b = 25.433(1)\,\text{\AA}$, and $c = 4.27(2)\,\text{\AA}$ at room temperature, in good agreement with previous work~\cite{Malliakas2006}. We found it challenging to obtain high-quality powder x-ray data from crushed single crystals, which include traces of Te flux on their surface and form thin flakes, even when thoroughly ground in a mortar. We also verified the stoichiometric chemical composition of our crystals by energy-dispersive x-ray spectroscopy (EDX). Cleaved single crystals have a reddish-brown surface; but even in vacuum, the colour of the surface changes to silver-metallic, and then to black, after two weeks or so. A red hue can be recovered by renewed surface cleaving.\\

\textbf{Magnetization measurements and crystal alignment}\\
We use a commercial magnetometer with $T = 2\,\mathrm{K}$ base temperature and a maximum magnetic field of $7\,\mathrm{T}$ (MPMS, Quantum Design, USA). 
The measurement is carried out using a rectangular-shaped single crystal of mass $m = 1.22\,\mathrm{mg}$, with carefully aligned edges along the $a$ and $c$ crystal axes. By means of a single crystal diffractometer (Malvern Panalytical Empyrean, Netherlands), we confirm the extinction rule $h+k=\mathrm{even}$ in space group $Cmcm$.
It is difficult to distinguish $a$ and $c$ axes in this orthorhombic, yet nearly tetragonal structure by eye or with the help of the Laue diffractogram. 
Temperature dependent susceptibility $\chi(T)$ is measured in a DC magnetometer with $1000\,$Oe applied field; there is no observable difference between field-cooled and zero-field cooled magnetization traces. 
A demagnetization correction is carried out according to the standard expression $\mathbf{H}_\mathrm{int} = \mathbf{H}_\mathrm{ext}-N\mathbf{M}$, where $\mathbf{H}_\mathrm{ext}$, $\mathbf{M}$, and $N$ are the externally applied magnetic field, the bulk magnetization, and the dimensionless demagnetization factor. 
The latter is calculated by approximating the crystal as an oblate ellipsoid~\cite{Osborn1945}. 
For the $H-T$ phase diagram in Fig. \ref{fig4}, the $[101]$ direction is aligned within $\pm3^\circ$ and bulk magnetization is measured in discrete field steps, for selected temperatures. 
Fig. \ref{fig4} shows data for decreasing magnetic field $\partial H/\partial t < 0$. 
Note that hysteresis occurs at all phase transitions shown in Fig. \ref{fig4}\,\textbf{a}, indicating their first-order nature.
The magnetic anisotropy energy is expressed as $E / V_\mathrm{uc}=K_0+K_1\cos^2(\theta)+\mathcal{O}[\sin^4(\theta)]$, where $\theta$ is the angle between $\mathbf{M}$ and the $b$-axis. Utilizing the free energy expression $F = F_0+\sum_{\alpha}{a_i(T)M_\alpha^2}+\mathcal{O}(\mathbf{M}^4)$ and the Curie-Weiss law $\chi_\alpha = C / (T-\Theta_\mathrm{CW}^\alpha)$, where $\Theta_\mathrm{CW}^\alpha$ is the Curie-Weiss temperature along the $\alpha\in(a,b,c)$ direction and $C=2.077(1)\,\mathrm{K}$ is the Curie constant of DyTe$_3$, we obtain $K_1 = (\mu_0 / 2)\left(M /V_\mathrm{uc}\right)^2(\Delta\Theta_\mathrm{CW}/C) = 22(1)\,\mathrm{kJ}\,\mathrm{m}^{-3}$, where $\Delta\Theta_\mathrm{CW} = \Theta_\mathrm{CW}^{ac}-\Theta_\mathrm{CW}^{b} = -2.0(1)\,\mathrm{K}$ is the difference between the Curie-Weiss temperatures in the $ac$-plane and along the $b$-axis (c.f. Fig. \ref{fig1}\,\textbf{d}, inset).
Specific heat was recorded using a relaxation technique in a Quantum Design PPMS cryostat, in zero magnetic field. For specific heat anomalies in applied magnetic field, we employed the AC calorimetry technique in a custom-built setup. Anisotropy of the resistivity, as in Fig. \ref{fig2}, was recorded on exfoliated flakes of thickness $\sim 100\,\mathrm{\mu m}$ using the Montgomery technique. Electric contacts are made with Ag paste (Dupont) and deteriorate with time. To maintain excellent contact resistance $\sim1\,\mathrm{\Omega}$, it is crucial to immediately cool the contacted crystal in vacuum, after depositing the silver paste. The sample and contact quality is robust at low temperatures for at least two weeks.\\

\textbf{Elastic neutron scattering}\\
We performed unpolarized and polarized neutron scattering experiments using the POlarized Neutron Triple-Axis spectrometer (PONTA) installed at the 5G beam hole of the Japan Research Reactor 3 (JRR-3). Two single crystals of DyTe$_3$ (Sample A and B) are cut into rectangular shapes with dimension $3.6\times 2.7 \times 0.9\,$mm and $1.7 \times 1.9 \times 0.8\,$mm, respectively. For both samples, the widest surface is normal to the $b$-axis, and the sides are parallel to the $a$ or $c$-axis. Each sample is set in an aluminum cell, which is sealed with $^4$He gas for thermal exchange. We employed a $^4$He closed-cycle refrigerator with base temperature of $2.2\,$K, and measured intensities on the $(0,K,L)$ horizontal scattering plane. Using a PG $(002)$ monochromator, the energy of the incoming neutron beam is set to $E_i=14.7\,$meV ($30.5\,$meV) for upolarized measurements of sample A (sample B). For the unpolarized measurements, the spectrometer is operated in two-axis mode with horizontal beam collimation of open-$80^\prime$-$80^\prime$. In both unpolarized and polarized experiments, sapphire and pyrolytic graphite (PG) filters are installed between the monochromator and the sample, to suppress higher-order reflections from the monochromator to less than $0.5\,\%$. The observed integrated intensities are converted to structure factors after applying the Lorentz factor and absorption corrections.

For sample B, we measured nuclear and magnetic Bragg reflections at $2.2\,$K by $\theta-2\theta$ scans. For the scattering profiles showing a well-defined Gaussian-shape peak, we estimated the background from both ends of the profile. For the magnetic reflections located near the powder diffraction lines of the Al sample holder, we carried out background scans at $10\,$K, and subtracted the intensities from those measured at $2.2\,$K. We also measured the background data at $10\,$K for relatively weak commensurate magnetic reflections in the $Q$-range of $\left|Q\right|>4.0\,\mathrm{\AA}^{-1}$, to check for possible $\lambda/2$ contamination from the nuclear reflections. As for the absorption correction, we calculated the scattering path length $l$ inside the sample, based on the dimensions of Sample B and on the incident and scattered directions of the neutrons. The neutron transmission is given by $\exp(-\mu l)$, where $\mu$ is the linear absorption coefficient. Taking into account the incident energy and the absorption and incoherent scattering cross-sections of DyTe$_3$, $\mu$ is calculated to be $8.392\,\mathrm{cm}^{-1}$.

The diffraction profiles and integrated intensities shown in Figs. \ref{fig2}, \ref{fig3} were measured using Sample A. Contrary to integrated intensities in the case of refinement, the temperature dependences in Figs. \ref{fig2}\,\textbf{c}, \ref{EfigPolSampleA} are obtained from $L$-scans of magnetic scattering. The calculation of the scattering intensity in Fig. \ref{fig2}\,\textbf{f}, which includes the third harmonic reflection, takes into account instrumental resolution broadening (Fig. \ref{EfigCalibrationPONTA}), anharmonicity of the cycloidal magnetic structure component, and the presence of two magnetic domains (Extended Section \ref{sec:Esec_structure_factors_incomm}). 

Sample A is also used for polarized neutron scattering, in which the spectrometer is operated in the triple-axis mode with horizontal beam collimation of open-$80^\prime$-$80^\prime$-open. A polarized neutron beam with $E_i=13.7\,$meV is obtained by a Heusler (111) crystal monochromator. The spin direction of incident neutrons is set to be perpendicular to the scattering plane. We thus applied weak vertical magnetic fields of approximately $5\,$mT throughout the beam path by guide magnets and a Helmholtz coil. We used a Mezei-type $\pi$ spin flipper placed between monochromator and sample, and employed a Heusler (111) crystal analyzer to select the energy and spin states of scattered neutrons, separating spin-flip (SF) and non-spin flip (NSF) intensities. The spin polarization of the incident neutron beam ($P_0$) is $0.823$, as measured using the $(002)$ nuclear Bragg reflection of the sample. \\

\textbf{Small angle neutron scattering in magnetic field}\\
SANS measurements were performed using the SANS-I instrument at Paul Scherrer Institute (PSI), Switzerland. A bulk single crystal of DyTe$_3$ (Sample E, $m=73.4\,\mathrm{mg}$) was carefully aligned (c.f. crystal alignment methods) and installed into a $1.8\,\mathrm{T}$ horizontal-field cryomagnet so that the $a$-axis is vertical, and the incident neutron beam is in the $bc$-plane. The magnetic field is applied parallel to the crystal $c$-axis, as shown in Fig. \ref{fig5}\,\textbf{a}. The incident neutron beam with $\lambda = 3.1\,\text{\AA}$ wavelength (15\% $\Delta\lambda / \lambda$) is collimated over a distance of $4.5\,\mathrm{m}$ before the sample, and the scattered neutrons are detected by a 1~m$^2$ two-dimensional multidetector (pixel size 7.5mm x 7.5mm) placed $1.7\,\mathrm{m}$ behind the sample. To cover a broader $q$-space up to $q_\mathrm{AFM}$ along the $(01L)$ direction, the detector was also translated $0.45\,\mathrm{m}$ in the horizontal plane. For all SANS data, background signals from the sample and the instrument are subtracted using the data of the nonmagnetic state at $T=10\,\mathrm{K}$ and $\mu_0H=0\,\mathrm{T}$. The field-dependent SANS measurements are performed during a field-increasing process, after an initial zero-field cool to the base temperature of $2\,\mathrm{K}$. For each measurement, rocking scans were performed, i.e. the cryomagnet is rotated together with the sample around the vertical crystal $a$-axes (rocking angle $\omega$) in a range from $-102^\circ$ to $55^\circ$ and steps of $2^\circ$ ($-38^\circ\leq\omega\leq28^\circ$) and $1^\circ$ (else). Here $\omega=0^\circ$ is carefully aligned and corresponds to the configuration where the beam $\mathbf{k}_\mathrm{in}$ is parallel to the crystallographic $b$-axis. The SANS maps shown in this paper are obtained by performing a 2D cut of the volume of reciprocal space measured through cumulative detector measurements taken at each angle of the rocking scan. For the SANS maps shown in Fig. \ref{fig5}, the integration width along the out-of-plane $(H 0 0)$ direction is $\pm0.15$ reciprocal lattice units. The line cuts along $(0K_\mathrm{fix}L)$ shown in Fig. \ref{fig5}\,\textbf{c}, \textbf{e} are extracted by integrating over a region of $\pm0.2$ reciprocal lattice units in the $(0K0)$ direction. Peak positions and integrated intensities are calculated using those linecuts and a multi-peak fitting.
\\

\textbf{Crystal electric field calculations}\\
We use the software package PyCrystalField~\cite{Scheie2021} for the calculation of crystal electric field energies via the point charge model in the limit of strong spin-orbit interactions. The calculation is based on published fractional coordinates of Dy and Te ions within the crystallographic unit cell~\cite{Malliakas2006, Ru2008}, with a $\sim0.15\,\%$ tensile strain along the $a$-axis, lifting tetragonal symmetry and yielding finite $K_2$. In Fig. \ref{EfigCEFvsCharge}, we vary the effective crystal electric field originating from Te-A (on the Te$_2$ slab) by changing its point charge, while keeping the total charge in the environment of Dy unchanged. An unperturbed, diagonal Hamiltonian matrix is constructed from the energies in Fig. \ref{EfigCEFvsCharge}, and the operator of total angular momentum $\mathbf{J} = \mathbf{L}+\mathbf{S}$ is also expressed in the basis of these CEF eigenstates. Adding an effective exchange field $E_{ex}J_\alpha$ ($\alpha = a, b, c$ are vector components), the total Hamiltonian is diagonalized and the expectation value of $J_a$, $J_b$, $J_c$ is evaluated in the respective ground state. The anisotropy constants are approximated, as
\begin{align}
K_1 &\propto \frac{1}{\left<J_b\right>_b^2} - \frac{1}{\left<J_a\right>_a^2}\\
K_2 &\propto \frac{1}{\left<J_c\right>_c^2} - \frac{1}{\left<J_a\right>_a^2}
\end{align}
so that $K_1<0$ for easy-axis anisotropy along the $b$-axis and $K_2>0$ if $a$-axis orientation is energetically preferred over the $c$-axis. Here, $\left<J_\alpha\right>_\alpha$ is shorthand for $\left<\psi_{0,\alpha}\left|J_\alpha\right|\psi_{0,\alpha}\right>$, where $\left|\psi_{0,\alpha}\right>$ is the ground state of the total Hamiltonian when an exchange field of magnitude $E_{ex}$ is applied along the $\alpha$-direction.

The anisotropic part of the charge density is exaggerated $20\times$ in Fig. \ref{fig4} according to the expression $20\cdot(R-R_0) + R_0$, where $R_0$ corresponds to $10$ Bohr radii. More details are given in Figs. \ref{EfigCEFvsCharge},\ \ref{EfigCEFvsChargeIncludingDistortion}.\\

\textbf{Spin model calculations}\\
A model Hamiltonian, Eq. (\ref{eq:main_hamiltonian}), is introduced from the viewpoint of symmetry in section \ref{sec:ESec_Spin_Hamiltonian} of Extended Data, to explain the essential experimental results. Here, magnetic frustration is lifted and the separation of spin components by modulation ($\mathbf{q}$-)vectors is explained naturally by the off-diagonal $E^{ab}_\mathrm{CDW}$ and $E^{ac}_\mathrm{CDW}$ terms. An analytic solution is obtained in Fourier space, and variational calculations are carried out based on the spin ansatz and ignoring higher harmonics, for simplicity,
\begin{equation}
   \left(S^a_n,S^b_n,S^c_n\right)=\left((-1)^nm,\sqrt{1-m^2}\cos[(\pi+q_\mathrm{CDW})z_n],\sqrt{1-m^2}\sin[(\pi+q_\mathrm{CDW})z_n]\right),
\end{equation}
In these terms, the energy is given by
\begin{align}
E[m] = -J_2^\mathrm{AFM}m^2-\frac{E^{ab}_\mathrm{CDW}+E^{ac}_\mathrm{CDW}}{2}m\sqrt{1-m^2}
\end{align}
and easily optimized at $m = m^*$ satisfying $\delta E[m^*] = 0$. The optimal antiferromagnetic moment $m^*$ gives the squared intensities $I_\mathrm{AFM} = \left|S^a(q=\pi)\right|^2= m^2$ and $I_\mathrm{cyc} = \left|S^b(q=\pi+q_\mathrm{CDW})\right|^2+\left|S^c(q=\pi+q_\mathrm{CDW})\right|^2 = (1-m^2)/2$, depicted in Fig. \ref{fig4}.\\

\textbf{Acknowledgments}\\
We thank M. Nakano for permission to use his single-crystal x-ray diffractometer, and for support during the measurement. Moreover, we thank S. Kitou for initial advice on crystal field calculations, M. Kriener for support with experiments on magnetization and calorimetry, and Y. Kato for critical advice on the theoretical spin model. We thank P.M. Neves for support with the analysis of the wide-rocking angle SANS diffraction data. This work is based partly on experiments performed at the Swiss spallation neutron source SINQ, Paul Scherrer Institute, Villigen, Switzerland. This work was supported by JSPS KAKENHI Grant Nos. JP22H04463, JP22F22742, JP22K13998, JP23H01119, JP23KJ0557 and JP22K20348 as well as JST CREST Grant Number JPMJCR1874 and JPMJCR20T1 (Japan), and JST FOREST JPMJFR2238 (Japan). M.M.H. was funded by the Deutsche Forschungsgemeinschaft (DFG, German Research Foundation) -- project number 518238332. The authors are grateful for support by the Fujimori Science and Technology Foundation, New Materials and Information Foundation, Murata Science Foundation, Mizuho Foundation for the Promotion of Sciences, Yamada Science Foundation, Hattori Hokokai Foundation, Iketani Science and Technology Foundation, Mazda Foundation, Casio Science Promotion Foundation, Inamori Foundation, Marubun Exchange Grant, and Kenjiro Takayanagi Foundation. \\

\textbf{Data availability}\\
The data supporting the findings of this study are available from the authors upon reasonable request.\\

\textbf{Author contributions}\\
Sh.A., M.H., and Y.O. synthesized and characterized the single-crystals. S.E., M.H., and Sh.A. carried out calorimetry and magnetic measurements, while T.N., Se.A., Sh.A., Ri.Y., S.E. and M.H. carried out and analyzed elastic neutron scattering measurements, with extensive guidance from T.-h.A. S.E., J.S.W. and M.H. carried out and analyzed small angle neutron scattering measurements. Electric transport measurements were carried out by S.E. and M.H., and theoretical modeling of crystal fields was conducted by S.E. under the guidance of T.-h.A. S.G., S.O., and Ry.Y. modeled the magnetic structure. M.H., M.M.H., and S.E. wrote the manuscript with contributions and comments from all co-authors.\\

\textbf{Competing interests}\\
The authors declare no competing interests.

%===========================================================================================%%
%% MAIN TEXT FIGURES
%%===========================================================================================%%
\clearpage
\begin{center}
\Large{Main text Figures}
\end{center}

\clearpage

%%===========================================================================================%%
%% EXTENDED FIGURES
%%===========================================================================================%%
\renewcommand\thefigure{E\arabic{figure}} 
\setcounter{figure}{0}

\renewcommand\thesection{E\arabic{section}} 
\setcounter{section}{0}   

\renewcommand\thetable{E\arabic{table}} 
\setcounter{table}{0}   

\begin{center}
\Large{Extended Data and Figures}
\end{center}

\begin{table}[h!]
    \centering 
    \caption{\textbf{Magnetic properties of van-der Waals systems with complex magnetic order.} Electrical transport properties are categorized into metals (M), insulators (I), and materials with metal-to-insulator transition (MIT), and spin textures are classified into coplanar (CP) and noncoplanar (NCP). DyTe$_3$ is the only metallic compound that is (a) incommensurate with the underlying lattice, and (b) has a component of the modulation vector $\mathbf{q}$ perpendicular to the stacking direction. Moreover, it is a rare example of noncoplanar (NCP) magnetism in a bulk vdW compound.}
    \label{ETab1}
    \vspace*{1em}
    \begin{tabular}{lllccl}
    \hline
    \hline
    Compound \hspace*{.7em}& Space group \hspace*{.7em}& $\mathbf{q}$-vector \hspace*{.7em}& Transport \hspace*{.7em}& Magnetism \hspace*{.7em}& Ref. \\
    \hline
    \multirow[t]{2}{*}{DyTe$_3$} & \multirow[t]{2}{*}{$Cmcm$} & $q_\mathrm{cyc}\sim$ (0~,~$b^*$~,~$0.207c^*$) & M & NCP & \multirow[t]{2}{*}{this work}\\
    & & $q_\mathrm{AFM}\sim$ (0~,~$b^*$~,~$0.5c^*$)& &\\
    \hline
    \multirow[t]{2}{*}{Fe$_{5-x}$GeTe$_2$} & \multirow[t]{5}{*}{$R\overline{3}m$} & $\pm\frac{1}{3}(1~,~1~,~3)$ & M & NCP &\multirow[t]{2}{*}{\cite{Gao2020,Ly2021,May2019}} \\
     &  & $\pm\frac{3}{10}(0~,~0~,~3)$ & & & \\
    AgCrSe2 & & (0.037~,~0.037~,~3/2) & I & CP &\cite{Baenitz2021, Gautam2002} \\
    NiI$_2$ & & (0.138~,~0~,~1.457) & MIT & CP &\cite{Kurumaji2013, Lebedev2023} \\
    NiBr$_2$ & & (0.027~,~0.027~,~3/2) & I & CP &\cite{Adam1980, Tokunaga2011, Ronda1987} \\
    \hline
    \multirow[t]{2}{*}{CoI$_2$}& \multirow[t]{3}{*}{$P\overline{3}m1$} & (1/12~,~1/12~,~1/2) & - & \multirow[t]{2}{*}{CP} &\multirow[t]{2}{*}{\cite{Kurumaji2013}} \\
    & & (1/8~,~0~,~1/2) & & &\\
    MnI$_2$ &  & (0.181~,~0~,~0.439) & I & CP & \cite{Kurumaji2011, Ronda1987} \\
    \hline
    Co$_{1/3}$NbS$_2$ & \multirow[t]{2}{*}{$P6_322$} & (0.5~,~0~,~0) & M & NCP & \cite{Ghimire2013, Lu2022, Tenasini2020} \\
    Co$_{1/3}$TaS$_2$ & & $(0.5~,~0~,~0)$ & M & NCP & \cite{Takagi2023} \\
    Cr$_{1/3}$NbS$_2$ & & $(0~,~0~,~0.025)$ & M & NCP & \cite{Kousaka2016, Miyadai1983,Wang2017} \\
    Cr$_{1/3}$TaS$_2$ & & $(0~,~0~,~0.081)$ & M & NCP & \cite{Kousaka2016, Obeysekera2021, Zhang2021, Zhang2022} \\
    \hline
    \hline
    \end{tabular}
\end{table}

%%%%%%%%%%%%%%%%%%%%%%%%%%%%%%%%%%%%%%%%%%%%%%%%%%%%%%%%%%%%%%%%%
%%%%%%%%%%%%%%%%%%%%%%%%%%%%%%%%%%%%%%%%%%%%%%%%%%%%%%%%%%%%%%%%%
%%                   NEW SECTION
%%%%%%%%%%%%%%%%%%%%%%%%%%%%%%%%%%%%%%%%%%%%%%%%%%%%%%%%%%%%%%%%%
%%%%%%%%%%%%%%%%%%%%%%%%%%%%%%%%%%%%%%%%%%%%%%%%%%%%%%%%%%%%%%%%%
\section{Spin Hamiltonian}
\label{sec:ESec_Spin_Hamiltonian}

\textbf{Model Hamiltonian.} To describe the ground state and the field-induced transition in DyTe$_3$ for $\mathbf{B}\parallel c$, consider a one-dimensional (1D) chain, where each unit cell contains a single magnetic site, of index $n$, in the paramagnetic state. For this model, the lattice constant is set to $c=1$, the reciprocal lattice constant is $c^* = 2\pi$, the wavevector $q$ is dimensionless, and the CDW wavenumber (for DyTe$_3$) is $q_\mathrm{CDW} = 2\pi\cdot 0.293$~\cite{Malliakas2006}. As compared to the effective zigzag chain in DyTe$_3$, the number of magnetic sites is halved, i.e. the lower part of the chain is omitted. Hence, $ud$ order for the component $S^a$ of the local, quasi-classical spin $\mathbf{S}$ is equivalent to $uudd$ order on the zigzag chain.

We start with a real-space ansatz comprising antiferromagnetic exchange, a spatially modulated on-site coupling induced by the lattice distortion from the CDW, and a Zeeman term for magnetic field applied along the chain axis ($c$-axis),
\begin{equation}
    \mathcal{H} = \sum_{n}\Big[J_2 S_{n}^aS_{n+1}^a -E^{ab}_\mathrm{CDW}\cos\left(q_\mathrm{CDW}z_{n}\right)S_{n}^a S_{n}^b -E^{ac}_\mathrm{CDW}\sin\left(q_\mathrm{CDW}z_{n}\right)S_{n}^a S_{n}^c + BS_n^c\Big]
\label{eq:hamiltonian_real_space_ansatz}
\end{equation}
where $n$ labels magnetic moments on a single layer of a single zigzag chain. Here, $n$, $n+1$ represent nearest neighbours in the crystal lattice of DyTe$_3$, so that their coupling $J_2$ can be expected to be stronger than the coupling $J_1$ between the sheets (see next section).

The oscillating off-diagonal terms, $E^{ab}_\mathrm{CDW}\cos\left(q_\mathrm{CDW}z_{n}\right)$ and $E^{ac}_\mathrm{CDW}\sin\left(q_\mathrm{CDW}z_{n}\right)$ are permitted since the global $\mathcal{M}_b$ and the local $\mathcal{M}_c$-mirror are broken by the lattice distortion due to the CDW, respectively.
Moving to Fourier space according to the conventions $\mathbf{S}_n = (1/\sqrt{N})\sum_q \mathbf{S}_q\,\exp(i q z_n)$ and $\sum_n \exp(i q z_n) = N \delta(q)$, where $N$ is the number of sites on the chain, we have 
\begin{align}
    \mathcal{H} = \sum_q \Bigg[J_2\cos(q)S_{q}^aS_{-q}^a &- \frac{E^{ab}_\mathrm{CDW}}{2}S_{q}^a \left(S_{-q+q_\mathrm{CDW}}^b+S_{-q-q_\mathrm{CDW}}^b\right)\nonumber\\ 
    &-\frac{E^{ac}_\mathrm{CDW}}{2i}S_{q}^a \left(S_{-q+q_\mathrm{CDW}}^c-S_{-q-q_\mathrm{CDW}}^c\right)\Bigg]+ \frac{B}{\sqrt{N}}\,S_{q=0}^c
\end{align}
For a Heisenberg Hamiltonian without off-diagonal terms, the Luttinger-Tisza rule dictates the choice of a single, optimal $q$ for the long-range ordered state; yet here, given the off-diagonal coupling of spin components, we naturally select different wavenumbers for different spin components, explaining the clear separation of spin components by $\mathbf{q}$-vector observed experimentally in Fig. \ref{fig2} of the main text.
Hereafter, we set $E^{ab}_\mathrm{CDW}=E^{ac}_\mathrm{CDW}=E_\mathrm{CDW}$ for simplicity.

If $J_2>0$, the minimum of $J_2\cos(q)$ is at $q = \pi$ and $S_{q=\pi}^a$ is realized. If further $E_\mathrm{CDW}>0$, $S_q^c$ is directly locked to the dominant $S_q^a$ and 
\begin{align}
    S_q^b\propto \delta(q=\pi+q_\mathrm{CDW}) + \delta(q=\pi-q_\mathrm{CDW})\\
    S_q^c\propto \delta(q=\pi+q_\mathrm{CDW}) - \delta(q=\pi-q_\mathrm{CDW})
\end{align}
realizing the coupled wavevectors that are experimentally observed in phase I of DyTe$_3$. Application of a magnetic field $B$ larger than $J_2$ enforces $S_q^c\propto \delta(q=0)$ and the cross-terms induce $S_q^a \propto \delta(q=\pm q_\mathrm{CDW})$ and the second harmonics $S_q^b \propto \delta(q=\pm 2q_\mathrm{CDW})$ demonstrated experimentally in Fig.~5. Indeed, the Zeeman energy per magnetic moment of $m_\mathrm{Dy} = 10\,\mu_\mathrm{B}$ (Bohr magneton) at the critical field $B_\mathrm{c} = 0.5\,$T in Fig. \ref{fig5} is roughly $3.6\,$K, very close to the value of the N{\'e}el temperature $T_\mathrm{N}$. 

If, in contrast, $E_\mathrm{CDW}<0$, there is no reason why wavenumbers such as $\pi\pm q_\mathrm{CDW}$ should appear in the ground state. As compared to anisotropic exchange interactions, for example of the type $S_n^cS_{n+1}^a + S_n^aS_{n+1}^c$, the present $E_\mathrm{CDW}$ terms cannot induce spontaneous magnetic order by themselves, but rather they create a 'parasitic' spin modulation -- driven by the charge-density wave of $R$Te$_3$, $R = \,$rare earth -- on the back of either AFM order below $B_\mathrm{c}$ or of the field-polarized moment above $B_\mathrm{c}$.
\\

\begin{figure}[h]%
\centering
\includegraphics[width=0.95\textwidth]{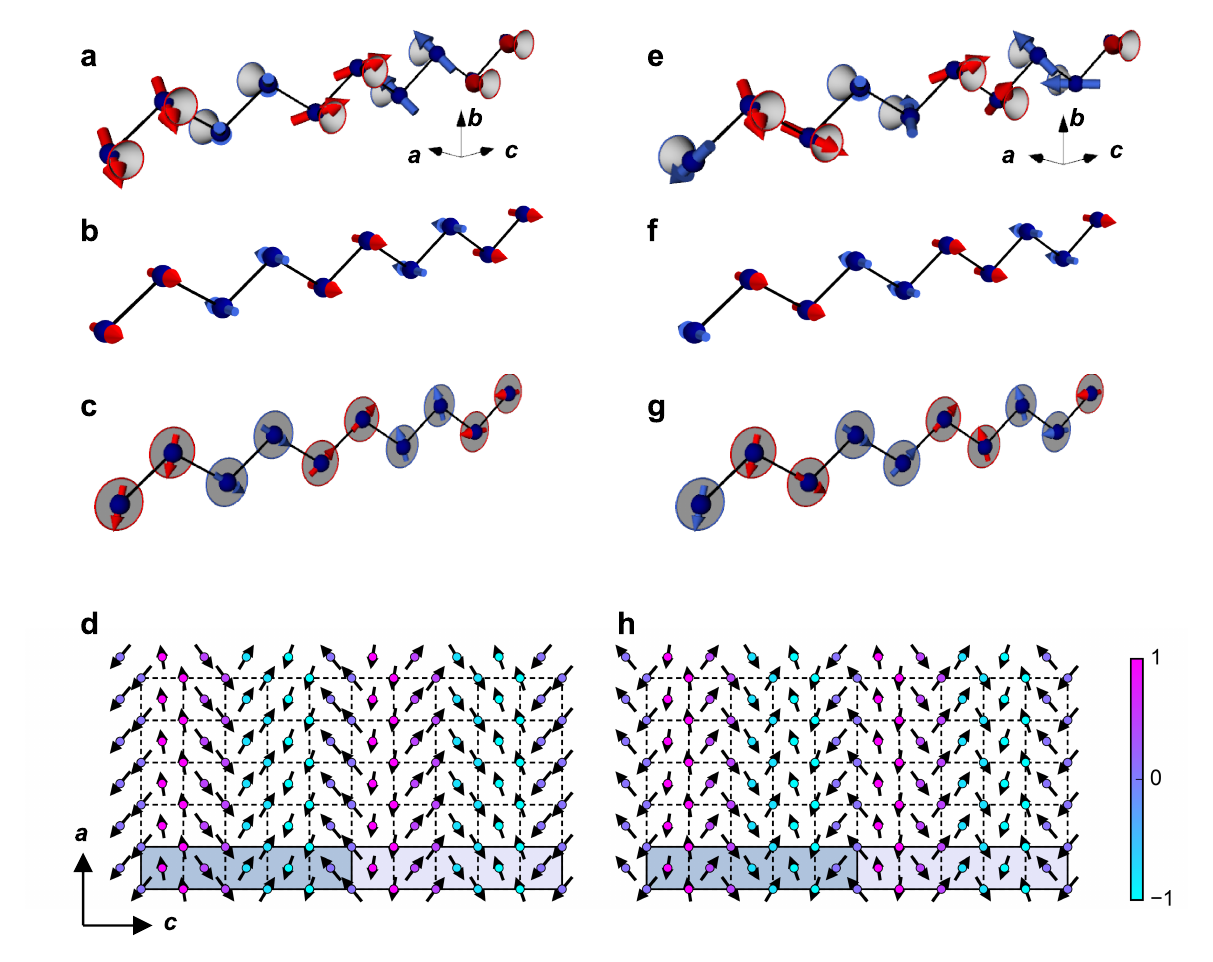}
\caption{\textbf{Two possible magnetic structures in the ground state of DyTe$_3$.} Neutron scattering does not constrain the combination of cycloidal ($S_b$, $S_c$) and antiferromagnetic (AFM, $S_a$) components, which can be either in-phase (left) or out of phase (right). The AFM and cycloidal components of in-phase (out of phase) combinations are depicted, in panels \textbf{b},\textbf{c} (in panels \textbf{f}, \textbf{g}), respectively. Only a single zigzag chain is shown, although there is antiferromagnetic coupling between subsequent zigzag chains along the $b$-axis, c.f. Fig. 1. \textbf{d},\textbf{h} Magnetic texture in a single DyTe magnetic square net bilayer, corresponding to the simplified zigzag-chain picture in panels a,e. Colour on each ionic site (arrows) illustrate the $b$-axis (the $ac$-plane) component of the magnetic moments. Dashed lines are guides to the eye, describing the square net in the upper sheet of the DyTe bilayer slab. The bright (dark) highlight marks the size of the magnetic unit cell (the size of half a magnetic unit cell, as shown in panels a-c, e-g, and in the figures of the main text). For clarity, the relative phase $\delta$ between the layers is chosen so that pairs of spins are fully collinear (instead of the experimentally determined $79^\circ$ phase shift, section \ref{Esec:magnetic_structure_analysis}).}\label{EfigMstructure}
\end{figure}

\textbf{Additional Heisenberg coupling on zigzag chain.}
We introduce an additional $J_1$ between neighbouring sites on the zigzag chain as 
\begin{equation}
\label{eq:Heisenberg_NN_zigzag}
\mathcal{H}^{\prime\prime} = J_1 \sum_n \left(\mathbf{S}_{n1}\cdot\mathbf{S}_{n2} + \mathbf{S}_{n2}\cdot\mathbf{S}_{n+1,1}\right)
\end{equation}
which can be separated into three independent equations $\mathcal{H}_a^{\prime\prime}$, $\mathcal{H}_b^{\prime\prime}$, $\mathcal{H}_c^{\prime\prime}$ -- one for each spin component. Note that $\mathbf{S}_{n1}$, $\mathbf{S}_{n+1,1}$ are \textit{not} nearest neighbours in the lattice of DyTe$_3$, so that their coupling $J_1$ is expected to be weaker than the dominant antiferromagnetic $J_2$ in Eq. (\ref{eq:hamiltonian_real_space_ansatz}). We use the trial functions
\begin{align}
S_{n\alpha}^a &= \bar{S}_a \cos\left(2\pi\cdot\frac12\cdot z_{n\alpha} +\varphi_\alpha^\mathrm{AFM}\right)\\
S_{n\alpha}^b &= \bar{S}_b \sin\left(h_\alpha\cdot q_\mathrm{cyc}\cdot z_{n\alpha} +\varphi_\alpha\right)\label{eq:spin_texture_1dchain_b}\\
S_{n\alpha}^c &= \bar{S}_c \cos\left(h_\alpha\cdot q_\mathrm{cyc}\cdot z_{n\alpha} +\varphi_\alpha\right)\label{eq:spin_texture_1dchain_c}
\end{align}
with a helicity parameter $h_\alpha$ indicating the propagation \textit{direction} of the texture, the incommensurate wavevector $q_\mathrm{cyc} = 2\pi\, L_\mathrm{cyc}$ in reciprocal lattice units (r.l.u.), as well as the phases $\varphi_\alpha$ and $\varphi_\alpha^\mathrm{AFM}$. The latter can only vary in steps of $\pi$, if we assume a fixed length of the magnetic moment. Based on this,
\begin{align}
\mathcal{H}_b^{\prime\prime} & = J_1 \bar{S}_b^2\cos\left(h_1 q_\mathrm{cyc}/2\right)\sum_n \left[\cos\left((h_2-h_1)q_
\mathrm{cyc}(n + 1/2)+\varphi_2-\varphi_1\right)-\right.\notag\\
&\qquad\qquad\qquad\qquad\qquad\qquad
\left.-\cos\left((h_2+h_1)q_\mathrm{cyc}(n+1/2)+\varphi_2+\varphi_1\right)\right]\\
\mathcal{H}_c^{\prime\prime} & = J_1 \bar{S}_c^2\cos\left(h_1 q_\mathrm{cyc}/2\right)\sum_n \left[\cos\left((h_2-h_1)q_
\mathrm{cyc}(n + 1/2)+\varphi_2-\varphi_1\right)+\right.\notag\\
&\qquad\qquad\qquad\qquad\qquad\qquad
\left.+\cos\left((h_2+h_1)q_\mathrm{cyc}(n+1/2)+\varphi_2+\varphi_1\right)\right]
\end{align}
and the two cases $p = h_1\cdot h_2 = +1$ and $p = -1$ yield
\begin{align}
E^{\prime\prime}_{p\, =\, +1} &= J_1\,N\left(\bar{S}_c^2 +\bar{S}_b^2\right)\cos\left(q_\mathrm{cyc}/2\right)\cos\left(\varphi_2-\varphi_1\right) \label{eq:interchain_energy_same}\\
E^{\prime\prime}_{p\, =\, -1} &= J_1\,N\left(\bar{S}_c^2 -\bar{S}_b^2\right)\cos\left(q_\mathrm{cyc}/2\right)\cos\left(\varphi_2+\varphi_1\right)\label{eq:interchain_energy_opposite}
\end{align}
These terms in Eq. (\ref{eq:interchain_energy_same}, \ref{eq:interchain_energy_opposite}) are independent of $z_n$; especially for $p = +1$, optimizing $\delta = \varphi_2 - \varphi_1$ yields a favorable energy contribution for any given $J_1$, $E_\mathrm{CDW}$.

%%%%%%%%%%%%%%%%%%%%%%%%%%%%%%%%%%%%%%%%%%%%%%%%%%%%%%%%%%%%%%%%%
%%%%%%%%%%%%%%%%%%%%%%%%%%%%%%%%%%%%%%%%%%%%%%%%%%%%%%%%%%%%%%%%%
%%                   NEW SECTION
%%%%%%%%%%%%%%%%%%%%%%%%%%%%%%%%%%%%%%%%%%%%%%%%%%%%%%%%%%%%%%%%%
%%%%%%%%%%%%%%%%%%%%%%%%%%%%%%%%%%%%%%%%%%%%%%%%%%%%%%%%%%%%%%%%%

\section{Expressions for scattering intensities}
\label{Esec:scattering_intensities}
We review the expressions used to calculate neutron scattering intensities from atomic and magnetic structures of DyTe$_3$, which serve to define a variety of parameters (such as the phase shift $\delta$) used in the discussion of the main text. Working with the triple-axis diffractometer PONTA-5G at JRR-3 research reactor, we fit $\omega$ or $\theta-2\theta$ scans of neutron intensities with Gaussian profiles and calculate the total observed intensity $I_\mathrm{obs}$ for each reflection, taking the $\mathbf{Q}$-dependence of the peak shape into account (Fig. \ref{EfigCalibrationPONTA}). From this, we calculate the observed structure factor as
\begin{equation}
F_\mathrm{obs}(\mathbf{Q}) = \sqrt{I_\mathrm{obs}(\mathbf{Q})/\mathcal{L}(2\theta)}
\end{equation}
where $\mathcal{L}(2\theta) = \lambda^3/\sin(2\theta)$ is the Lorentz factor, $\lambda$ is the wavelength of the monochromatized neutrons, and $2\theta$ is the scattering angle.

To reproduce this $F_\mathrm{obs}(\mathbf{Q})$ quantitatively, we start from the expression for the differential cross-sections for nuclear and magnetic scattering, i.e. the beam intensity scattered into a solid angle $d\Omega$ corresponding to the direction of the momentum transfer $\mathbf{Q}$ of magnitude $Q = \left|\mathbf{Q}\right|$ and direction $\hat{\mathbf{Q}} = \mathbf{Q}/Q$
\begin{align}
\left(\frac{d\sigma}{d\Omega}\right)^N_\mathrm{cal} &\equiv I^N_\mathrm{cal} = \Phi \left|\sum_{j\in\mathrm{lattice}} b_j \exp\left(\imath\mathbf{Q}\cdot \mathbf{r}_j\right)\right|^2 \equiv \Phi\,\left|F^N_\mathrm{cal}(\mathbf{Q})\right|^2\, \sum_{\mathbf{l}, \mathbf{l}^\prime}\exp\left[\imath\mathbf{Q}\cdot(\mathbf{l}-\mathbf{l}^\prime)\right]\\
\left(\frac{d\sigma}{d\Omega}\right)^M_\mathrm{cal} &\equiv I^M_\mathrm{cal}= \Phi \left|-2.7\sum_{j\in\mathrm{lattice}} f_\mathrm{mag}(Q)\,\mathbf{m}_{\perp,j} \exp\left(\imath\mathbf{Q}\cdot \mathbf{r}_j\right)\right|^2 \label{eq:dsigmadomega_mag}
\end{align}
Here, we have introduced the neutron flux $\Phi$, the position of each atom on the lattice $\mathbf{r}_j$, the nuclear scattering length $b_j$, the unit cell coordinates $\mathbf{l}$, $\mathbf{l}^\prime$, the magnetic form factor $f_\mathrm{mag}(Q)$ evaluated from an analytic expression~\cite{ILLffacts2023}, and the component of the magnetic moment at site $j$ that is perpendicular to the momentum transfer
\begin{equation}
\mathbf{m}_{\perp,j}(\mathbf{Q}) = \mathbf{m}_j - \left(\mathbf{m}_j\cdot \hat{\mathbf{Q}}\right)\cdot \hat{\mathbf{Q}} \label{eq:m_perp_definition}
\end{equation}
Note the scattering lengths for Dy$^{3+}$ ($16.9\,$fm) and Te ($5.8\,$fm), and that only magnetic ions contribute to Eq. (\ref{eq:dsigmadomega_mag}). 

For nuclear scattering, we use~\cite{Squires2012}
\begin{align}
&F_\mathrm{cal}^N(\mathbf{Q}) = \sum_{\mathbf{d}\in\mathrm{c.u.c.}} b_j \exp\left(\imath\,\mathbf{Q}\cdot\mathbf{d}\right) \\
&\Phi \sum_{\mathbf{l}, \mathbf{l}^\prime}\exp\left[\imath\mathbf{Q}\cdot(\mathbf{l}-\mathbf{l}^\prime)\right] = \Phi \, N \frac{(2\pi)^3}{\nu_0}\,\delta^{(3)}(\mathbf{Q}-\mathbf{G}) \label{eq:lattice_sum}
\end{align}
for a reciprocal lattice vector $\mathbf{G}$ and using $\mathbf{r}_j = \mathbf{d} + \mathbf{l}$, where the former is a coordinate within the crystallographic unit cell, and the latter labels the origin of each unit cell. $\Phi$, $N$, $F_\mathrm{cal}^N$, are the flux of incident neutrons, the number of crystallographic unit cells (c.u.c., of volume $\nu_0$) in the sample, and the nuclear structure factor. We find good agreement of the experimental scattering data and model when using the atomic positions from the high-temperature $Cmcm$ space group of DyTe$_3$~\cite{Slovyanskikh1985}. In reality, the formation of charge order below $T_\mathrm{CDW}\approx 320\,$K lowers the symmetry, as discussed by Malliakas \textit{et al.} in Refs. ~\cite{Malliakas2005,Malliakas2006}.

Then, a scale factor $s$ is defined by equating to the experimentally observed intensity,
\begin{equation}
\left|F_\mathrm{obs}^N(\mathbf{Q})\right|^2 = \Phi\, N\,\frac{(2\pi)^3}{\nu_0}\left|F_\mathrm{cal}^N(\mathbf{Q})\right|^2 \equiv s\left|F_\mathrm{cal}^N(\mathbf{Q})\right|^2
\end{equation}
as shown in Fig. \ref{EfigNuclearRefinement}.

Next, for magnetic scattering from a structure with lattice-commensurate magnetic order, we use Eq. (\ref{eq:lattice_sum}) with a larger unit cell (reduced set of $\mathbf{G}$) and equate
\begin{align}
\left|F_\mathrm{obs}^M(\mathbf{Q})\right|^2 &= I_\mathrm{obs}^M(\mathbf{Q})/\mathcal{L}(2\theta) = I^M_\mathrm{cal}(\mathbf{Q}) = \Phi\, N_M\,\frac{(2\pi)^3}{\nu_M}\,\left|\mathbf{F}_\mathrm{cal}^M(\mathbf{Q})\right|^2\equiv s \left|\frac{\mathbf{F}_\mathrm{cal}^M}{x}\right|^2\\
\mathbf{F}_\mathrm{cal}^M(\mathbf{Q}) &= -2.7f_\mathrm{mag}(Q)\,\sum_{j\in\mathrm{m.u.c.}}\mathbf{m}_{\perp,j}\,\exp(\imath\mathbf{Q}\cdot \mathbf{r}_j)
\end{align}
We introduced the calculated magnetic structure factor $\mathbf{F}_\mathrm{cal}^M$, which has three complex components $F_{\mathrm{cal},x}^M$, $F_{\mathrm{cal},y}^M$, $F_{\mathrm{cal},z}^M$; the volume of the magnetic unit cell (m.u.c.) $\nu_M = x\cdot \nu_0$, the magnetic form factor of Dy$^{3+}$, $f_\mathrm{mag}$, and the number $N_M=N/x$ of m.u.c. in the sample. The prefactor $(-2.7)$ describes the scattering length of the electron, and the sum is now over all magnetic (dysprosium) ions in the m.u.c.

In case of two domains of this commensurate order, as relevant for the analysis in DyTe$_3$, we have
\begin{equation}
I^M_\mathrm{cal} = \Phi\,\frac{N}{2}\,\frac{(2\pi)^3}{\nu_0}\,\left|\frac{\mathbf{F}_\mathrm{cal}^{M1}}{x}\right|^2 + \Phi\,\frac{N}{2}\,\frac{(2\pi)^3}{\nu_0}\,\left|\frac{\mathbf{F}_\mathrm{cal}^{M2}}{x}\right|^2 =\frac{s}{2}\left(\left|\frac{\mathbf{F}_\mathrm{cal}^{M1}}{x}\right|^2+\left|\frac{\mathbf{F}_\mathrm{cal}^{M2}}{x}\right|^2\right)\label{eq:two_domains}
\end{equation}
where $\mathbf{F}_\mathrm{cal}^{Mk}$, $k = 1, 2$ are the magnetic structure factors for two domains.

%%%%%%%%%%%%%%%%%%%%%%%%%%%%%%%%%%%%%%%%%%%%%%%%%%%%%%%%%%%%%%%%%
%%%%%%%%%%%%%%%%%%%%%%%%%%%%%%%%%%%%%%%%%%%%%%%%%%%%%%%%%%%%%%%%%
%%                   NEW SECTION
%%%%%%%%%%%%%%%%%%%%%%%%%%%%%%%%%%%%%%%%%%%%%%%%%%%%%%%%%%%%%%%%%
%%%%%%%%%%%%%%%%%%%%%%%%%%%%%%%%%%%%%%%%%%%%%%%%%%%%%%%%%%%%%%%%%
\section{Symmetry and structure factor: commensurate component in phase I}
\label{sec:Esec_structure_factors_comm}

\begin{figure}[h]%
\centering
\includegraphics[trim=1.2cm 0.5cm 7.cm 0.1cm, clip,width=0.8\textwidth]{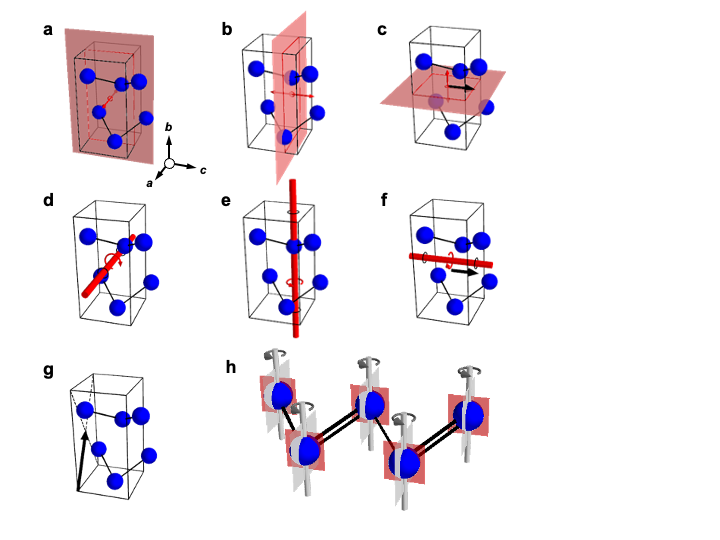}
\caption{\textbf{Space group $Cmcm$ and its symmetries in the $(000)+$ set, as well as breaking of bond-symmetry by AFM commensurate order.} Only dysprosium ions are shown (blue), including two ions outside the crystallographic unit cell (right) to illustrate the structural zigzag chain feature. \textbf{a}, Mirror plane $\mathcal{M}_a$. \textbf{b}, Mirror plane $\mathcal{M}_c$. \textbf{c}, Glide mirror plane $\mathcal{G}_b$ with $c/2$ translation along the $c$-axis (black arrow). \textbf{d}, Two-fold rotation $C_{2a}$ through the inversion center $\mathcal{I}=(\frac12, \frac12, \frac12)$. \textbf{e}, Two-fold rotation $C_{2b}$ through a position shifted by $\left(0, 0, \frac14\right)$ from $\mathcal{I}$. \textbf{f}, Screw rotation $C_{2c}\times \left(0, 0, \frac12\right)$ with $c/2$ shift along the $c$-axis. \textbf{g} Base centering operation $C$, i.e., a translation by $\left(\frac12, \frac12, 0\right)$. \textbf{h} In the presence of commensurate order $\mathbf{q}_\mathrm{AFM} = (0, b^*, 0.5c^*)$, the magnetic space group symmetry is lowered to $C_c2/m$: $\mathcal{M}_c$ and $C_{2b}$ are broken (grey), while $\mathcal{M}_a$ survives (red). The equivalence of bonds on the effective zigzag chain (main text) is lifted in presence of $uudd$ or $uddu$ order (black double and single bonds).}\label{Efig_Cmcm_space_group_symmetries}
\end{figure}

\subsection{Symmetry consideration (commensurate)}
\label{sec:ESubsec_symmetry_comm}

Section \ref{sec:ESec_Spin_Hamiltonian} discusses oscillatory terms that are allowed in the Hamiltonian due to local symmetry breaking from the charge-density wave (CDW). In this section, we focus on \textit{global} symmetries of DyTe$_3$ and their lowering by magnetic order. We start from space group $Cmcm$, ignoring the incommensurate CDW at first, and briefly consider the effect of the CDW on global symmetries at the end of the section.

\begin{figure}[h]%
\centering
\includegraphics[width=0.9\textwidth]{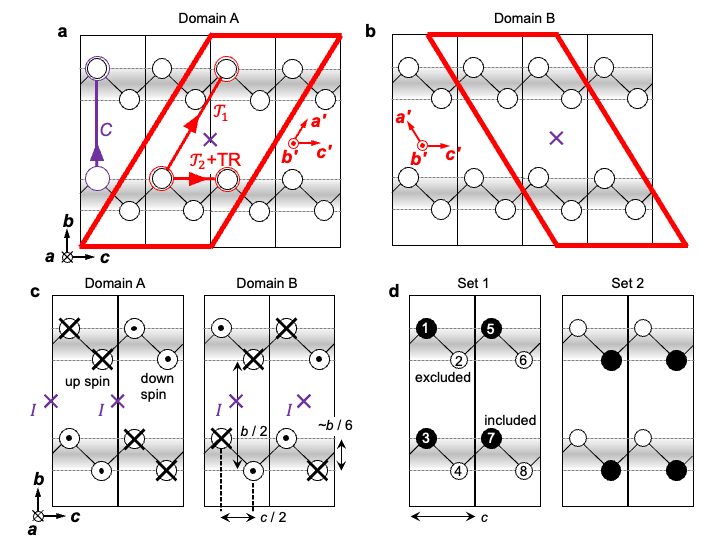}
\caption{\textbf{Unit cell, magnetic domains for commensurate spin component, and illustration for structure factor calculation in phase I of DyTe$_3$ (antiferromagnetic part, AFM).} \textbf{a}, \textbf{b}, Primitive magnetic unit cell (red boundary, monoclinic) in two AFM domains of DyTe$_3$, with base centering translation vector $\mathcal{T}_1$ and $C_c$ translation vector (plus time reversal) $\mathcal{T}_2$. The base centering operation of the parent structure $Cmcm$ ($C$), and its inversion center ($I$), are also indicated by a violet arrow and cross, respectively. Only magnetic sites are shown, and sites with different $x$-positions have been projected onto a single plane. See section \ref{sec:Esec_structure_factors_comm}. \textbf{c}, Conventional unit cell for two AFM domains in phase I. Circles with dot (with cross) signify magnetic moment pointing into (out of) the plane of the figure. The $b$-axis spacing of magnetic sites in a bilayer (grey shading) is approximately $b/6$. Inversion centers $\mathcal{I}$ are indicated by violet crosses. \textbf{d}, Two sets of sites (black, white circles) of magnetic ions in the unit cell. In the calculation of the structure factor for the combined scattering of two equally populated domains A, B according to Eq. (\ref{eq:struct_fact_afm}), the only contributions are from terms $\sim\exp(\imath\mathbf{Q}\cdot(\mathbf{r_i-\mathbf{r}_j}))$ for sites $i,j$ in the same set.}\label{Efig_AFM_domain_cartoon}
\end{figure}

In polarized neutron scattering (PNS), orthorhombic symmetry of the crystal structure allows us to set the scattering plane to $HK0$, with separation of three orthogonal magnetization components $m_a$, $m_b$, and $m_c$. The PNS data strictly constrains the commensurate moment to be along the $a$-axis, and the observation of reflections of the type $(0, \mathrm{odd}, 0.5)$ -- while $(0, \mathrm{even}, 0.5)$ are absent --, establishes a unit cell with eight magnetic dysprosium ions, vanishing net magnetization along the $a$-axis, as well as opposite directions for moments separated by a distance $b/2$ along the $b$-axis (Fig. \ref{Efig_AFM_domain_cartoon}c). Four free parameters remain, namely the lengths of four magnetic moments $m_a$ in a DyTe slab spanning two crystallographic unit cells, i.e. the upper zigzag chain in Fig. \ref{Efig_AFM_domain_cartoon}.

The full Hermann-Mauguin symbol for orthorhombic space group $Cmcm$ (SG63) is $C \,\,2/m\,\, 2/c\,\, 2_1/m$, and $\mathbf{q}_\mathrm{AFM} = (0, 1, 0.5)$ does not break any of these symmetries. Using the $\mathbf{k}$-Subgroupsmag tool of the Bilbao crystallographic server~\cite{Bilbao}, we determined the highest-symmetry magnetic subgroups of $Cmcm$ that are consistent with $\mathbf{q}_\mathrm{AFM}$. In (black-and-white-type) Belov-Neronova-Smirnova notation: $I_ama2$, $C_c2/c$, and $C_c2/m$. From the transformation matrices provided by this tool, we also find the lattice vectors $\mathbf{a}^\prime$, $\mathbf{b}^\prime$, $\mathbf{c}^\prime$ in terms of the lattice vectors $\mathbf{a}$, $\mathbf{b}$, $\mathbf{c}$ of the 'parent' $Cmcm$: $(-2\mathbf{c}, \mathbf{a}, -\mathbf{b})$, $(-\mathbf{b}-2\mathbf{c}, \mathbf{a}, 2\mathbf{c})$, and $(\mathbf{b}+2\mathbf{c}, -\mathbf{a}, 2\mathbf{c})$, for the three abovementioned symbols. Here, for example, $\mathbf{b} = \mathbf{e}_b\cdot b$ for basis (unit) vector $\mathbf{e}_b$ and lattice constant $b$ (and so on).

We rule out $I_ama2$ and $C_c2/c$ for the commensurate order in DyTe$_3$. The former has a $\mathcal{M}_{a^\prime}$ mirror plane, which is perpendicular to $\mathbf{e}_c$ of the $Cmcm$ cell and located on the Dy-sites. A moment $m_a$ on the Dy-site is inconsistent with this mirror plane. The latter space group has a glide mirror plane $\mathcal{G}_{b^\prime}$, consisting of a mirror operation perpendicular to $\mathbf{e}_a$ combined with a lattice translation by $\mathbf{c}$. The moment direction $m_a$ at a given site is unchanged under $\mathcal{G}_{b^\prime}$, but translated by half the length of the magnetic unit cell. Given there are merely four sites in the upper DyTe slab of the magnetic unit cell, this operation requires orders of the type udud, and is thus inconsistent with the expansion of the unit cell along $\mathbf{e}_c$ that is implied by $\mathbf{q}_\mathrm{AFM}$.

We focus on monoclinic, centrosymmetric $C_c2/m$, where $C_c$ includes base centering as a translation $\mathcal{T}_1 = (1/2, 1/2, 0)$ and $\mathcal{T}_2^\prime$, i.e. the translation $\mathcal{T}_2 = (0, 0, 1/2)$ combined with time reversal. There are two domains A and B with basis vectors $(\mathbf{b}+2\mathbf{c}, -\mathbf{a}, 2\mathbf{c})$ and $(-\mathbf{b}+2\mathbf{c}, \mathbf{a}, 2\mathbf{c})$; i.e., they are characterized by a reversal of the monoclinic tilt (Fig. \ref{Efig_AFM_domain_cartoon}). Only the mirror plane perpendicular to $\mathbf{e}_a$ remains intact, the number of inversion centers is halfed as compared to $Cmcm$, and the broken $c$-mirror relates the two possible domains A and B depicted in Fig. \ref{Efig_AFM_domain_cartoon}c.

Note: A previous x-ray scattering study reports the superspace group $C2cm(00\gamma)000$ for the CDW state in DyTe$_3$~\cite{Malliakas2005}. In average space group $C2cm$ (No. 40), the $\mathcal{M}_{a}$ mirror of $Cmcm$ is already broken. Starting from this lower-symmetry symbol, analogous discussion leads to $C_c 2$ for the commensurate component of the magnetic order in phase I.

We consider the possibility of further symmetry lowering, as caused by non-uniform magnetic moment length $m_a$. Magnetic space group $C_c2/m$ has $\mathcal{T}_2'$, which ensures that moments $m_a$ in the same layer $\alpha$ point in opposite directions, but have the same length. 
Instead of $uudd$ (up-up-down-down), this leaves $(u_\mathrm{short}u_\mathrm{long}d_\mathrm{short}d_\mathrm{long})$ as a viable configuration, but $C_{2a}$ rotation symmetry interrelates layers $\alpha = 1, 4$ and eliminates this possibility. The neutron data are well described by $C_c2/m$ (or $C_c2$, which leads to the same commensurate structure models), so that a further symmetry lowering -- relaxing the uniformity constraint on $m_a$ -- is deemed unnecessary.

\subsection{Structure factor calculation (commensurate)}
\label{sec:ESubsec_structure_factors_comm}

We now derive explicit expressions for the scattered intensity at commensurate (AFM) reflections in momentum space, in phase I of DyTe$_3$.

For momentum transfer $\mathbf{Q}$ in the $(0KL)$ plane, we define a normalized moment for each domain $\mathbf{m}_{\perp,j}=m_a\,f_{A,B}(\mathbf{r}_j)\,\mathbf{e}_a$, where $f_{A,B}=\pm 1$. Assuming equal population of domains A and B, Eq. (\ref{eq:two_domains}) gives 
\begin{align}
I^M_\mathrm{cal}&=\frac{s}{8}\left(-2.7\,f_\mathrm{mag}(Q)\,m_a\right)^2\cdot\notag\\
&\sum_{i,j=1,...,8}\exp\left[\imath\mathbf{Q}\cdot(\mathbf{r}_i-\mathbf{r}_j)\right]\cdot\left[f_A(\mathbf{r}_i)f_A^*(\mathbf{r}_j)+f_B(\mathbf{r}_i)f_B^*(\mathbf{r}_j)\right]\label{eq:struct_fact_afm}
\end{align}
where the magnetic sites $i,j$ are labeled in Fig. \ref{Efig_AFM_domain_cartoon}\,\textbf{d} (not Fig. \ref{fig2}). In the latter edgy brackets, the two terms cancel (sum to $\pm2$) when the relationship between magnetic moments at site $i,j$ is opposite (the same) in domains A, B. In addition to four pairs of type $i=j$, examination of Fig.~\ref{Efig_AFM_domain_cartoon}\,\textbf{d} shows that, for the purpose of evaluating Eq. (\ref{eq:struct_fact_afm}), the magnetic sites can be split into two sets 1 and 2: the non-vanishing terms in the sum correspond to pairs of magnetic moments that are just above / below one another, or two sites apart along the $c$-direction.

The relative distances between sites are the same for set 1 and 2, and there are three types of site pairs $i,j$ in each set: 
\begin{align}
\mathbf{r}_i-\mathbf{r}_j &= \pm c\,\mathbf{e}_c\\
\mathbf{r}_i-\mathbf{r}_j &= \pm (b/2)\,\mathbf{e}_b\\
\mathbf{r}_i-\mathbf{r}_j &= \pm (c\,\mathbf{e}_c-(b/2)\,\mathbf{e}_b)\label{eq:special_pair}
\end{align}
with both $\pm$ appearing in the sum of Eq. (\ref{eq:struct_fact_afm}). The moments are antiparallel for all pairs of sites, $\left[f_A(\mathbf{r}_i)f_A^*(\mathbf{r}_j)+f_B(\mathbf{r}_i)f_B^*(\mathbf{r}_j)\right] = -2$, except the pairs in Eq. (\ref{eq:special_pair}). We specialize to the $(0, Kb^*, Lc^*+q_\mathrm{AFM})$ scattering plane, so that
\begin{equation}
\mathbf{Q}=K \frac{2\pi}{b}\,\mathbf{e}_b+(L+0.5)\frac{2\pi}{c}\,\mathbf{e}_c
\end{equation}
as in our experiments, and obtain
\begin{align}
\sum_{i\neq j, set1} \exp(\imath\mathbf{Q}\cdot(\mathbf{r}_i-\mathbf{r}_j))&=4\cos(Q_c\,c) + 4\cos\left(Q_b\,\frac{b}{2}\right)-4\cos(Q_c\,c)\cos\left(Q_b\frac{b}{2}\right)\notag\\
    &=-4+4\cdot(-1)^K+4\cdot(-1)^K=4\left(2(-1)^K-1\right)
\end{align}
which takes the values $-12$ ($+4$) for $K=\,$odd (even), independent of $L$. Then, $\left|F_\mathrm{obs}^M\right|\equiv 0$ when $K$ is even, and for $K=\,$odd,
\begin{align}
I^M_\mathrm{cal}&=\frac{s}{8}\left(-2.7f_\mathrm{mag}(Q)\,m_a\right)^2\,\left(2\cdot 8-2\cdot2\cdot(-12)\right)\notag\\
&=8s\left(-2.7f_\mathrm{mag}(Q)\,m_a\right)^2\label{eq:magnetic_structure_factor_C}
\end{align}
without any explicit dependence on $K$ and $L$ both (but note the magnetic form factor). The above discussion shows that contributions from domains A, B partially cancel each other, meaning that the ratio of domains can be refined from the data.

%%%%%%%%%%%%%%%%%%%%%%%%%%%%%%%%%%%%%%%%%%%%%%%%%%%%%%%%%%%%%%%%%
%%%%%%%%%%%%%%%%%%%%%%%%%%%%%%%%%%%%%%%%%%%%%%%%%%%%%%%%%%%%%%%%%
%%                   NEW SECTION
%%%%%%%%%%%%%%%%%%%%%%%%%%%%%%%%%%%%%%%%%%%%%%%%%%%%%%%%%%%%%%%%%
%%%%%%%%%%%%%%%%%%%%%%%%%%%%%%%%%%%%%%%%%%%%%%%%%%%%%%%%%%%%%%%%%
\section{Symmetry and structure factor: incommensurate component in phase I}
\label{sec:Esec_structure_factors_incomm}

\subsection{Symmetry consideration (incommensurate)}
\label{sec:ESubsec_symmetry_incomm}

\begin{figure}[h]%
\centering
\includegraphics[width=0.8\textwidth]{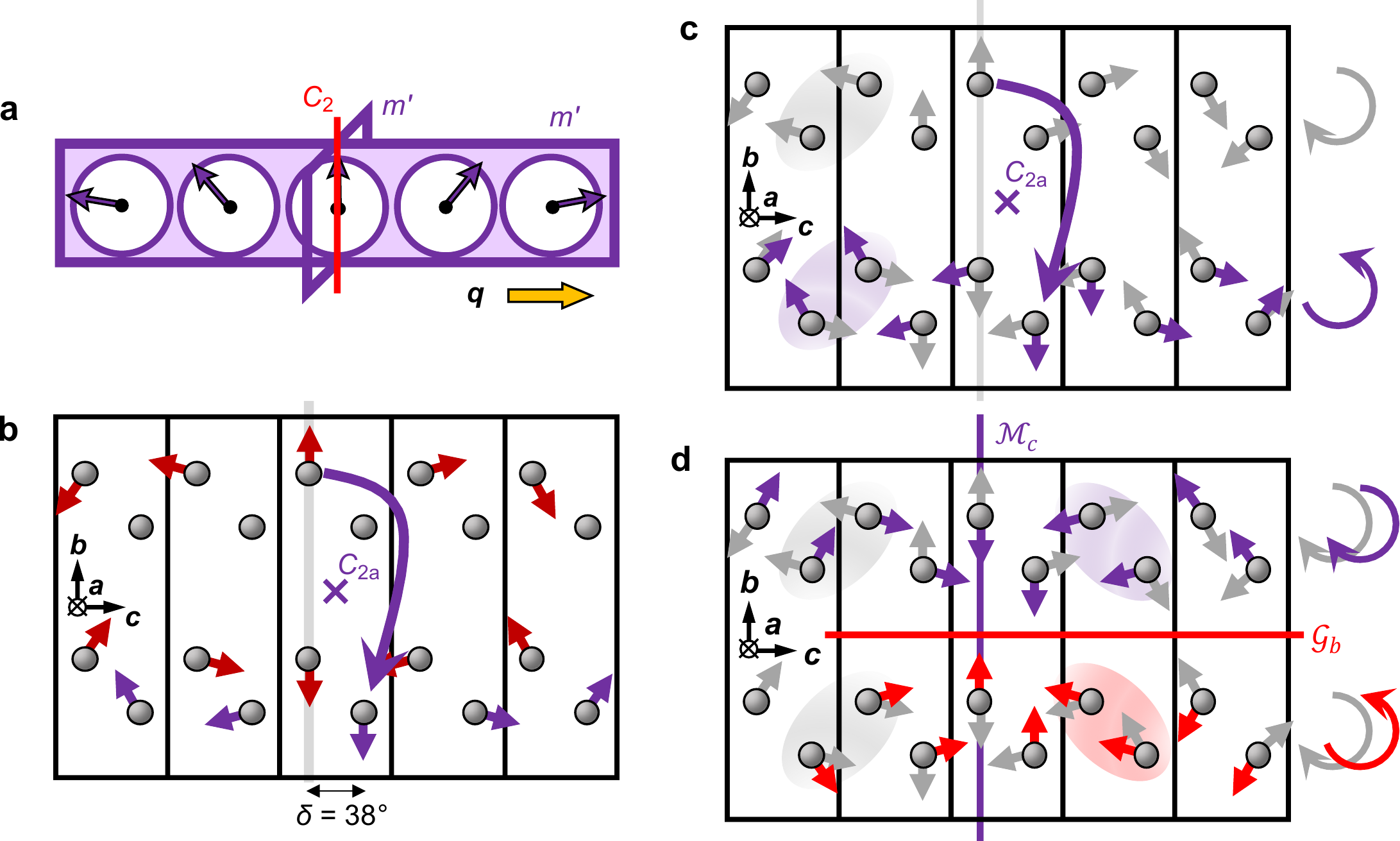}
\caption{\textbf{Symmetry considerations regarding incommensurate magnetic order in phase I, for DyTe$_3$.} \textbf{a}, Magnetic point group symmetries of a cycloid. \textbf{b}, A magnetic texture with $P2$ symmetry and ordering vector $\mathbf{q}_\mathrm{cyc}$. Red arrows: Setting a cycloid in the uppermost layer $\alpha = 1$ with helicity $h_1 = +1$, the $\alpha=3$ layer's structure with $h_3 = +1$ is imposed by antiferromagnetic coupling ($K=1$). Violet arrows: Applying $C_{2a}$ symmetry enforces a counter-propagating cycloid in layers $\alpha = 2, 4$, with $h_2 = h_4 = -1$, where the phase $\delta = \varphi_2 - \varphi_1$ is fixed to $38^\circ$. The texture in layer $\alpha=2$ can be inferred from AFM coupling to $\alpha=4$. 
\textbf{c}, Switching between helicity domains in the $Cc$ model with uniform helicity $\mathbf{h} = \pm(1, 1, 1, 1)$; for simplicity of the illustration, the phase difference is set to $\delta = 38\,^\circ$. Grey arrows: assuming broken $C_{2a}$ symmetry, uniform helicity and adjustable phase shift $\delta$ are allowed. The $C_{2a}$ operation (violet arrow) maps $\mathbf{h}\leftrightarrow -\mathbf{h}$, while leaving $\delta$ unchanged. (Nearly) collinear blocks, (purple highlight) remain coupled. \textbf{d}, Simultaneous switching of $\delta$ and $\mathbf{h}$ via glide mirror $\mathcal{G}_b$, and switching of $\delta$ without $\mathbf{h}$ by mirror $\mathcal{M}_c$. Highlighted ovals mark pairs of spins with nearly collinear alignment in the $bc$ plane. Again, $\delta=\pm 38$ is used for the purpose of this illustration, although the experimental value is $79^\circ$.}\label{fig:Efig_cycloid_C2_symmetry}
\end{figure}

\begin{figure}[h]%
\centering
\includegraphics[width=0.98\textwidth]{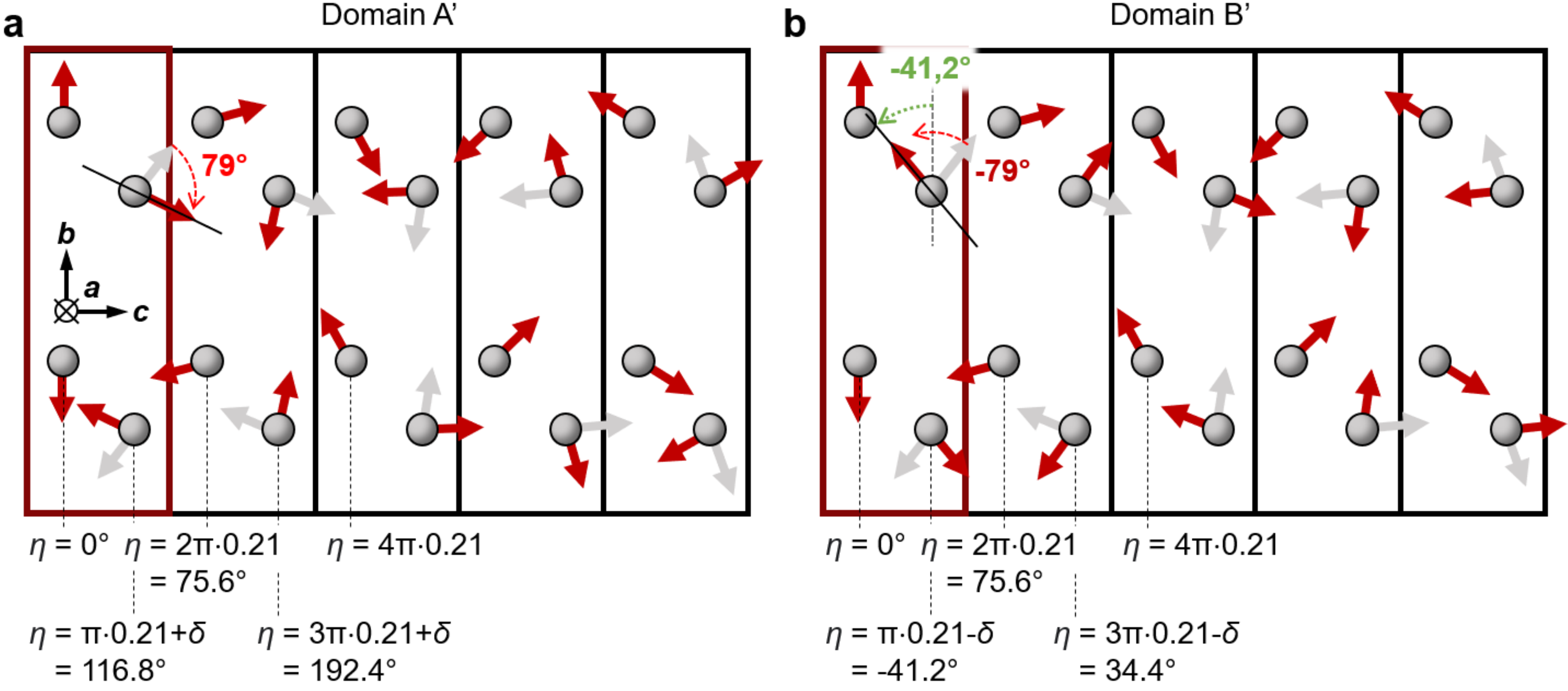}
\caption{\textbf{Two '$\delta$-domains' (A$^\prime$ and B$^\prime$), consistent with the $\mathbf{q}_\mathrm{AFM} = (0, b^*, 0.5c^*)$ ordering vector, for the incommensurate part of the magnetic order in phase I of DyTe$_\mathbf{3}$.} The value of the phase shift $\delta = 79^\circ$ between sheets in a bilayer is adjusted to minimize the reliability index $R$ for the structure factor $F$. Grey arrows indicate the original moment direction that is expected without any phase shift $\delta$. Note that there is a phase shift of $\pi$ between upper and lower Dy-bilayer, and that reversing the sign of the helicity vector $\mathbf{h} = (h_1, h_2, h_3, h_4)$, as defined in section \ref{sec:Esec_structure_factors_incomm}, does not affect the value of scattering intensities; $\eta$ is the angle between the $bc$-component of the magnetic moment and the crystallographic $b$-axis.}\label{Efig_cycloid_delta_domains}
\end{figure}

Starting from Eqs. (\ref{eq:spin_texture_1dchain_b}, \ref{eq:spin_texture_1dchain_c}), and given antiferromagnetic coupling between structural bilayers in DyTe$_3$, as enforced by $K=\,$odd in $(0, K, L_\mathrm{cyc})$, the phase relation between layers $\alpha =1,3$ and $2,4$ is fixed to $\pi$ and $h_1 = h_3$, $h_2 = h_4$. The remaining free parameters, besides $\bar{S}_b$, $\bar{S}_c$, are a single phase $\delta = \varphi_1 - \varphi_2 = \varphi_3 - \varphi_4$ and the parameter $p = h_1 \cdot h_2 = \pm 1$, which characterizes the relative helicity in two adjacent layers. We set a single (possibly distorted), incommensurate cylcoid into the uppermost layer $\alpha = 1$ with $\varphi_1 = 0$, and -- enforced by $K = \,$odd -- a copy with $\phi_3 = \pi$ into the third layer. We then approximate the incommensurate order by an (arbitrarily large) supercell, and discuss the symmetries that leave the magnetic moments unchanged. From this, we deduce constraints on $p$ and $\delta$.

Starting from $C_c2/m$, the cycloid in $\alpha = 1$ removes the $\mathcal{M}_a$ mirror operation. Figure \ref{fig:Efig_cycloid_C2_symmetry}a shows the point group symmetries of a cycloid, with one of the $m^\prime$ operations perpendicular to $\mathbf{e}_a$ in DyTe$_3$; but the (mirror $\times$ time reversal) operation is not consistent with the commensurate component of phase I. Instead, a $c$-glide appears in the place of $\mathcal{M}_a$ (in the frame of $Cmcm$). The translations $\mathcal{T}_1$ and $\mathcal{T}_2^\prime$ of $C_c$ (Fig. \ref{Efig_AFM_domain_cartoon}) are broken by the cycloids in $\alpha = 1, 3$. Without further symmetry lowering, the average magnetic space group is $C2/c$, where -- somewhere along the $c$-direction -- a global inversion center survives. Due to the remaining $C_{2a}$ symmetry, the four layers are stacked as $\mathbf{h}_a = (1, -1, 1, -1)$, where $\mathbf{h} = (h_1, h_2, h_3, h_4)$; recall that $h_1 = h_3$ and $h_2 = h_4$ (Fig. \ref{fig:Efig_cycloid_C2_symmetry}b). Importantly, the configuration $\mathbf{h}_b = (-1, 1, -1, 1)$ is not related to $\mathbf{h}_a$ by a broken symmetry and does not have the same energy as $\mathbf{h}_a$; the energy difference between these two is largely determined by the local Dzyaloshinskii-Moriya (local DMI) interaction on the $J_2$ bond of the zigzag chain (Fig. \ref{fig1}), i.e. the local DMI between nearest neighbors on a single square net of the DyTe slab. The competition between inter-layer Heisenberg exchange $J_1$, Eq. (\ref{eq:Heisenberg_NN_zigzag}), and intra-layer local DMI determines the choice of co-propagating or counter-propagating cycloids, and hence the average magnetic space group.

For the $\mathbf{h}_a = (1, -1, 1, -1)$ stack, a degree of freedom remains regarding the phase relation of cycloids in two layers of a DyTe slab. In particular, the phase shift angle $\delta$ determines whether blocks of parallel moments point along $\mathbf{e}_b$ and blocks of antiparallel moments point along $\mathbf{e}_c$, or vice versa. Absent further symmetry lowering, the remaining $C_{2a}$ symmetry in $C2/c$
fixes $h_4 = -h_1$, with $\delta_0 = \pm q_\mathrm{cyc}\cdot 0.5 c \approx \pm 0.207 \pi = \pm37.8^\circ$, where the sign depends on the position of the inversion center in $C_c 2/m$, i.e., on domain A or B of the AFM order. We have derived an analytic expression for the magnetic structure factor of counter-propagating cycloids in DyTe$_3$, section \ref{Esec:counter_propagating}, and consider the alternative models of high-symmetry, counter-propagating cycloids in that section.

For $\mathbf{h} = (1, -1, 1, -1)$, we can further lower symmetry to $Cc$ (on average, space group number 9) by shifting $\delta$ away from $\delta_0$, thus breaking $C_{2a}$. However, the spontaneous formation of helimagnetism with a unique sense of rotation is common in zigzag chain magnets, and more generally in systems where two or more sublattices are connected by space inversion, a twofold screw axis along the chain, and / or a glide mirror. Examples are Ni$_3$V$_2$O$_8$~\cite{Kenzelmann2006}, CuO~\cite{Brown1991}, and the zigzag chain magnet MnWO$_4$~\cite{Lautenschlaeger1993}, which all host single-sense helices under these conditions. Thus, we consider the cycloid of uniform helicity $\mathbf{h}^{(1)} = (1, 1, 1, 1)$ (average space group $Cc$), which allows the spin system to adjust $\delta$ to minimize inter-chain exchange energy (section \ref{sec:ESec_Spin_Hamiltonian}). We are left with two helicity-domains, $\mathbf{h}^{(1)}$ and $\mathbf{h}^{(2)} = -\mathbf{h}^{(1)}$ for each AFM domain ($uudd$ or $uddu$). These are related by $C_{2a}$, which reverses the rotation sense of the cycloids, but maintains the same AFM domain and the same phase relation $\delta$ for a given bond (Fig. \ref{fig:Efig_cycloid_C2_symmetry}c). As the symmetry is reduced to $Cc$ in each domain, there is no constraint on the number value of $\delta$. The discussion remains qualitatively unchanged if the commensurate part has magnetic space group $C_c 2$.

Figure \ref{Efig_Cmcm_space_group_symmetries}h demonstrates the inequivalence of bonds on the zigzag chain, caused by the commensurate magnetic structure component, which is essential for symmetry reduction and for allowing off-diagonal anisotropy terms such as $S_n^aS_n^c$ in the Hamiltonian of section \ref{sec:ESec_Spin_Hamiltonian}. In particular, $\delta$ becomes a refinable parameter only when the \textit{average} symmetry is lowered to $Cc$.

\subsection{Structure factor calculation (incommensurate)}
\label{sec:ESubsec_Struct_Factor_incomm}

We translate the the spin-chain model, Eqs. (\ref{eq:spin_texture_1dchain_b}, \ref{eq:spin_texture_1dchain_c}), into expressions suitable for structure factor calculations. Instead of the full, three-dimensional propagation vector $\mathbf{q}_\mathrm{cyc} = (0, b^*, 0.207\,c^*)$, we use $\mathbf{q}_\mathrm{cyc}^\prime = (0, 0, 0.207\,c^*)$, which helps to define the equations and relative phases of magnetic layers in language that is consistent with the main text and with section \ref{sec:ESubsec_symmetry_incomm}. The magnetic sites are labeled by index $j$ as in Fig. \ref{fig2}, $\mathbf{r}_j = \mathbf{l} + \mathbf{d}$, where $\mathbf{l}$ and $\mathbf{d}$ label the origin of the unit cell and the intra-cell coordinate of the site, respectively. As there is only one magnetic (dysprosium) site per layer $\alpha$ [defined in Eqs. (\ref{eq:spin_texture_1dchain_b}, \ref{eq:spin_texture_1dchain_c})], we replace the layer index $\alpha$ by $d$ and write
\begin{align}
\mathbf{m}_j &= 2\mathbf{X}\cos\left[h_d\cdot\mathbf{q}_\mathrm{cyc}^\prime\cdot(\mathbf{d}+\mathbf{l})+\varphi_d\right]-2\mathbf{Y}\sin\left[h_d\cdot\mathbf{q}_\mathrm{cyc}^\prime\cdot(\mathbf{d}+\mathbf{l})+\varphi_d\right]\label{eq:struct_helix}\\
&=\left(\mathbf{X}+\imath\,\mathbf{Y}\right)\exp\left(\imath\,h_d\cdot \mathbf{q}_\mathrm{cyc}^\prime\cdot(\mathbf{d}+\mathbf{l})+\imath\varphi_d\right)+\notag\\
&+\left(\mathbf{X}-\imath\,\mathbf{Y}\right)\exp\left(-\imath\,h_d\cdot \mathbf{q}_\mathrm{cyc}^\prime\cdot(\mathbf{d}+\mathbf{l})-\imath\varphi_d\right)\notag
\end{align}
where $\mathbf{X} = X\,\mathbf{e}_X$, $\mathbf{Y} = Y\,\mathbf{e}_Y$ with two orthogonal unit vectors $\mathbf{e}_X$, $\mathbf{e}_Y$. As before, the phase degree of freedom $\varphi_d$ is specific to each layer index $d(\mathbf{r}_j) = 1, 2, 3, 4$, and the prefactors $h_d=\pm1$ indicate the helicity (sense of rotation, or handedness) in a layer. Based on the discussion in section \ref{sec:ESubsec_symmetry_incomm}, we use $\varphi_1 =0$, $\varphi_2 = \pm \delta$, $\varphi_3= \pi$, and $\varphi_4 = \pi \pm \delta$, where $\pm$ indicates two magnetic domains of the phase shift angle $\delta$. 

We insert this into Eq. (\ref{eq:dsigmadomega_mag}), drop the helicity factor $h_d$, and define $\mathbf{X}_\perp = \mathbf{X} - \mathbf{X}\cdot \hat{\mathbf{Q}}$ with a normal vector $\hat{\mathbf{Q}} = \mathbf{Q} / \left|\mathbf{Q}\right|$. The quantity $\mathbf{Y}_\perp$ is defined analogously, and we are careful to consider the structure factor in the incommensurate case as a sum over the full lattice, not over a single magnetic unit cell,
\begin{align}
&I^M_\mathrm{cal} = \Phi \left|-2.7\,f_\mathrm{mag}(Q)\right|^2\left(\mathbf{X}_\perp^2+\mathbf{Y}_\perp^2\right)\cdot\\
&\cdot\left[\left|\tilde{F}^+(\mathbf{Q})\right|^2 \sum_{\mathbf{l},\mathbf{l}^\prime}\exp\left(\imath(\mathbf{Q}+\mathbf{q}_\mathrm{cyc}^\prime)\cdot(\mathbf{l}-\mathbf{l}^\prime)\right)+\left|\tilde{F}^-(\mathbf{Q})\right|^2\sum_{\mathbf{l},\mathbf{l}^\prime}\exp\left(\imath(\mathbf{Q}-\mathbf{q}_\mathrm{cyc}^\prime)\cdot(\mathbf{l}-\mathbf{l}^\prime)\right)\right]\notag\\
&\qquad\qquad\qquad\qquad\tilde{F}^+(\mathbf{Q}) = \sum_{\mathbf{d}\in \mathrm{c.u.c.}} \exp\left(\imath(\mathbf{Q}+\mathbf{q}_\mathrm{cyc}^\prime)\cdot\mathbf{d}+\imath\varphi_d\right)\label{eq:Fplus}\\
&\qquad\qquad\qquad\qquad\tilde{F}^-(\mathbf{Q}) = \sum_{\mathbf{d}\in \mathrm{c.u.c.}} \exp\left(\imath(\mathbf{Q}-\mathbf{q}_\mathrm{cyc}^\prime)\cdot\mathbf{d}-\imath\varphi_d\right)\label{eq:Fminus}
\end{align}
Now, the sums in Eq. (\ref{eq:Fplus}, \ref{eq:Fminus}) are over a single chemical unit cell (c.u.c.). This can be rewritten, due to the periodicity of the lattice, as
\begin{align}
I^M_\mathrm{cal} &= \Phi \left|-2.7\,f_\mathrm{mag}(Q)\right|^2\left(\mathbf{X}_\perp^2+\mathbf{Y}_\perp^2\right)\left(N\frac{(2\pi)^3}{\nu_0}\right)\cdot\\
&\cdot\left[\left|\tilde{F}^+(\mathbf{Q})\right|^2 \sum_\mathbf{G}\delta(\mathbf{Q}+\mathbf{q}_\mathrm{cyc}^\prime-\mathbf{G})+\left|\tilde{F}^-(\mathbf{Q})\right|^2 \,\sum_\mathbf{G}\delta(\mathbf{Q}-\mathbf{q}_\mathrm{cyc}^\prime-\mathbf{G})\right]\notag
\end{align}
where $\mathbf{G}$ are the reciprocal lattice vectors. Enforced by the $\delta$-functions, the $\tilde{F}^\pm$ depend only on the reciprocal lattice vector $\mathbf{G}_0(\mathbf{Q})$ from which the incommensurate reflection 'originates', i.e. $\mathbf{Q} = \mathbf{G}_0 + \mathbf{q}_\mathrm{cyc}^\prime$ and $\tilde{F}^\pm(\mathbf{Q})\equiv \tilde{F}^\pm(\mathbf{G}_0)$, with
\begin{equation}
\mathbf{G}_0 = H_0\frac{2\pi}{a}\,\mathbf{e}_a + K_0\frac{2\pi}{b}\,\mathbf{e}_b + L_0\frac{2\pi}{c}\,\mathbf{e}_c
\end{equation}
and $H_0$, $K_0$, $L_0$ integers.

Specializing to the $(0KL)$ scattering plane, $\tilde{F}^\pm$ can be calculated explicitly by considering four types of pairs $\mathbf{d}=\mathbf{d}^\prime$, as well as four types of non-identical partners (we omit the $a$-component of each vector),
\begin{align}
\mathbf{d}_1-\mathbf{d}_3 = \mathbf{d}_2-\mathbf{d}_4 & = \frac{b}{2}\,\mathbf{e}_b,& \,\,\,\,\,\,\Delta \varphi_{1,3} &= -\pi\\
\mathbf{d}_2-\mathbf{d}_1 = \mathbf{d}_4-\mathbf{d}_3 & = \frac{c}{2}\,\mathbf{e}_c- b_0\,\mathbf{e}_b,& \,\,\,\,\,\,\Delta \varphi_{2,1} &= \delta\\
\mathbf{d}_4-\mathbf{d}_1 & = \frac{c}{2}\,\mathbf{e}_c- \left(\frac{b}{2}+b_0\right)\,\mathbf{e}_b,& \,\,\,\,\,\,\Delta \varphi_{4,1} &= \delta + \pi\\
\mathbf{d}_2-\mathbf{d}_3 & = \frac{c}{2}\,\mathbf{e}_c+ \left(\frac{b}{2}-b_0\right)\,\mathbf{e}_b,& \,\,\,\,\,\,\Delta \varphi_{2,3} &= \delta -\pi
\end{align}
with $b_0$ the spacing, close to $b/6$, between two layers in a covalently bonded bilayer (Fig. \ref{Efig_AFM_domain_cartoon}). Due to the above definition of $\mathbf{q}_\mathrm{cyc}^\prime$, these phases are identical for the calculation of both $\tilde{F}^\pm(\mathbf{G}_0)$. The parameter $\delta$ by itself is either negative and positive, depending on the domain ($A^\prime$, $B^\prime$).

Executing Eqs. (\ref{eq:Fplus}, \ref{eq:Fminus}) in terms of this limited set of atom pairs,
\begin{equation}
\left|\tilde{F}^\pm_{\delta}(\mathbf{G}_0)\right|^2 = 4\left[1-\cos(\pi K_0)\right]\cdot\left[1+\cos(\pi L_0)\cos\left(-\frac{2\pi b_0}{b}\,K_0\pm\delta\right)\right]\label{eq:int_incomm_one_domain}
\end{equation}
The total scattering intensity for two equally populated domains of the $\delta$ angle is hence zero for $K_0=\,$ even and, for $K_0 =\,$odd,
\begin{align}
I^M_\mathrm{cal} &= \Phi \left(2.7\,f_\mathrm{mag}(Q)\right)^2\left(\mathbf{X}_\perp^2+\mathbf{Y}_\perp^2\right)\left(N\frac{(2\pi)^3}{\nu_0}\right)\cdot \frac{1}{2}\left[\left|\tilde{F}^\pm_{+\delta}(\mathbf{G}_0)\right|^2 + \left|\tilde{F}^\pm_{-\delta}(\mathbf{G}_0)\right|^2\right]\label{eq:int_incomm_two_domain}\\
&\qquad\qquad\Big[\ldots\Big] =8\left(1+\cos(\pi L_0)\cdot\cos\left(\frac{2\pi b_0}{b}K_0\right)\cdot\cos(\pm\delta)\right) 
\end{align}
independent of whether we are looking at the left / right reflection; i.e., the $\pm$ sign in $\tilde{F}_\pm$ is completely cancelled out. It is, finally, possible to simplify the magnetic moment projection as
\begin{align}
\mathbf{X}_\perp^2 = \mathbf{X}^2 &- \left(\mathbf{X}\cdot \hat{\mathbf{Q}}\right)^2\\
\left(\mathbf{X}_\perp^2+\mathbf{Y}_\perp^2\right)/X^2 &=\sin^2\beta + \gamma^2 \cos^2\beta
\end{align}
with $\gamma^2 = Y^2 / X^2$, $\beta = \angle(\hat{\mathbf{Q}}, \mathbf{e}_c)$, and we set $\mathbf{e}_X = \mathbf{e}_c$, $\mathbf{e}_Y = \mathbf{e}_b$.

%% %%%%%%%%%%%%%%%%%%%%%%%%%%%%%%%%%%%%%%%%%%%%%%%%%%%%%%%
%%          SUBSECTION
%% %%%%%%%%%%%%%%%%%%%%%%%%%%%%%%%%%%%%%%%%%%%%%%%%%%%%%%%
\subsection{Example: intensity ratio of two reflections}

\begin{table}[htb]
    \centering 
    \caption{\textbf{Calculation of expected intensity ratio of scattered neutron intensity at $(0, 9, 0.207)$ and $(0, 1, 1.207)$, based on magnetic structure model.} The intensity ratio $r$ is calculated from Eq. (\ref{eq:peak_intensity_ratio}), using these parameters. For convenience, we use $\mathcal{L} = 1/\sin(2\theta)$ (neutron wavelength $\lambda= 1$) in the calculation of the Lorentz factor.}
    \label{ETab2}
    \vspace*{1em}
    \begin{tabular}{ccc|cccc|c|c|c|c}
    \hline
    \hline
    \hspace*{.7em}$H$\hspace*{.7em} & \hspace*{.7em}$K$\hspace*{.7em} & \hspace*{.7em}$L$\hspace*{1.1em} &\hspace*{.7em} $Q_a$\hspace*{.7em} & \hspace*{.7em}$Q_b$\hspace*{.7em} & \hspace*{.7em}$Q_c$\hspace*{.7em} &\hspace*{.7em} $|Q|$\hspace*{.7em} & \hspace*{.9em}$2\theta$\hspace*{.9em} & \hspace*{.9em}$\mathcal{L}(\theta)$\hspace*{.9em} &\hspace*{.9em} $f_\mathrm{mag}$ \hspace*{.9em}&\hspace*{.9em} $\beta$\hspace*{.9em} \\
    \multicolumn{3}{c|}{(r.l.u.)}& \multicolumn{4}{c|}{$\mathrm{(\AA^{-1})}$} & (deg.) & $\,$ &  $\,$ & (deg.)\\
    \hline
    \hline
    0 & 9	& 0.207 &	0	& 2.228 & 	0.302 &	2.248 &	51.853 & 1.272 & 0.825 & 82.27 \\
    0 & 1 & 1.207 & 0 & 0.248 &	1.763 &	1.780 &	40.505 &	1.533 & 0.885 & 7.99\\
    \hline
    \hline
    \end{tabular}
\end{table}

Specifically for two reflections $\mathbf{Q}_1$ and $\mathbf{Q}_2$ in reciprocal space, with Miller indices $(0, 9, 0.207)$ and $(0, 1, 1.207)$, the corresponding Miller indices for the $\mathbf{G}_0$ positions are $(090)$ and $(011)$, respectively. According to Eq. (\ref{eq:int_incomm_two_domain}) with $\delta = 79^\circ$ from the refinement in section \ref{Esec:magnetic_structure_analysis}, the intensity ratio is
\begin{equation}
r = \frac{I^M_\mathrm{cal}(\mathbf{Q}_1)}{I^M_\mathrm{cal}(\mathbf{Q}_2)} = \frac{\mathcal{L}_1}{\mathcal{L}_2}\cdot\frac{\sin^2\beta_1 + \gamma^2 \cos^2\beta_1}{\sin^2\beta_2 + \gamma^2 \cos^2\beta_2}\cdot \frac{f_\mathrm{mag}(Q_1)^2}{f_\mathrm{mag}(Q_2)^2}\cdot \frac{1+\cos\left(18\pi\frac{b_0}{b}\right)\cos\delta}{1-\cos\delta}\label{eq:peak_intensity_ratio}
\end{equation}
where $\mathcal{L} = 1/\sin(2\theta)$ (scattering angle $2\theta$) is the Lorentz factor, which corrects for the scattering geometry. Table \ref{ETab2} shows the numerical values for various steps in the calculation. From the observed ratio of peak intensities in Fig. \ref{fig2}, $r =0.77$   
and $\gamma = Y/X \approx 0.97$ for these two reflections measured on sample A. This $\gamma$ is somewhat larger than the result for sample B with full refinement in section \ref{Esec:magnetic_structure_analysis}, but sample A is a larger crystal, with anisotropic shape, where absorption correction is not applied. In particular, the observed intensities are expected to be larger along $(0, 1, L)$, close to transmission geometry. Such limitations of the data quality for sample A do not affect polarization analysis and the qualitative evolution of line scan intensities with temperature.

%% %%%%%%%%%%%%%%%%%%%%%%%%%%%%%%%%%%%%%%%%%%%%%%%%%%%%%%%
%%          SUBSECTION
%% %%%%%%%%%%%%%%%%%%%%%%%%%%%%%%%%%%%%%%%%%%%%%%%%%%%%%%%
\subsection{Structure factor for counter-propagating cycloids}
\label{Esec:counter_propagating}

\begin{figure}[h]%
\centering
\includegraphics[width=0.55\textwidth]{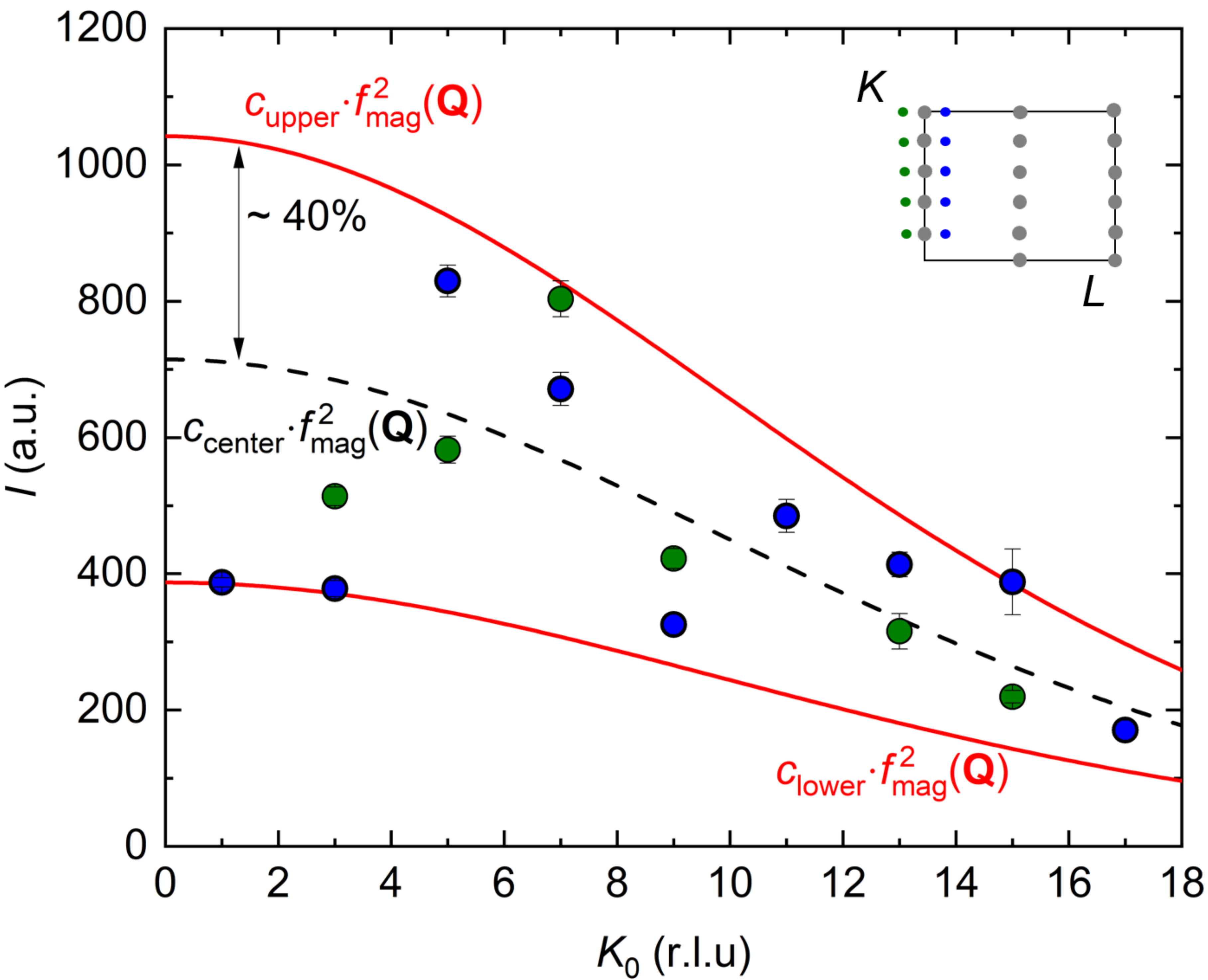}
\caption{\textbf{Testing analytic model for the magnetic structure factor of the magnetic ground state of DyTe$_\mathbf{3}$.} Observed magnetic structure factor, after correction for the Lorentz factor, for incommensurate reflections with $L_0 = 0$. The inset depicts the relevant region of momentum space in the $(0KL)$ scattering plane. The red curves (black dashed line) are envelopes to the data (is the center of the envelopes), bounding the data which oscillates according to Eq. (\ref{eq:oscillating_delta}). Blue and green are reflections with $\pm \mathbf{q}^\prime$, respectively. The oscillation amplitude rules out $|\cos\delta| > 0.4$, as discussed in section \ref{Esec:counter_propagating}.}\label{EfigEvolutionStructureFactor}
\end{figure}

Previously, a stack of counter-propagating helices was reported in $\gamma$-Li$_2$IrO$_3$ and ascribed to the presence of Kitaev-type anisotropic exchange interactions~\cite{Biffin2014}. Analogously, we consider the structure factor for the helicity stack $\mathbf{h} = (+1, -1, +1, -1)$ as defined in section \ref{sec:ESubsec_symmetry_incomm}. In Eq. (\ref{eq:struct_helix}), $\delta = \pm \pi L_\mathrm{cyc}$ ($\delta = \pi \pm \pi L_\mathrm{cyc}$) can describe counter-propagating cycloids where the textures of two layers $d = 1,2$ meet, at certain points, to create pairs of moments that are parallel to the $b$-axis (parallel to the $c$-axis). Recognizing that 
\begin{align}
\mathbf{m}_\perp & = \mathbf{X}_\perp \cos\left(h_d \mathbf{q}^\prime \cdot \mathbf{r}_j +\varphi_d\right) + \mathbf{Y}_\perp \sin\left(h_d \mathbf{q}^\prime \cdot \mathbf{r}_j +\varphi_d\right)=\notag\\
&=\left(\mathbf{X}_\perp + ih_d \mathbf{Y}_\perp\right)\exp\left(i\mathbf{q}^\prime \cdot \mathbf{r}_j +ih_d \varphi_d\right)+\notag\\
&+\left(\mathbf{X}_\perp - ih_d \mathbf{Y}_\perp\right)\exp\left(-i\mathbf{q}^\prime \cdot \mathbf{r}_j -ih_d \varphi_d\right),
\label{eq:counter-prop_m}
\end{align}
algebra in line with section \ref{sec:ESubsec_Struct_Factor_incomm}, for one domain of the $\delta$-angle, yields 
\begin{align}
I^M_\mathrm{cal,\pm} &= \Phi \left(2.7\,f_\mathrm{mag}(Q)\right)^2\left(N\frac{(2\pi)^3}{\nu_0}\right)\cdot 4\left[1-\cos(\pi K_0)\right]\cdot\notag\\
&\qquad\qquad\cdot\left[\mathbf{X}_\perp^2+\mathbf{Y}_\perp^2+ \left(\mathbf{X}_\perp^2-\mathbf{Y}_\perp^2\right)\cos\tau\mp 2\mathbf{X}_\perp\cdot \mathbf{Y}_\perp \cdot\sin\tau\right]\\
\tau&= \frac{2\pi b_0}{b}K_0 -\pi L_0 \pm \delta
\end{align}
Special cases can be considered. When $\mathbf{Q}$ is (nearly) parallel to $\mathbf{e}_b$, $\mathbf{Y}_\perp \approx 0$ and the co- and counter-propagating models deliver the same result (now including two $\delta$-domains, introducing a separate $\cos\delta$ factor):
\begin{align}
I^M_\mathrm{cal,\pm} & \propto (1-\cos(\pi K_0))\,\, \mathbf{X}_\perp^2 \cdot \left(1+\cos(\pi L_0)\cos\delta \cos\left(\frac{2\pi b_0}{b}K_0\right)\right) 
\label{eq:oscillating_delta}
\end{align}
This is expected from Eq. (\ref{eq:counter-prop_m}), where additional $h_d$'s appear only before $\mathbf{Y}_\perp$. Figure \ref{EfigEvolutionStructureFactor} shows the intensities for $L_0 = 0$, as a function of $K_0$, with an upper and lower envelope function defined by the magnetic form factor. The maximum amplitude of the oscillation with $K$, $\sim 40\,\%$, suggests that $\abs{\cos \delta} \le 0.4 $, excluding $\delta = \pm \pi L_\mathrm{cyc} = \pm38^\circ$ and $\delta = \pi \pm 38^\circ$ corresponding to counter-propagating cycloids with high symmetry (see above).

%%%%%%%%%%%%%%%%%%%%%%%%%%%%%%%%%%%%%%%%%%%%%%%%%%%%%%%%%%%%%%%%%
%%%%%%%%%%%%%%%%%%%%%%%%%%%%%%%%%%%%%%%%%%%%%%%%%%%%%%%%%%%%%%%%%
%%                   NEW SECTION
%%%%%%%%%%%%%%%%%%%%%%%%%%%%%%%%%%%%%%%%%%%%%%%%%%%%%%%%%%%%%%%%%
%%%%%%%%%%%%%%%%%%%%%%%%%%%%%%%%%%%%%%%%%%%%%%%%%%%%%%%%%%%%%%%%%

\section{Analysis of nuclear elastic neutron scattering}
\label{Esec:nuclear_neutron_scattering}

\begin{figure}[h]%
\centering
\includegraphics[width=0.9\textwidth]{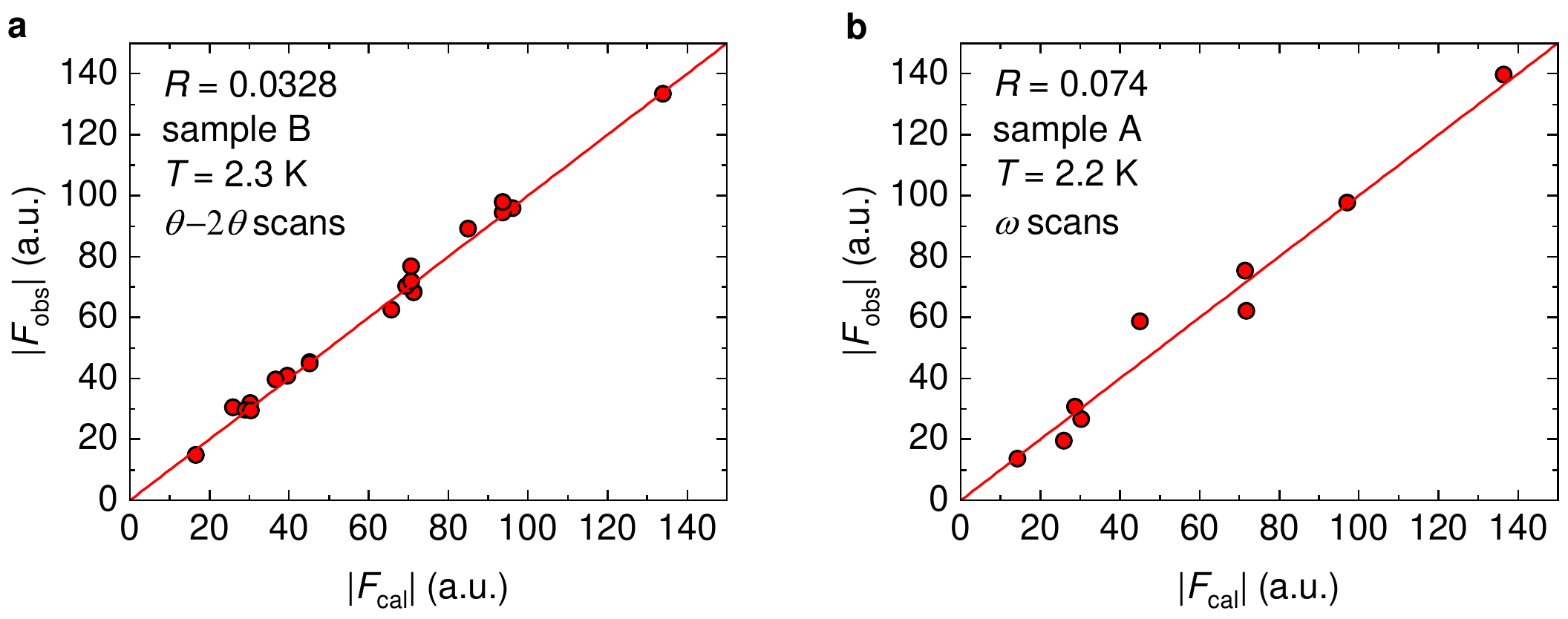}
\caption{\textbf{Nuclear reflections observed in neutron scattering, consistent with orthorhombic crystal structure of DyTe$_\mathbf{3}$.} The observed structure factor $F_\mathrm{obs}$ is compared to model calculations, $F_\mathrm{cal}$, in space group $Cmcm$ as detailed in section \ref{Esec:scattering_intensities}. We use the reliability index for the structure factor, $R = \sum{\left| F_\mathrm{obs}-F_\mathrm{cal}\right|} / \sum{\left| F_\mathrm{obs}\right|}$, where the sums are over all data points; $R = 0.0328$ for sample B (\textbf{a}) and $R = 0.074$ for sample A (\textbf{b}) indicate good agreement of model and fit. Note that $\omega$-scans without absorption correction are used to obtain the integrated intensities for sample A.}\label{EfigNuclearRefinement}
\end{figure}

For sample B, we collected $20$ nuclear reflections, which contain $17$ independent reflections. Observed and calculated nuclear structure factors are compared to determine the scale factor, a parameter for the extinction correction, and isotropic atomic displacement factors $B_{iso}$. These three parameters are later used in the magnetic structure analysis (section \ref{Esec:magnetic_structure_analysis}). For simplicity, we assumed that all the tellurium sites have the same $B_{iso}$. We performed a least-squares fit and found that the observed and calculated structure factors agree well; $R(F)$ is $3.28\,\%$. The values of $B_{iso}$ for Dy and Te are determined to be $0.1(2)$ and $0.8(2)$, respectively. We also performed the same analysis after averaging the structure factors of the equivalent reflections, and obtained the internal $R(F)$ value of $3.35\,\%$.

For sample A, we collected $9$ independent nuclear reflections. Observed and calculated nuclear structure factors are compared to determine the scale factor, while fractional coordinates of the atoms and $B_{iso}$ are fixed at the values reported in Ref. 
\cite{Pardo1967}. 
We performed a least-squares fit and found that the observed and calculated structure factors agree well: $R(F)=7.4\,\%$ (Fig. \ref{EfigMagneticRefinement}).\\

%%%%%%%%%%%%%%%%%%%%%%%%%%%%%%%%%%%%%%%%%%%%%%%%%%%%%%%%%%%%%%%%%
%%%%%%%%%%%%%%%%%%%%%%%%%%%%%%%%%%%%%%%%%%%%%%%%%%%%%%%%%%%%%%%%%
%%                   NEW SECTION
%%%%%%%%%%%%%%%%%%%%%%%%%%%%%%%%%%%%%%%%%%%%%%%%%%%%%%%%%%%%%%%%%
%%%%%%%%%%%%%%%%%%%%%%%%%%%%%%%%%%%%%%%%%%%%%%%%%%%%%%%%%%%%%%%%%
\section{Results of magnetic structure analysis}
\label{Esec:magnetic_structure_analysis}

\begin{figure}[h]%
\centering
\includegraphics[width=0.9\textwidth]{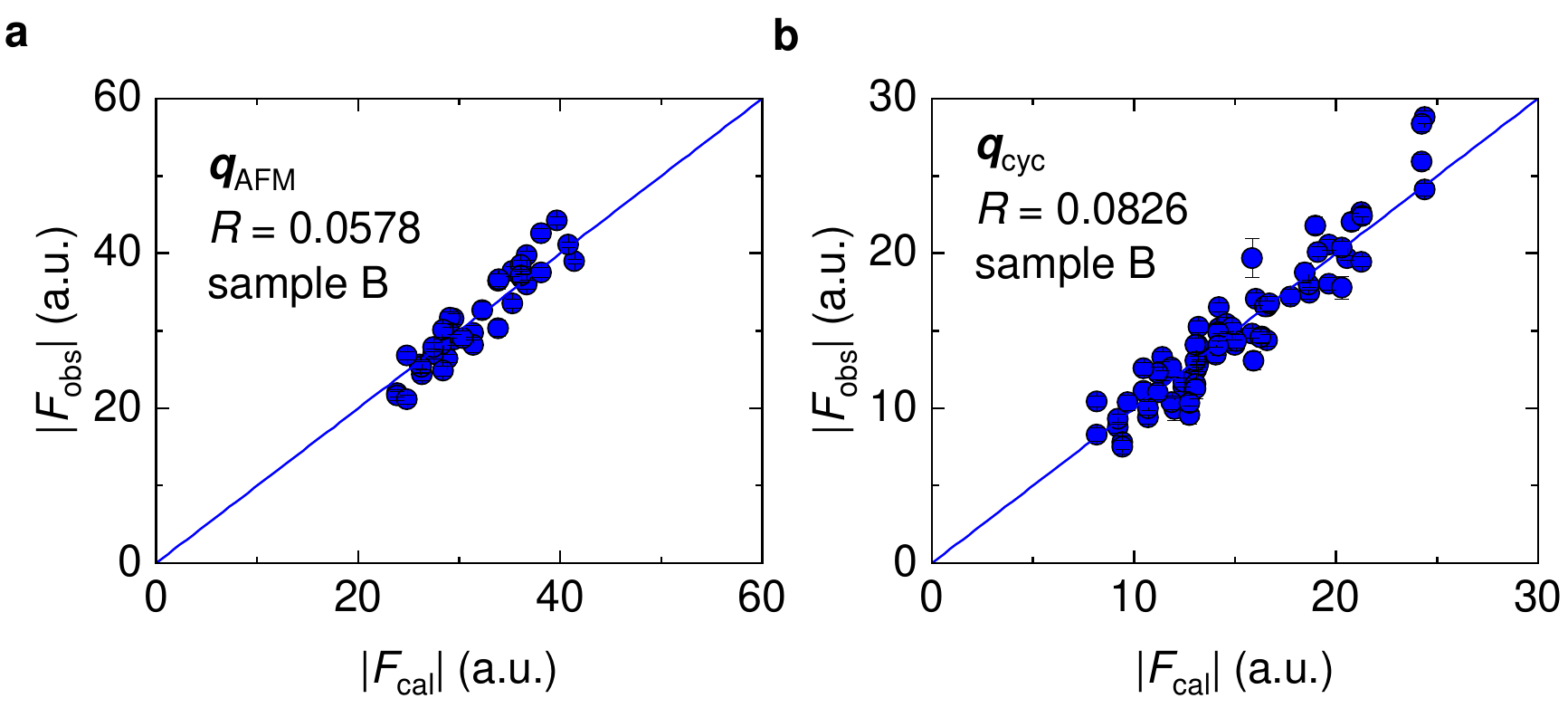}
\caption{\textbf{Magnetic structure refinement for the ground state of DyTe$_\mathbf{3}$.} Antiferromagnetic (commensurate, $\mathbf{q}_\mathrm{AFM}$) and cycloidal (incommensurate $\mathbf{q}_\mathrm{cyc}$) components are refined separately from neutron scattering data, without polarization analyzer, at $T = 2.2\,K$sample B. \textbf{a}, Averaging two AFM domains with equal weight, we find good agreement [reliability factor $R = 0.0578$] between the measured neutron structure factor $F_\mathrm{obs}$ and model calculations $F_\mathrm{cal}$. \textbf{b}, Likewise, good agreement is found when summing two $\delta$-domains for the incommensurate order $\mathbf{q}_{cyc}$. Here, we minimized $R$ by adjusting the value of the phase shift $\delta$ between sheets in a bilayer, as defined in Fig. \ref{fig3}\,g, and the elliptic distortion $Y/X$ of the cycloid (section \ref{Esec:magnetic_structure_analysis}).}\label{EfigMagneticRefinement}
\end{figure}

For sample B, we collected $38$ ($70$) Bragg reflections for the commensurate (incommensurate) magnetic reflections, all of which are independent. 
The polarized neutron scattering experiment in Fig. \ref{EfigFlippingRatios} reveals that the magnetic moments corresponding to the commensurate component are parallel to the $a$-axis. Assuming the volume fractions of two domains in Fig. \ref{Efig_AFM_domain_cartoon} to be equal, we performed a least-squares fit to the nonpolarized neutron scattering data from sample B and found that the magnitude of the commensurate component is $5.52(2)\,\mu_B$, with $R(F)=5.78\,\%$. 
 
For the incommensurate magnetic component, polarized neutron scattering shows that the magnetic moments are confined in the $bc$-plane. Further considering the existence of the third higher-harmonic magnetic reflections, we assume an elliptic (distorted) cycloidal magnetic modulation, with principal axes parallel to the $b$- and $c$-axes, and moment amplitudes $m_b$ and $m_c$ common to all four Dy atoms in a chemical unit cell. We assume equal volume fraction of two domains for the phase shift $\delta$, described in section \ref{sec:Esec_structure_factors_incomm}, and refine $m_b$, $m_c$, and $\delta$ by least-squares analysis. This yields $3.93(3)\,\mu_B$, $6.61(3)\,\mu_B$, and $\pm 79.2(5)\,$deg, respectively. The $R(F)$ value is $8.26\,\%$ (Fig. \ref{EfigMagneticRefinement}) and combined, the total moment length at the Dy site is 
\begin{equation}
m_\mathrm{tot} = \sqrt{m_{a}^2+\max{\left(m_{b}^2,m_{c}^2\right)}} = 8.61\pm 0.1\,\mu_\mathrm{B}/\mathrm{Dy}
\end{equation}
at $T = 2.2\,$ K, which is about $T/T_\mathrm{N} = 0.6$, i.e. still rather close to the critical temperature; the ordered moment at absolute zero temperature is expected to be close to the local-ion value of $10\,\mu_\mathrm{B}$ per dysprosium, also due to an underestimation in our calculation by neglecting higher harmonics contributions. The difference in value between $m_b$ and $m_c$ is consistent with the presence of third harmonic magnetic reflections; such distortions and higher harmonics have also been observed in insulating multiferroics~\cite{Kenzelmann2005}.
 
Our model demonstrates that, at least, the incommensurate component involves both $m_b$ and $m_c$ components, of different amplitudes. Further experiments using spherical neutron polarimetry will be suitable to investigate, in more detail, the directions of the minor and major axes for the spin ellipsis in the $bc$ plane.

The lattice constants $a = c=4.302\,\mathrm{\AA}$, $b=25.381\,\mathrm{\AA}$ are determined from neutron scattering at low temperature. These values are close to the literature values for the orthorhombic, yet nearly tetragonal structure of DyTe$_3$, $a=c=4.296\,\mathrm{\AA}$, $b=25.450\,\mathrm{\AA}$~\cite{Pardo1967}. Our neutron measurements are not able to pick up minute atomic displacements due to the CDW, which can be seen in careful x-ray studies~\cite{Malliakas2005}; we used the fractional atomic positions of the average structure from Ref.~\cite{Pardo1967} in our analysis of the nuclear and magnetic scattering.\\

%%%%%%%%%%%%%%%%%%%%%%%%%%%%%%%%%%%%%%%%%%%%%%%%%%%%%%%%%%%%%%%%%
%%%%%%%%%%%%%%%%%%%%%%%%%%%%%%%%%%%%%%%%%%%%%%%%%%%%%%%%%%%%%%%%%
%%                   NEW SECTION
%%%%%%%%%%%%%%%%%%%%%%%%%%%%%%%%%%%%%%%%%%%%%%%%%%%%%%%%%%%%%%%%%
%%%%%%%%%%%%%%%%%%%%%%%%%%%%%%%%%%%%%%%%%%%%%%%%%%%%%%%%%%%%%%%%%

\section{Anisotropic electronic transport properties}
\label{sec:ESec_Montgomery}

\begin{figure}[h]%
\centering
\includegraphics[width=0.95\textwidth]{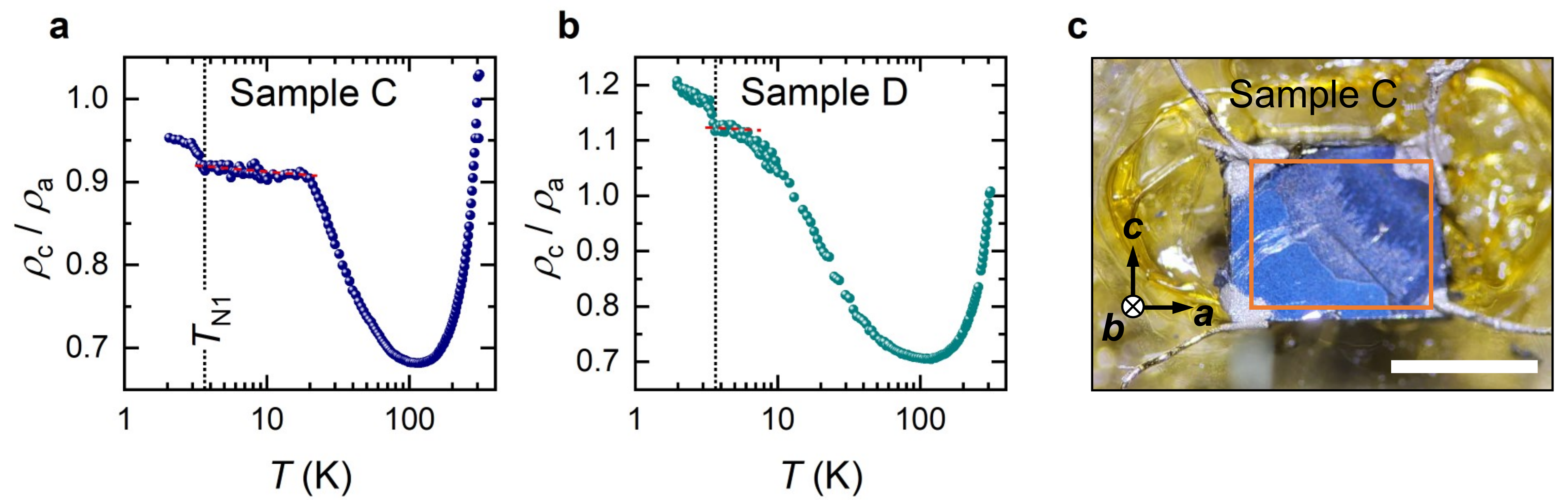}
\caption{\textbf{Anisotropy of electrical transport properties in $ac$ basal plane of DyTe$_\mathbf{3}$.} \textbf{a,b}, $ac$ basal plane resistance ratio $\rho_c/\rho_a$ as a function of temperature for different samples. In Sample D, the assumed rectangular sample shape (c.f. panel c) is slightly -- in particular 4\,\% along the $a$ direction -- adjusted to match $\rho_c/\rho_a \sim 1$ at $T>T_\mathrm{CDW}$, c.f. Eq. (\ref{Eeq:Montgomery_ratio}). Both samples show a clear kink at the magnetic phase transition $T_\mathrm{N1}$, followed by a broad plateau region, indicated by a red dashed line, with nearly constant ratio. \textbf{c}, Sample C mounted in Montgomery geometry. The orange rectangle depicts the assumed sample shape that is used to estimate the length ratio $l_a/l_c = 1.266$, where the white scale bar on the bottom right corresponds to $1\,$mm.}\label{EfigAnisoTransport}
\end{figure}

Angle-resolved photoemission (ARPES) studies combined with a tight binding (TB) model in Ref. \cite{Brouet2008} reveals that, in the family of the rear-earth tritellurides $R$Te$_3$, the states at the Fermi level are mainly formed by the in-plane $p_x$ and $p_z$ orbitals of the Te-A ions in the Te square net, c.f. Fig. \ref{fig1}a. Indeed, they are well separated by more than 1\,eV from other bands, already indicating anisotropic bonding and transport properties. With a standard four-probe method and a modified Montgomery geometry Ru \textit{et al}. determined both in-plane ($\rho_{ac}$) and out-of-plane resistance ($\rho_b$) of various $R$Te$_3$ compounds \cite{Ru2008}. They differ by at least one order of magnitude ($\rho_b\gg \rho_{ac}$), consistent with a metallic Te square net as well as covalently bonded $R$Te slabs.

To further extract the in-plane anisotropy ratio $\rho_c/\rho_a$ of DyTe$_3$, we use the Montgomery technique and follow the method described in Ref. \cite{DosSantos2011}. From temperature dependent resistance measurements $R_a(T)$ and $R_c(T)$, we calculate the shape of a hypothetical isotropic sample with dimensions $L_a\times L_b \times L_c$ and same absolute resistance values $R_a^\prime$ and $R_c^\prime$
\begin{equation}
    L = \frac{L_c}{L_a} = \frac{1}{2}\,\left[\,\frac{1}{\pi}\,\ln\left(\frac{R_c}{R_a}\right)+\sqrt{\left(\frac{1}{\pi}\,\ln\left(\frac{R_c}{R_a}\right)\right)^2+4}\,\right]
\end{equation}
Based on published data from Ref. \cite{Ru2008} and our sample geometry, a thin exfoliated flake, we estimated that the calculation of the anisotropy ratio $\rho_c/\rho_a$ is possible in the thin-layer limit. Following Ref. \cite{DosSantos2011} and using the real sample dimensions $l_a\times l_b \times l_c$ this leads to
\begin{align}
    \rho_a &= \frac{\pi}{8}\,\frac{l_b\,l_c}{l_a}\,L^{-1}\,R_a\,\sinh{\left(\pi L\right)}\\
    \rho_c &= \frac{\pi}{8}\,\frac{l_b\,l_a}{l_c}\,L\,R_a\,\sinh{\left(\pi L\right)}
\end{align}
and finally to
\begin{equation}
    \frac{\rho_c}{\rho_a} = \left(\frac{l_a}{l_c}\,L\right)^2 \label{Eeq:Montgomery_ratio}
\end{equation}

There is some ambiguity in the interpretation of resistance anisotropy changes at $T_N$, which can be ascribed either to opening of a partial charge gap in the ordered state, or to fluctuations in the paramagnetic regime and their suppression below $T_N$. In DyTe$_3$, the sign of the observed change in $\rho_c/\rho_a$ implies that, when moving from the paramagnetic into the ordered regime, the resistance along the $c$-axis becomes larger than the resistance along the $a$-axis. This result can be neatly explained by attributing partial gap opening to $\mathbf{q}_\mathrm{AFM}= (0, b^*, 0.5c^*)$, $\mathbf{q}_\mathrm{cyc} = (0, b^*, 0.207c^*)$, while unidirectional fluctuations should enhance $\rho_c/\rho_a$ above $T_N$, and suppress it below $T_N$.

\begin{figure}[h]%
\centering
\includegraphics[width=0.6\textwidth]{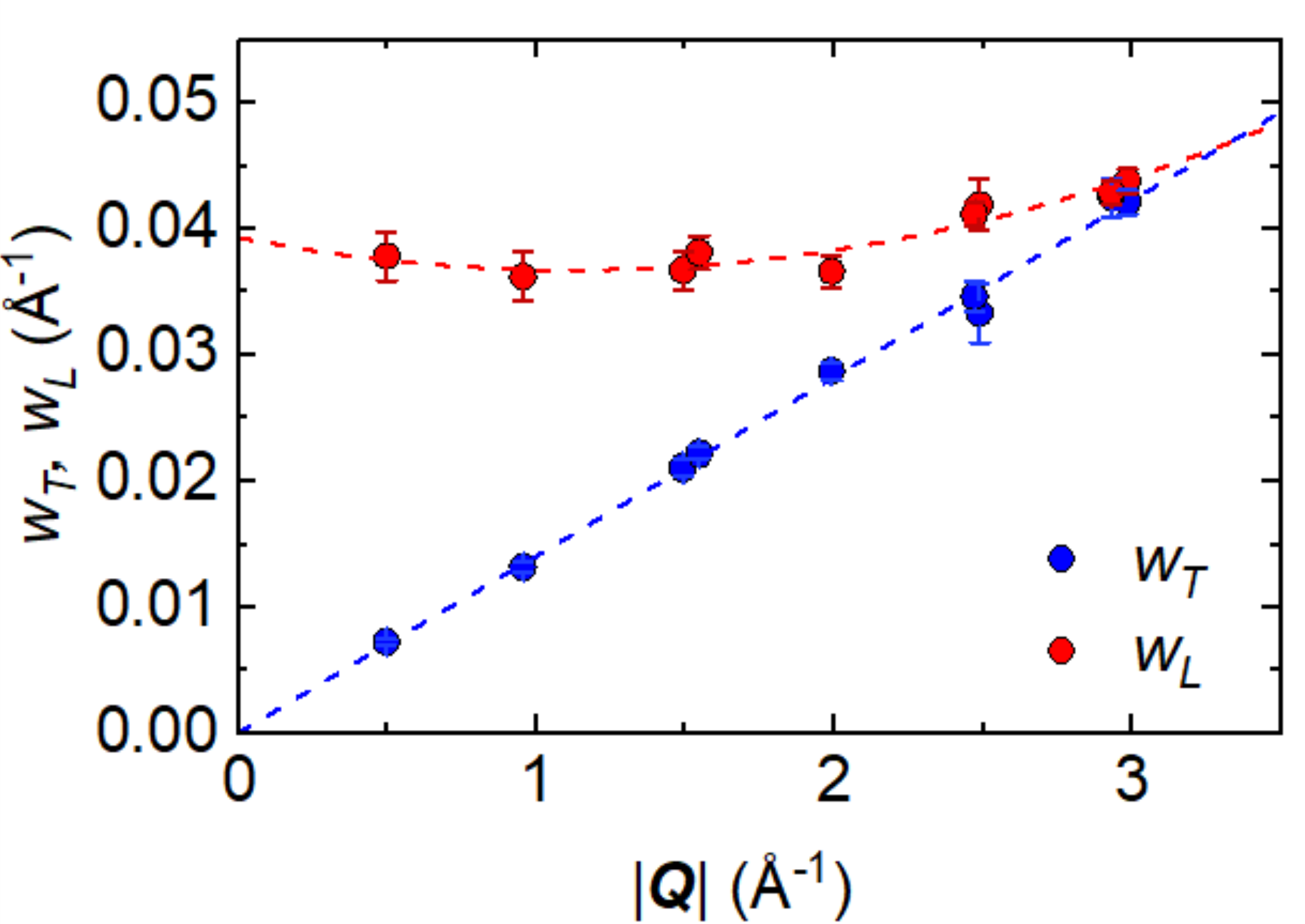}
\caption{\textbf{Calibration of instrument resolution at 5G-PONTA at the JRR-3 neutron reactor source, for sample A.} Each pair of (red, blue) data points corresponds to the full width at half maximum (FWHM) of a nuclear lattice reflection. The shape of the resolution ellipsoid in the $b^*$-$c^*$ plane is defined by the width of an $\omega$-scan ($w_T$, transverse width) and the width of a $\omega-2\theta$ scan ($w_L$, longitudinal width), respectively. These parameters depend on the sample shape, crystal quality, the momentum transfer $\left|\mathbf{Q}\right|$, and the performance of the instrument. Blue and red dashed lines are a linear fit, and a second order polynomial fit to the data, respectively. These fits can be used for estimation of the instrument resolution at arbitrary positions in the $bc$ scattering plane, e.g. in Fig. \ref{fig4}\,c. Error bars correspond to statistical uncertainties of Gaussian fits to the nuclear reflections.}\label{EfigCalibrationPONTA}
\end{figure}

\begin{figure}[h]%
\centering
\includegraphics[width=0.9\textwidth]{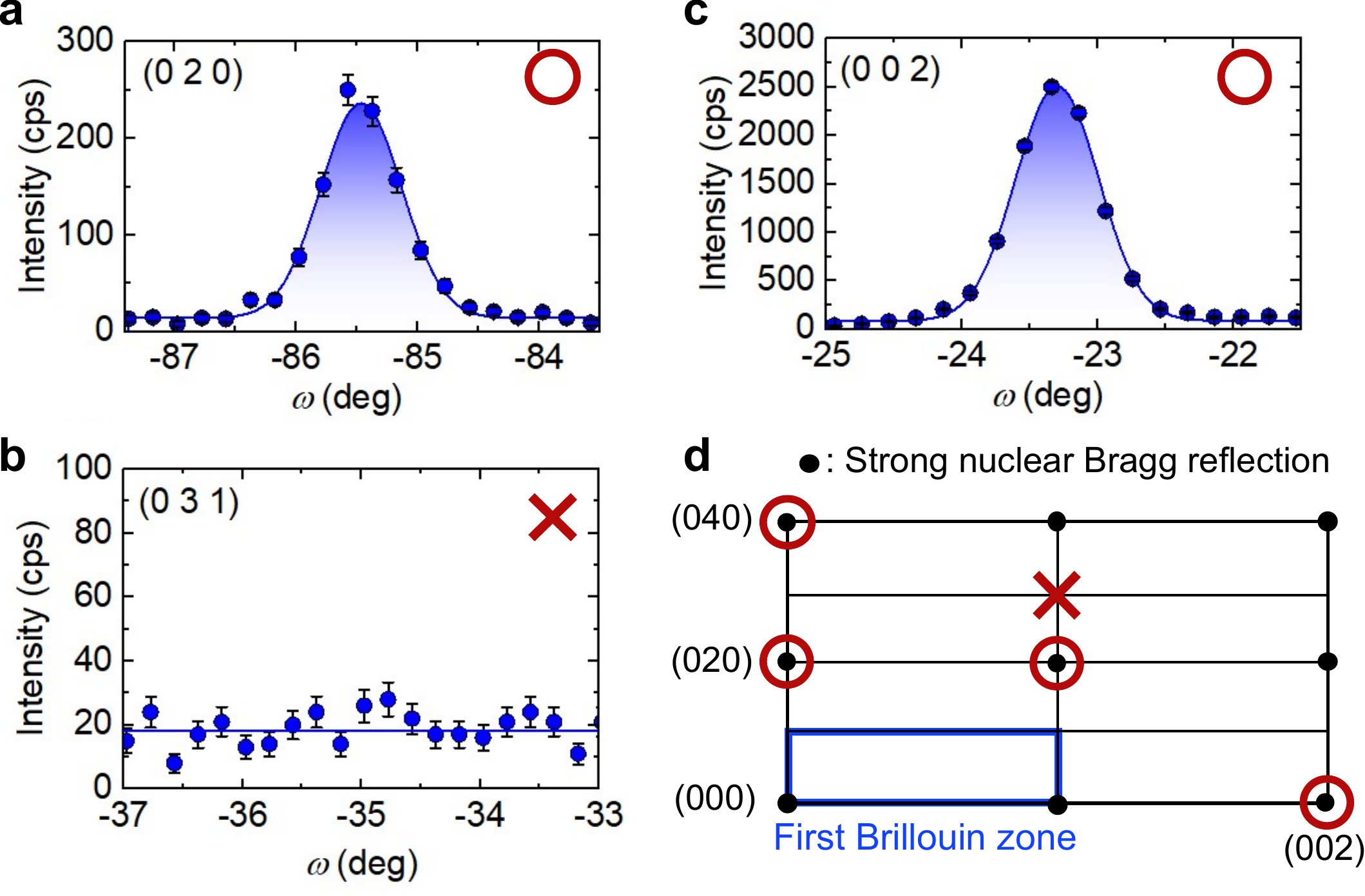}
\caption{\textbf{Confirmation of crystallographic extinction rule in DyTe$_\mathbf{3}$ in sample A.} Reflections $h+k = \mathrm{odd}$ are forbidden in $Cmcm$ (space group 63), or specifically $k = 2n$ is required in the $(0KL)$ scattering plane, which is used for the present experiment. We observe zero intensity at $(031)$ and $800\,$counts per second (cps) at $(021)$; the latter reflection is not shown in this figure. When instead rotating the sample into the $(HK0)$ scattering plane, $(130)$ has $1200\,$cps and $(120)$ exhibits zero intensity. Specifically, the comparison of $(130)$ and $(031)$ allows us to confirm the alignment of the crystal in our neutron experiment.}\label{EfigExtinctionSampleA}
\end{figure}

\begin{figure}[h]%
\centering
\includegraphics[width=0.9\textwidth]{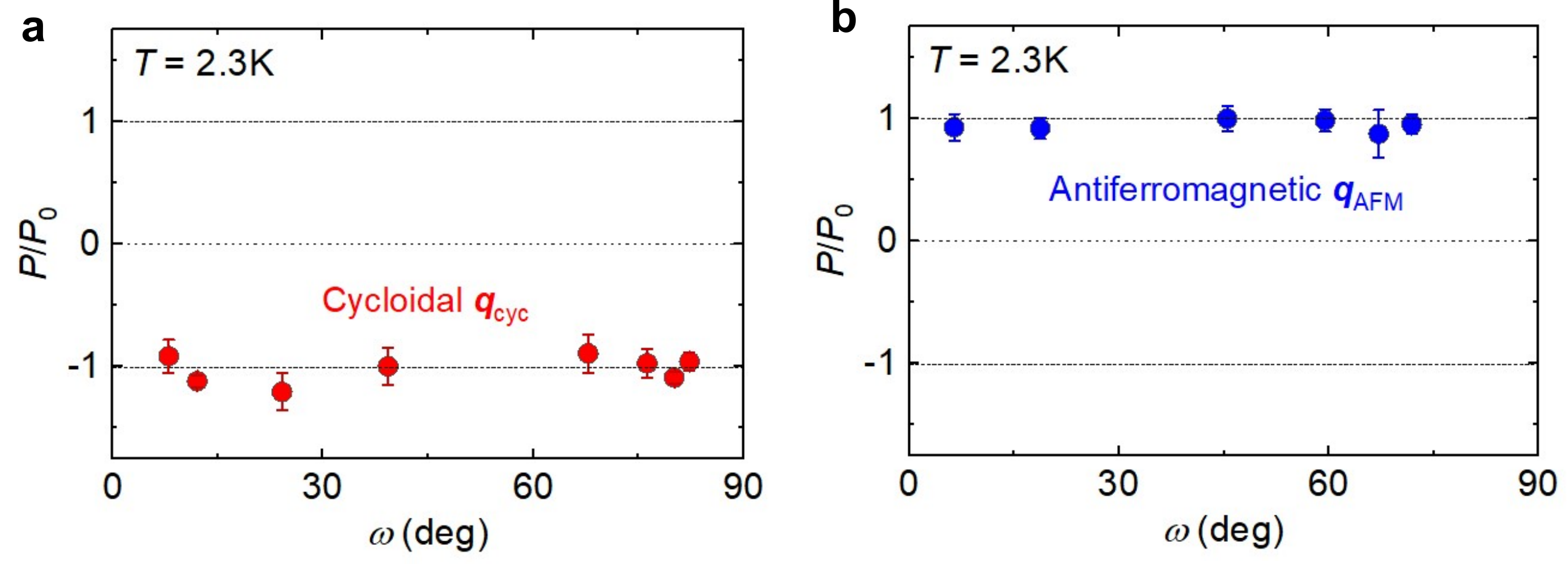}
\caption{\textbf{Neutron flipping ratio in sample A for two components of the magnetic order in DyTe$_3$'s ground state.} Here, the flipping ratio $P = (I_\mathrm{NSF}-I_\mathrm{SF})/(I_\mathrm{NSF}+I_\mathrm{SF})$ is the normalized ratio of spin flip (SF) and non-spin flip (NSF) intensities for a magnetic reflection. $P_0$ is the flipping ratio at the nuclear reflection $(002)$ (Methods). In our geometry, SF intensity is dominated by the $b$-component (the $c$-component) of the magnetization at $\omega \approx 0^\circ$ (at $\omega \approx 90^\circ$), although the structure and magnetic form factors have to be carefully taken into account when comparing intensities of various reflections (section \ref{Esec:magnetic_structure_analysis}). \textbf{a}, The incommensurate magnetization component at $\mathbf{q}_\mathrm{cyc}$ gives dominant SF scattering, with $P/P_0\equiv -1$ independent of the $\omega$ angle. This implies presence of both $m_b$ and $m_c$ for the incommensurate reflection. \textbf{b}, The antiferromagnetic reflection $\mathbf{q}_\mathrm{AFM}$ has dominant NSF scattering, which is again independent of $\omega$ and consistent with magnetization component exclusively along the $a$-axis; hence leading to $P/P_0\equiv +1$ for all reflections measured.}\label{EfigFlippingRatios}
\end{figure}

\begin{figure}[h]%
\centering
\includegraphics[width=0.9\textwidth]{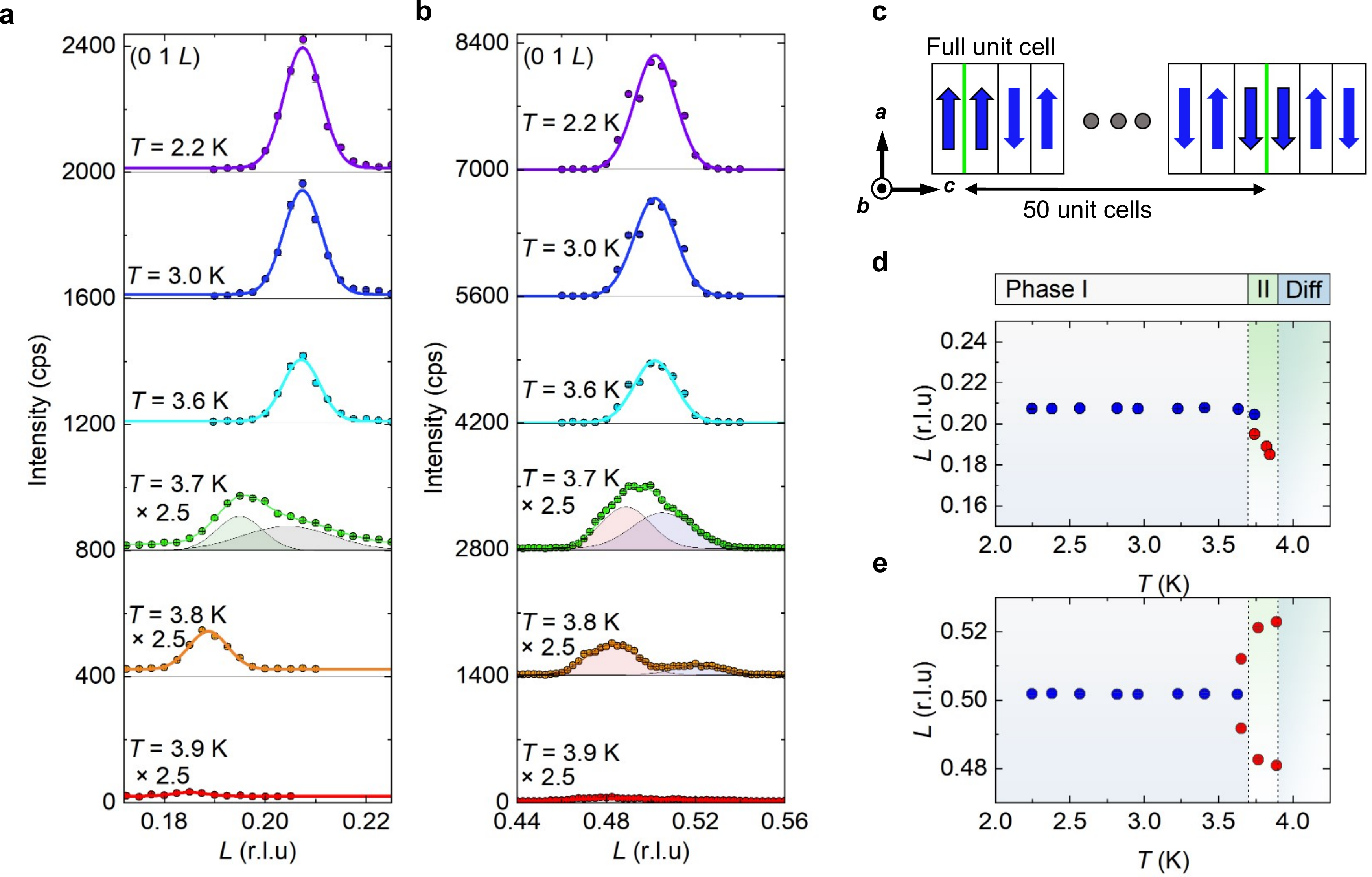}
\caption{\textbf{Incommensurate and commensurate antiferromagnetic component in phase II of DyTe$_\mathbf{3}$ (sample A).} \textbf{a},\textbf{b}, Line scans of magnetic intensity through the incommensurate (a) and commensurate (b) reflection on the $(01L)$ line, with clear temperature dependence. Shaded Gaussian curves at $T=3.7\,$K indicate a double-Gaussian fit in the regime of phase coexistence between phases I and II. High temperature data is multiplied by a scale factor to enhance visibility. The magnetic intensity vanishes in the incommensurate line entirely at $T_\mathrm{N2}=3.85\,$K, with no indications of diffuse scattering in the thermally disordered regime. Above $T_\mathrm{N2}$, we observe a weak, diffuse neutron signal along the $(0,K, 1/2)$ line, with rapid decay of the coherence length. \textbf{c}, Illustration of discommensuration-driven shift of $q_\mathrm{AFM}$ in $\mathbf{q}_\mathrm{AFM}=(0, b^*, q_\mathrm{AFM})$ for phase II, where a defect every $\sim50$ unit cells is introduced to release magnetoelastic strain built up between the crystal lattice and the collinear antiferromagnetic structure. \textbf{c},\textbf{d}, Temperature dependence of $\mathbf{q}_\mathrm{cyc}$ and $\mathbf{q}_\mathrm{AFM}$,  respectively. The different regimes, phase I, II and the diffuse scattering regime (Diff), are highlighted by different color shadings.}\label{EfigPolSampleA}
\end{figure}

\begin{figure}[h]%
\centering
\includegraphics[width=0.9\textwidth]{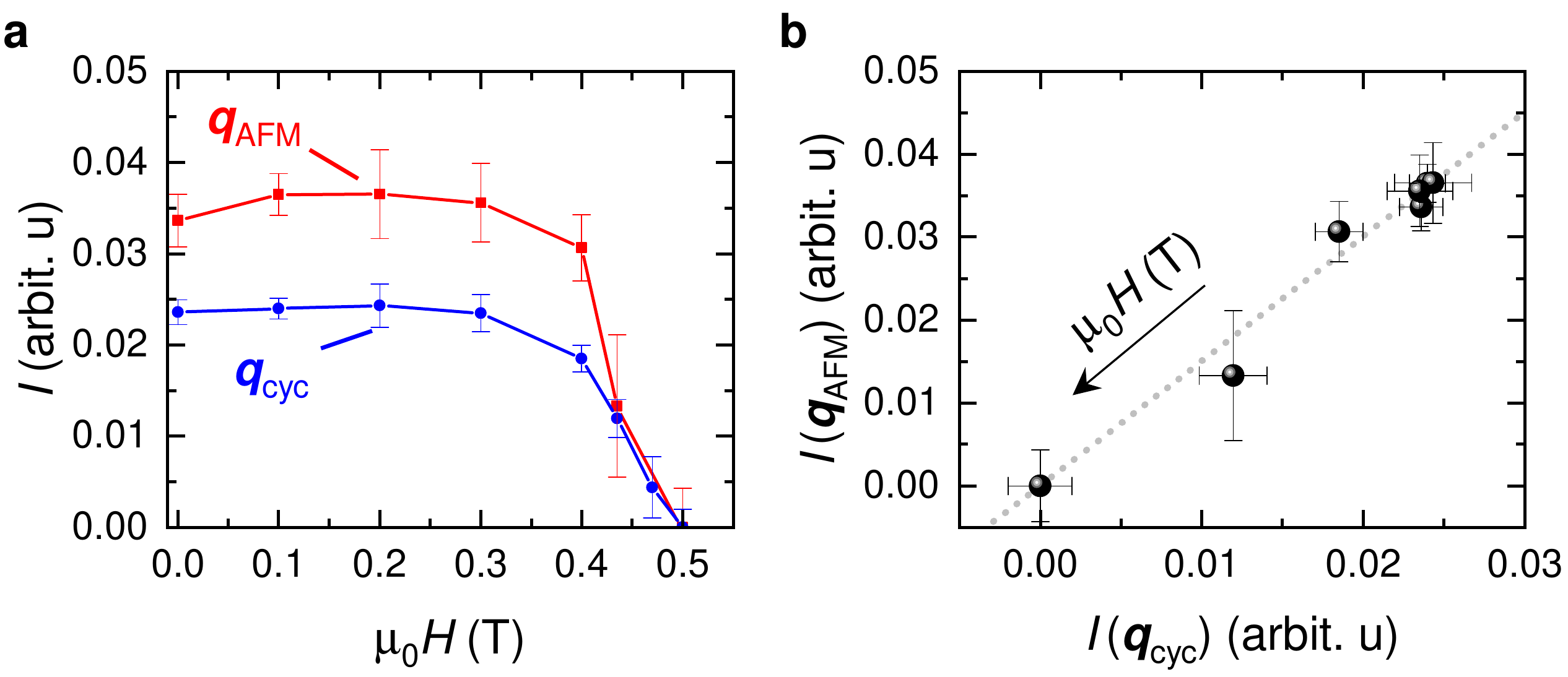}
\caption{\textbf{Coupling of commensurate and incommensurate magnetic order in DyTe$_3$, tracked by small-angle neutron scattering (SANS) in a magnetic field along the $c$-axis, for sample E.} \textbf{a}, Integrated intensity of commensurate ($\mathbf{q}=\mathbf{q}_\mathrm{AFM}$) and incommensurate ($\mathbf{q}=\mathbf{q}_\mathrm{cyc}$) reflections of type $(0, -1, -L)$ as a function of magnetic field and at a temperature of $T=2\,\mathrm{K}$. The observed intensity of both drops simultaneously at around $\mu_0H = 0.4\,\mathrm{T}$, with a fixed intensity ratio $I(\mathbf{q}_\mathrm{AFM})/I(\mathbf{q}_\mathrm{cyc}) \sim 1.5$, as indicated by a grey dashed line in panel \textbf{b}. The error bars correspond to Poisson counting errors of the integrated neutron scattering intensity.}\label{EfigSANSIRatio}
\end{figure}

\begin{figure}[h]%
\centering
\includegraphics[width=0.9\textwidth]{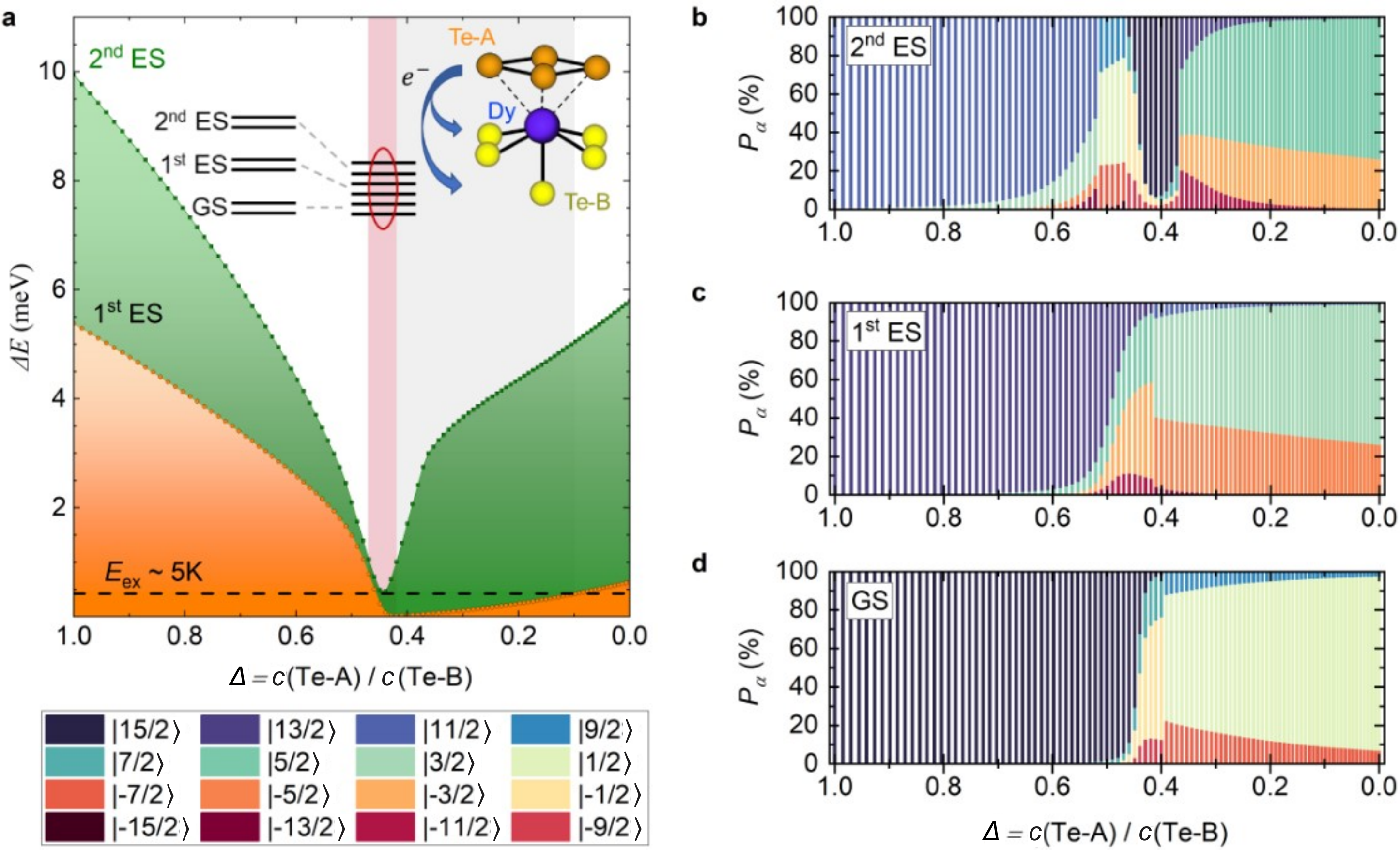}
\caption{\textbf{Crystal electric field calculations and local environment of Dy in DyTe$_\mathbf{3}$.} \textbf{a}, Evolution of the energy gap $\Delta E_i$ between ground state doublet (GS, $\left|\psi_0^\pm \right>$) and the first (1$^\mathrm{st}$ ES, $\left|\psi_1^\pm\right>$) or second (2$^\mathrm{nd}$ ES, $\left|\psi_2^\pm \right>$) excited state Kramers doublets. We consider a (virtual) charge transfer from the metallic Te$_2$ square net (Te-A) to the covalent bonded Tellurium (Te-B) ions, as sketched in the inset. The $x$-axis is labeled by the ratio of effective point charges $c$ on the two types of Te ions. Within the red (grey) shaded area, the excitation gaps to the the first and second excited states are (to the first excited state is) on the order of the exchange interaction energy $E_\mathrm{ex}\sim 2\,\mathrm{K}$ (dashed horizontal line). Panels \textbf{b}-\textbf{d} show the composition of $\left|\psi_{0,1,2}^+\right>$ in terms of eigenstates of ($b$-component of) total angular momentum $\ket{J_b = \pm n / 2}$ with $n\leq 15$, as suitable for the $4f^9$ shell of dysprosium. The $P_\beta$ are probabilities, i.e. absolute squares of amplitudes, for each contribution to the total wavefunction.}\label{EfigCEFvsCharge}
\end{figure}

\begin{figure}[h]%
\centering
\includegraphics[width=0.9\textwidth]{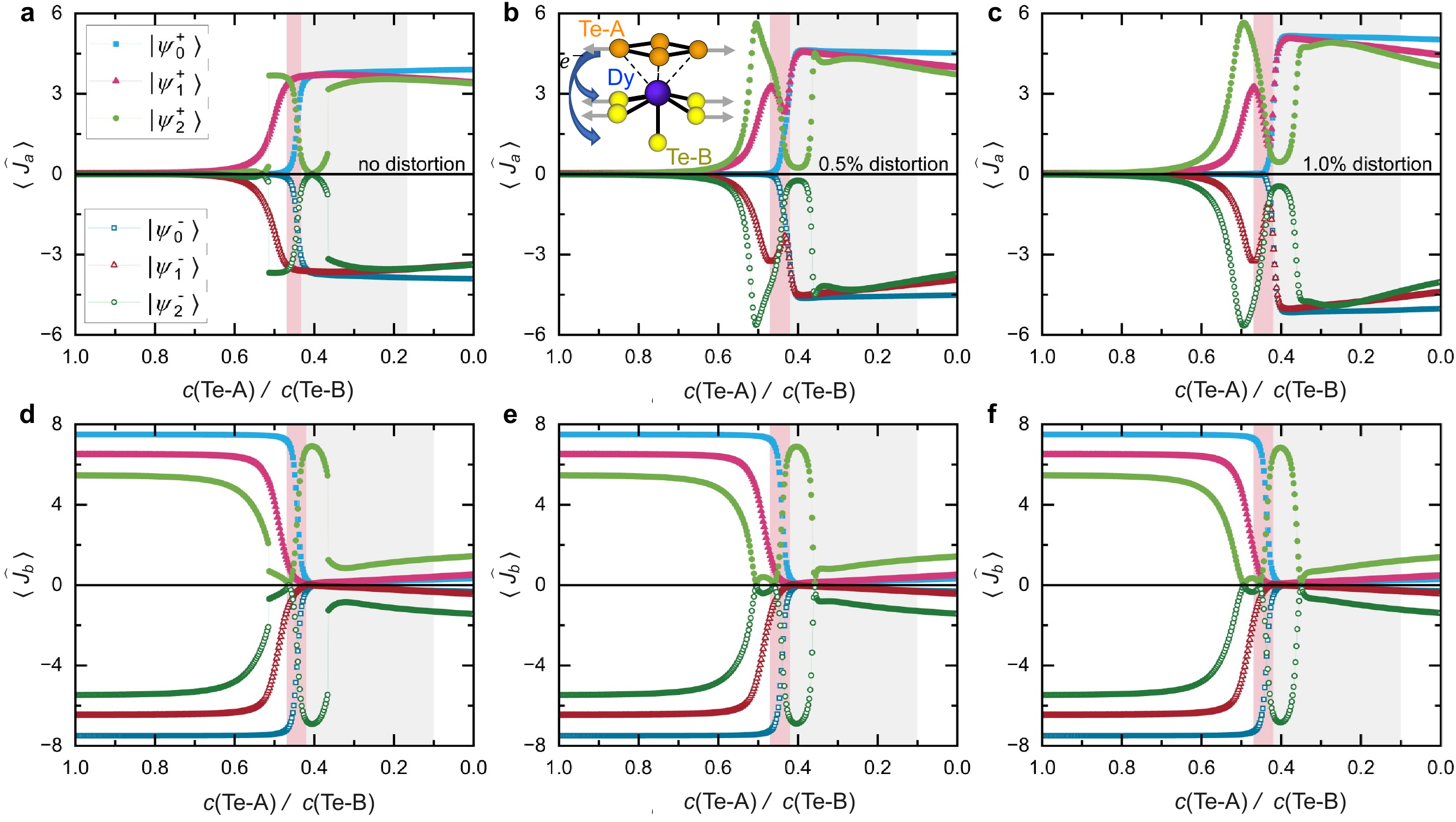}
\caption{\textbf{Mixing of crystal field states in DyTe$_\mathbf{3}$ by exchange interactions.} \textbf{a-c}, The maximal and minimal eigenvalues of $\hat{J}_\mathrm{a}$ and \textbf{d-f} of $\hat{J}_\mathrm{b}$ calculated for the crystal field doublets $\left|\psi_{0,1,2}^\pm \right>$. The charge ratio ($x$-coordinate) describes the ratio of effective charges assigned to Te-A and Te-B tellurium ions corresponding to metallic and ionic bonds around Dy, respectively. A matrix representation of the operators $\hat{J}_a$ and $\hat{J}_b$ is calculated on each respective subspace spanned by a Kramers pair of states, and subsequently diagonalized. There is a transition from dominant $\left<\psi_0\left|\hat{J}_b\right|\psi_0\right> = \pm 1/2$ (easy-plane) to $\pm 15/2$ (easy-axis) for the ground state $\left|\psi_0\right>$, when reducing the effective crystal electric field charge for Te-A on metallic bonds. The first excited state behaves similarly, while the second excited state generally covers a broader range of $\hat{J}_b$ eigenvalues. Red and grey shaded areas are defined as in Fig. \ref{EfigCEFvsCharge}. In the red region, the character of the ground state (GS, $\left|\psi_0^\pm\right>$, left column) changes from predominant $\left|J_b = \pm 15/2\right>$ to mainly $\left|J_b=\pm1/2\right>$ character. It is not possible to generate sizable in-plane ($b$-axis) magnetic moment using any linear combination of $\left|\psi_0^+\right>$, $\left|\psi_0^-\right>$ at relative weight of CEF charges $q$ on Te-A and Te-B sites $>0.5$ ($<0.4$).}\label{EfigCEFvsChargeIncludingDistortion}
\end{figure}


\begin{thebibliography}{83}%
\makeatletter
\providecommand \@ifxundefined [1]{%
 \@ifx{#1\undefined}
}%
\providecommand \@ifnum [1]{%
 \ifnum #1\expandafter \@firstoftwo
 \else \expandafter \@secondoftwo
 \fi
}%
\providecommand \@ifx [1]{%
 \ifx #1\expandafter \@firstoftwo
 \else \expandafter \@secondoftwo
 \fi
}%
\providecommand \natexlab [1]{#1}%
\providecommand \enquote  [1]{``#1''}%
\providecommand \bibnamefont  [1]{#1}%
\providecommand \bibfnamefont [1]{#1}%
\providecommand \citenamefont [1]{#1}%
\providecommand \href@noop [0]{\@secondoftwo}%
\providecommand \href [0]{\begingroup \@sanitize@url \@href}%
\providecommand \@href[1]{\@@startlink{#1}\@@href}%
\providecommand \@@href[1]{\endgroup#1\@@endlink}%
\providecommand \@sanitize@url [0]{\catcode `\\12\catcode `\$12\catcode
  `\&12\catcode `\#12\catcode `\^12\catcode `\_12\catcode `\%12\relax}%
\providecommand \@@startlink[1]{}%
\providecommand \@@endlink[0]{}%
\providecommand \url  [0]{\begingroup\@sanitize@url \@url }%
\providecommand \@url [1]{\endgroup\@href {#1}{\urlprefix }}%
\providecommand \urlprefix  [0]{URL }%
\providecommand \Eprint [0]{\href }%
\providecommand \doibase [0]{https://doi.org/}%
\providecommand \selectlanguage [0]{\@gobble}%
\providecommand \bibinfo  [0]{\@secondoftwo}%
\providecommand \bibfield  [0]{\@secondoftwo}%
\providecommand \translation [1]{[#1]}%
\providecommand \BibitemOpen [0]{}%
\providecommand \bibitemStop [0]{}%
\providecommand \bibitemNoStop [0]{.\EOS\space}%
\providecommand \EOS [0]{\spacefactor3000\relax}%
\providecommand \BibitemShut  [1]{\csname bibitem#1\endcsname}%
\let\auto@bib@innerbib\@empty
%</preamble>
\bibitem [{\citenamefont {Huang}\ \emph {et~al.}(2017)\citenamefont {Huang},
  \citenamefont {Clark}, \citenamefont {Navarro-Moratalla}, \citenamefont
  {Klein}, \citenamefont {Cheng}, \citenamefont {Seyler}, \citenamefont
  {D.~Zhong}, \citenamefont {McGuire}, \citenamefont {Cobden}, \citenamefont
  {Yao}, \citenamefont {Xiao}, \citenamefont {Jarillo-Herrero},\ and\
  \citenamefont {Xu}}]{Huang2017}%
  \BibitemOpen
  \bibfield  {author} {\bibinfo {author} {\bibfnamefont {B.}~\bibnamefont
  {Huang}}, \bibinfo {author} {\bibfnamefont {G.}~\bibnamefont {Clark}},
  \bibinfo {author} {\bibfnamefont {E.}~\bibnamefont {Navarro-Moratalla}},
  \bibinfo {author} {\bibfnamefont {D.}~\bibnamefont {Klein}}, \bibinfo
  {author} {\bibfnamefont {R.}~\bibnamefont {Cheng}}, \bibinfo {author}
  {\bibfnamefont {K.}~\bibnamefont {Seyler}}, \bibinfo {author} {\bibfnamefont
  {E.~S.}\ \bibnamefont {D.~Zhong}}, \bibinfo {author} {\bibfnamefont
  {M.}~\bibnamefont {McGuire}}, \bibinfo {author} {\bibfnamefont
  {D.}~\bibnamefont {Cobden}}, \bibinfo {author} {\bibfnamefont
  {W.}~\bibnamefont {Yao}}, \bibinfo {author} {\bibfnamefont {D.}~\bibnamefont
  {Xiao}}, \bibinfo {author} {\bibfnamefont {P.}~\bibnamefont
  {Jarillo-Herrero}},\ and\ \bibinfo {author} {\bibfnamefont {X.}~\bibnamefont
  {Xu}},\ }\bibfield  {title} {\bibinfo {title} {{Layer-dependent
  ferromagnetism in a van der Waals crystal down to the monolayer limit}},\
  }\href@noop {} {\bibfield  {journal} {\bibinfo  {journal} {Nature}\ }\textbf
  {\bibinfo {volume} {546}},\ \bibinfo {pages} {270} (\bibinfo {year}
  {2017})}\BibitemShut {NoStop}%
\bibitem [{\citenamefont {Gong}\ \emph {et~al.}(2017)\citenamefont {Gong},
  \citenamefont {Li}, \citenamefont {Li}, \citenamefont {Ji}, \citenamefont
  {Stern}, \citenamefont {Xia}, \citenamefont {Cao}, \citenamefont {Bao},
  \citenamefont {Wang}, \citenamefont {Wang}, \citenamefont {Qiu},
  \citenamefont {Cava}, \citenamefont {Louie}, \citenamefont {Xia},\ and\
  \citenamefont {Zhang}}]{Gong2017}%
  \BibitemOpen
  \bibfield  {author} {\bibinfo {author} {\bibfnamefont {C.}~\bibnamefont
  {Gong}}, \bibinfo {author} {\bibfnamefont {L.}~\bibnamefont {Li}}, \bibinfo
  {author} {\bibfnamefont {Z.}~\bibnamefont {Li}}, \bibinfo {author}
  {\bibfnamefont {H.}~\bibnamefont {Ji}}, \bibinfo {author} {\bibfnamefont
  {A.}~\bibnamefont {Stern}}, \bibinfo {author} {\bibfnamefont
  {Y.}~\bibnamefont {Xia}}, \bibinfo {author} {\bibfnamefont {T.}~\bibnamefont
  {Cao}}, \bibinfo {author} {\bibfnamefont {W.}~\bibnamefont {Bao}}, \bibinfo
  {author} {\bibfnamefont {C.}~\bibnamefont {Wang}}, \bibinfo {author}
  {\bibfnamefont {Y.}~\bibnamefont {Wang}}, \bibinfo {author} {\bibfnamefont
  {Z.}~\bibnamefont {Qiu}}, \bibinfo {author} {\bibfnamefont {R.}~\bibnamefont
  {Cava}}, \bibinfo {author} {\bibfnamefont {S.}~\bibnamefont {Louie}},
  \bibinfo {author} {\bibfnamefont {J.}~\bibnamefont {Xia}},\ and\ \bibinfo
  {author} {\bibfnamefont {X.}~\bibnamefont {Zhang}},\ }\bibfield  {title}
  {\bibinfo {title} {{Discovery of intrinsic ferromagnetism in two-dimensional
  van der Waals crystals}},\ }\href@noop {} {\bibfield  {journal} {\bibinfo
  {journal} {Nature}\ }\textbf {\bibinfo {volume} {546}},\ \bibinfo {pages}
  {265} (\bibinfo {year} {2017})}\BibitemShut {NoStop}%
\bibitem [{\citenamefont {Burch}\ \emph {et~al.}(2018)\citenamefont {Burch},
  \citenamefont {Mandrus},\ and\ \citenamefont {Park}}]{Burch2018}%
  \BibitemOpen
  \bibfield  {author} {\bibinfo {author} {\bibfnamefont {K.}~\bibnamefont
  {Burch}}, \bibinfo {author} {\bibfnamefont {D.}~\bibnamefont {Mandrus}},\
  and\ \bibinfo {author} {\bibfnamefont {J.-G.}\ \bibnamefont {Park}},\
  }\bibfield  {title} {\bibinfo {title} {{Magnetism in two-dimensional van der
  Waals materials}},\ }\href@noop {} {\bibfield  {journal} {\bibinfo  {journal}
  {Nature}\ }\textbf {\bibinfo {volume} {563}},\ \bibinfo {pages} {47}
  (\bibinfo {year} {2018})}\BibitemShut {NoStop}%
\bibitem [{\citenamefont {Gong}\ and\ \citenamefont {Zhang}(2019)}]{Gong2019}%
  \BibitemOpen
  \bibfield  {author} {\bibinfo {author} {\bibfnamefont {C.}~\bibnamefont
  {Gong}}\ and\ \bibinfo {author} {\bibfnamefont {X.}~\bibnamefont {Zhang}},\
  }\bibfield  {title} {\bibinfo {title} {{Two-dimensional magnetic crystals and
  emergent heterostructure devices}},\ }\href@noop {} {\bibfield  {journal}
  {\bibinfo  {journal} {Science}\ }\textbf {\bibinfo {volume} {363}},\ \bibinfo
  {pages} {aav4450} (\bibinfo {year} {2019})}\BibitemShut {NoStop}%
\bibitem [{\citenamefont {Amoroso}\ \emph {et~al.}(2020)\citenamefont
  {Amoroso}, \citenamefont {Barone},\ and\ \citenamefont
  {Picozzi}}]{Amoroso2020}%
  \BibitemOpen
  \bibfield  {author} {\bibinfo {author} {\bibfnamefont {D.}~\bibnamefont
  {Amoroso}}, \bibinfo {author} {\bibfnamefont {P.}~\bibnamefont {Barone}},\
  and\ \bibinfo {author} {\bibfnamefont {S.}~\bibnamefont {Picozzi}},\
  }\bibfield  {title} {\bibinfo {title} {{Spontaneous skyrmionic lattice from
  anisotropic symmetric exchange in a Ni-halide monolayer}},\ }\href@noop {}
  {\bibfield  {journal} {\bibinfo  {journal} {Nature Communications}\ }\textbf
  {\bibinfo {volume} {11}},\ \bibinfo {pages} {5784} (\bibinfo {year}
  {2020})}\BibitemShut {NoStop}%
\bibitem [{\citenamefont {Shimizu}\ \emph {et~al.}(2021)\citenamefont
  {Shimizu}, \citenamefont {Okumura}, \citenamefont {Kato},\ and\ \citenamefont
  {Motome}}]{Shimizu2021}%
  \BibitemOpen
  \bibfield  {author} {\bibinfo {author} {\bibfnamefont {K.}~\bibnamefont
  {Shimizu}}, \bibinfo {author} {\bibfnamefont {S.}~\bibnamefont {Okumura}},
  \bibinfo {author} {\bibfnamefont {Y.}~\bibnamefont {Kato}},\ and\ \bibinfo
  {author} {\bibfnamefont {Y.}~\bibnamefont {Motome}},\ }\bibfield  {title}
  {\bibinfo {title} {{Spin moir{\'e} engineering of topological magnetism and
  emergent electromagnetic fields}},\ }\href@noop {} {\bibfield  {journal}
  {\bibinfo  {journal} {Physical Review B}\ }\textbf {\bibinfo {volume}
  {103}},\ \bibinfo {pages} {184421} (\bibinfo {year} {2021})}\BibitemShut
  {NoStop}%
\bibitem [{\citenamefont {Jiang}\ \emph {et~al.}(2020)\citenamefont {Jiang},
  \citenamefont {Nii}, \citenamefont {Arisawa}, \citenamefont {Saitoh},\ and\
  \citenamefont {Onose}}]{Jiang2020}%
  \BibitemOpen
  \bibfield  {author} {\bibinfo {author} {\bibfnamefont {N.}~\bibnamefont
  {Jiang}}, \bibinfo {author} {\bibfnamefont {Y.}~\bibnamefont {Nii}}, \bibinfo
  {author} {\bibfnamefont {H.}~\bibnamefont {Arisawa}}, \bibinfo {author}
  {\bibfnamefont {E.}~\bibnamefont {Saitoh}},\ and\ \bibinfo {author}
  {\bibfnamefont {Y.}~\bibnamefont {Onose}},\ }\bibfield  {title} {\bibinfo
  {title} {{Electric current control of spin helicity in an itinerant
  helimagnet}},\ }\href@noop {} {\bibfield  {journal} {\bibinfo  {journal}
  {Nature Communications}\ }\textbf {\bibinfo {volume} {11}},\ \bibinfo {pages}
  {1601} (\bibinfo {year} {2020})}\BibitemShut {NoStop}%
\bibitem [{\citenamefont {Ohe}\ and\ \citenamefont {Onose}(2021)}]{Ohe2021}%
  \BibitemOpen
  \bibfield  {author} {\bibinfo {author} {\bibfnamefont {J.}~\bibnamefont
  {Ohe}}\ and\ \bibinfo {author} {\bibfnamefont {Y.}~\bibnamefont {Onose}},\
  }\bibfield  {title} {\bibinfo {title} {{Chirality control of the spin
  structure in monoaxial helimagnets by charge current}},\ }\href@noop {}
  {\bibfield  {journal} {\bibinfo  {journal} {Applied Physics Letters}\
  }\textbf {\bibinfo {volume} {118}},\ \bibinfo {pages} {042407} (\bibinfo
  {year} {2021})}\BibitemShut {NoStop}%
\bibitem [{\citenamefont {Masuda}\ \emph {et~al.}(2022)\citenamefont {Masuda},
  \citenamefont {Seki}, \citenamefont {Ohe}, \citenamefont {Nii}, \citenamefont
  {Takanashi},\ and\ \citenamefont {Onose}}]{Masuda2022}%
  \BibitemOpen
  \bibfield  {author} {\bibinfo {author} {\bibfnamefont {H.}~\bibnamefont
  {Masuda}}, \bibinfo {author} {\bibfnamefont {T.}~\bibnamefont {Seki}},
  \bibinfo {author} {\bibfnamefont {J.}~\bibnamefont {Ohe}}, \bibinfo {author}
  {\bibfnamefont {Y.}~\bibnamefont {Nii}}, \bibinfo {author} {\bibfnamefont
  {K.}~\bibnamefont {Takanashi}},\ and\ \bibinfo {author} {\bibfnamefont
  {Y.}~\bibnamefont {Onose}},\ }\href@noop {} {\bibinfo {title}
  {{Chirality-dependent spin current generation in a helimagnet: zero-field
  probe of chirality}}},\ \bibinfo {howpublished} {arXiv:2212.10980} (\bibinfo
  {year} {2022})\BibitemShut {NoStop}%
\bibitem [{\citenamefont {Wang}\ \emph
  {et~al.}(2022{\natexlab{a}})\citenamefont {Wang}, \citenamefont {Xu},
  \citenamefont {Zhao}, \citenamefont {Ji}, \citenamefont {Cao}, \citenamefont
  {Li},\ and\ \citenamefont {Li}}]{Wang2022_NiI2}%
  \BibitemOpen
  \bibfield  {author} {\bibinfo {author} {\bibfnamefont {Y.}~\bibnamefont
  {Wang}}, \bibinfo {author} {\bibfnamefont {X.}~\bibnamefont {Xu}}, \bibinfo
  {author} {\bibfnamefont {X.}~\bibnamefont {Zhao}}, \bibinfo {author}
  {\bibfnamefont {W.}~\bibnamefont {Ji}}, \bibinfo {author} {\bibfnamefont
  {Q.}~\bibnamefont {Cao}}, \bibinfo {author} {\bibfnamefont {S.}~\bibnamefont
  {Li}},\ and\ \bibinfo {author} {\bibfnamefont {Y.}~\bibnamefont {Li}},\
  }\bibfield  {title} {\bibinfo {title} {{Switchable half-metallicity in
  $A$-type antiferromagnetic NiI$_2$ bilayer coupled with ferroelectric
  In$_2$Se$_3$}},\ }\href@noop {} {\bibfield  {journal} {\bibinfo  {journal}
  {npj Computational Materials}\ }\textbf {\bibinfo {volume} {8}},\ \bibinfo
  {pages} {218} (\bibinfo {year} {2022}{\natexlab{a}})}\BibitemShut {NoStop}%
\bibitem [{\citenamefont {Schmitt}\ \emph {et~al.}(2011)\citenamefont
  {Schmitt}, \citenamefont {Kirchmann}, \citenamefont {Bovensiepen},
  \citenamefont {Moore}, \citenamefont {Chu}, \citenamefont {Lu}, \citenamefont
  {Rettig}, \citenamefont {Wolf}, \citenamefont {Fisher},\ and\ \citenamefont
  {Shen}}]{Schmitt2011}%
  \BibitemOpen
  \bibfield  {author} {\bibinfo {author} {\bibfnamefont {F.}~\bibnamefont
  {Schmitt}}, \bibinfo {author} {\bibfnamefont {P.}~\bibnamefont {Kirchmann}},
  \bibinfo {author} {\bibfnamefont {U.}~\bibnamefont {Bovensiepen}}, \bibinfo
  {author} {\bibfnamefont {R.}~\bibnamefont {Moore}}, \bibinfo {author}
  {\bibfnamefont {J.-H.}\ \bibnamefont {Chu}}, \bibinfo {author} {\bibfnamefont
  {D.}~\bibnamefont {Lu}}, \bibinfo {author} {\bibfnamefont {L.}~\bibnamefont
  {Rettig}}, \bibinfo {author} {\bibfnamefont {M.}~\bibnamefont {Wolf}},
  \bibinfo {author} {\bibfnamefont {I.}~\bibnamefont {Fisher}},\ and\ \bibinfo
  {author} {\bibfnamefont {Z.-X.}\ \bibnamefont {Shen}},\ }\bibfield  {title}
  {\bibinfo {title} {{Ultrafast electron dynamics in the charge density wave
  material TbTe$_3$}},\ }\href@noop {} {\bibfield  {journal} {\bibinfo
  {journal} {New Journal of Physics}\ }\textbf {\bibinfo {volume} {13}},\
  \bibinfo {pages} {063022} (\bibinfo {year} {2011})}\BibitemShut {NoStop}%
\bibitem [{\citenamefont {Kogar}\ \emph {et~al.}(2020)\citenamefont {Kogar},
  \citenamefont {ad~P.E.~Dolgirev}, \citenamefont {Shen}, \citenamefont
  {Straquadine}, \citenamefont {Bie}, \citenamefont {Wang}, \citenamefont
  {Rohwer}, \citenamefont {Tung}, \citenamefont {Yang}, \citenamefont {Li},
  \citenamefont {Yang}, \citenamefont {Weathersby}, \citenamefont {Park},
  \citenamefont {Kozina}, \citenamefont {Sie}, \citenamefont {Wen},
  \citenamefont {Jarillo-Herrero}, \citenamefont {Fisher}, \citenamefont
  {Wang},\ and\ \citenamefont {Gedik}}]{Kogar2020}%
  \BibitemOpen
  \bibfield  {author} {\bibinfo {author} {\bibfnamefont {A.}~\bibnamefont
  {Kogar}}, \bibinfo {author} {\bibfnamefont {A.~Z.}\ \bibnamefont
  {ad~P.E.~Dolgirev}}, \bibinfo {author} {\bibfnamefont {X.}~\bibnamefont
  {Shen}}, \bibinfo {author} {\bibfnamefont {J.}~\bibnamefont {Straquadine}},
  \bibinfo {author} {\bibfnamefont {Y.-Q.}\ \bibnamefont {Bie}}, \bibinfo
  {author} {\bibfnamefont {X.}~\bibnamefont {Wang}}, \bibinfo {author}
  {\bibfnamefont {T.}~\bibnamefont {Rohwer}}, \bibinfo {author} {\bibfnamefont
  {I.-C.}\ \bibnamefont {Tung}}, \bibinfo {author} {\bibfnamefont
  {Y.}~\bibnamefont {Yang}}, \bibinfo {author} {\bibfnamefont {R.}~\bibnamefont
  {Li}}, \bibinfo {author} {\bibfnamefont {J.}~\bibnamefont {Yang}}, \bibinfo
  {author} {\bibfnamefont {S.}~\bibnamefont {Weathersby}}, \bibinfo {author}
  {\bibfnamefont {S.}~\bibnamefont {Park}}, \bibinfo {author} {\bibfnamefont
  {M.}~\bibnamefont {Kozina}}, \bibinfo {author} {\bibfnamefont
  {E.}~\bibnamefont {Sie}}, \bibinfo {author} {\bibfnamefont {H.}~\bibnamefont
  {Wen}}, \bibinfo {author} {\bibfnamefont {P.}~\bibnamefont
  {Jarillo-Herrero}}, \bibinfo {author} {\bibfnamefont {I.}~\bibnamefont
  {Fisher}}, \bibinfo {author} {\bibfnamefont {X.}~\bibnamefont {Wang}},\ and\
  \bibinfo {author} {\bibfnamefont {N.}~\bibnamefont {Gedik}},\ }\bibfield
  {title} {\bibinfo {title} {{Light-induced charge density wave in LaTe$_3$}},\
  }\href@noop {} {\bibfield  {journal} {\bibinfo  {journal} {Nature Physics}\
  }\textbf {\bibinfo {volume} {16}},\ \bibinfo {pages} {159} (\bibinfo {year}
  {2020})}\BibitemShut {NoStop}%
\bibitem [{\citenamefont {Dolgirev}\ \emph {et~al.}(2020)\citenamefont
  {Dolgirev}, \citenamefont {Rozhkov}, \citenamefont {Zong}, \citenamefont
  {Kogar}, \citenamefont {Gedik},\ and\ \citenamefont {Fine}}]{Dolgirev2020}%
  \BibitemOpen
  \bibfield  {author} {\bibinfo {author} {\bibfnamefont {P.}~\bibnamefont
  {Dolgirev}}, \bibinfo {author} {\bibfnamefont {A.}~\bibnamefont {Rozhkov}},
  \bibinfo {author} {\bibfnamefont {A.}~\bibnamefont {Zong}}, \bibinfo {author}
  {\bibfnamefont {A.}~\bibnamefont {Kogar}}, \bibinfo {author} {\bibfnamefont
  {N.}~\bibnamefont {Gedik}},\ and\ \bibinfo {author} {\bibfnamefont
  {B.}~\bibnamefont {Fine}},\ }\bibfield  {title} {\bibinfo {title} {{Amplitude
  dynamics of the charge density wave in LaTe$_3$: Theoretical description of
  pump-probe experiments}},\ }\href@noop {} {\bibfield  {journal} {\bibinfo
  {journal} {Physical Review B}\ }\textbf {\bibinfo {volume} {101}},\ \bibinfo
  {pages} {054203} (\bibinfo {year} {2020})}\BibitemShut {NoStop}%
\bibitem [{\citenamefont {Gonzalez-Vallejo}\ \emph {et~al.}(2022)\citenamefont
  {Gonzalez-Vallejo}, \citenamefont {Jacques}, \citenamefont {Boschetto},
  \citenamefont {Rizza}, \citenamefont {Hadj-Azzem}, \citenamefont {Faure},\
  and\ \citenamefont {Bolloc'h}}]{Gonzalez2022}%
  \BibitemOpen
  \bibfield  {author} {\bibinfo {author} {\bibfnamefont {I.}~\bibnamefont
  {Gonzalez-Vallejo}}, \bibinfo {author} {\bibfnamefont {V.}~\bibnamefont
  {Jacques}}, \bibinfo {author} {\bibfnamefont {D.}~\bibnamefont {Boschetto}},
  \bibinfo {author} {\bibfnamefont {G.}~\bibnamefont {Rizza}}, \bibinfo
  {author} {\bibfnamefont {A.}~\bibnamefont {Hadj-Azzem}}, \bibinfo {author}
  {\bibfnamefont {J.}~\bibnamefont {Faure}},\ and\ \bibinfo {author}
  {\bibfnamefont {D.~L.}\ \bibnamefont {Bolloc'h}},\ }\bibfield  {title}
  {\bibinfo {title} {{Time-resolved structural dynamics of the
  out-of-equilibrium charge density wave phase transition in GdTe$_3$}},\
  }\href@noop {} {\bibfield  {journal} {\bibinfo  {journal} {Structural
  Dynamics}\ }\textbf {\bibinfo {volume} {9}},\ \bibinfo {pages} {014502}
  (\bibinfo {year} {2022})}\BibitemShut {NoStop}%
\bibitem [{\citenamefont {Wang}\ \emph
  {et~al.}(2022{\natexlab{b}})\citenamefont {Wang}, \citenamefont {Petrides},
  \citenamefont {McNamara}, \citenamefont {Hosen}, \citenamefont {Lei},
  \citenamefont {Wu}, \citenamefont {Hart}, \citenamefont {Lv}, \citenamefont
  {Yan}, \citenamefont {Xiao}, \citenamefont {Cha}, \citenamefont {Narang},
  \citenamefont {Schoop},\ and\ \citenamefont {Burch}}]{Wang2022_RTe3}%
  \BibitemOpen
  \bibfield  {author} {\bibinfo {author} {\bibfnamefont {Y.}~\bibnamefont
  {Wang}}, \bibinfo {author} {\bibfnamefont {I.}~\bibnamefont {Petrides}},
  \bibinfo {author} {\bibfnamefont {G.}~\bibnamefont {McNamara}}, \bibinfo
  {author} {\bibfnamefont {M.}~\bibnamefont {Hosen}}, \bibinfo {author}
  {\bibfnamefont {S.}~\bibnamefont {Lei}}, \bibinfo {author} {\bibfnamefont
  {Y.-C.}\ \bibnamefont {Wu}}, \bibinfo {author} {\bibfnamefont
  {J.}~\bibnamefont {Hart}}, \bibinfo {author} {\bibfnamefont {H.}~\bibnamefont
  {Lv}}, \bibinfo {author} {\bibfnamefont {J.}~\bibnamefont {Yan}}, \bibinfo
  {author} {\bibfnamefont {D.}~\bibnamefont {Xiao}}, \bibinfo {author}
  {\bibfnamefont {J.}~\bibnamefont {Cha}}, \bibinfo {author} {\bibfnamefont
  {P.}~\bibnamefont {Narang}}, \bibinfo {author} {\bibfnamefont
  {L.}~\bibnamefont {Schoop}},\ and\ \bibinfo {author} {\bibfnamefont
  {K.}~\bibnamefont {Burch}},\ }\bibfield  {title} {\bibinfo {title} {{Axial
  Higgs mode detected by quantum pathway interference in RTe$_3$}},\
  }\href@noop {} {\bibfield  {journal} {\bibinfo  {journal} {Nature}\ }\textbf
  {\bibinfo {volume} {606}},\ \bibinfo {pages} {896} (\bibinfo {year}
  {2022}{\natexlab{b}})}\BibitemShut {NoStop}%
\bibitem [{\citenamefont {Lei}\ \emph {et~al.}(2020)\citenamefont {Lei},
  \citenamefont {Lin}, \citenamefont {Jia}, \citenamefont {Gray}, \citenamefont
  {Topp}, \citenamefont {Farahi}, \citenamefont {Klemenz}, \citenamefont {Gao},
  \citenamefont {Rodolakis}, \citenamefont {McChesney}, \citenamefont {Ast},
  \citenamefont {Yazdani}, \citenamefont {Burch}, \citenamefont {Wu},
  \citenamefont {Ong},\ and\ \citenamefont {Schoop}}]{Lei2020}%
  \BibitemOpen
  \bibfield  {author} {\bibinfo {author} {\bibfnamefont {S.}~\bibnamefont
  {Lei}}, \bibinfo {author} {\bibfnamefont {J.}~\bibnamefont {Lin}}, \bibinfo
  {author} {\bibfnamefont {Y.}~\bibnamefont {Jia}}, \bibinfo {author}
  {\bibfnamefont {M.}~\bibnamefont {Gray}}, \bibinfo {author} {\bibfnamefont
  {A.}~\bibnamefont {Topp}}, \bibinfo {author} {\bibfnamefont {G.}~\bibnamefont
  {Farahi}}, \bibinfo {author} {\bibfnamefont {S.}~\bibnamefont {Klemenz}},
  \bibinfo {author} {\bibfnamefont {T.}~\bibnamefont {Gao}}, \bibinfo {author}
  {\bibfnamefont {F.}~\bibnamefont {Rodolakis}}, \bibinfo {author}
  {\bibfnamefont {J.}~\bibnamefont {McChesney}}, \bibinfo {author}
  {\bibfnamefont {C.}~\bibnamefont {Ast}}, \bibinfo {author} {\bibfnamefont
  {A.}~\bibnamefont {Yazdani}}, \bibinfo {author} {\bibfnamefont
  {K.}~\bibnamefont {Burch}}, \bibinfo {author} {\bibfnamefont
  {S.}~\bibnamefont {Wu}}, \bibinfo {author} {\bibfnamefont {N.}~\bibnamefont
  {Ong}},\ and\ \bibinfo {author} {\bibfnamefont {L.}~\bibnamefont {Schoop}},\
  }\bibfield  {title} {\bibinfo {title} {{High mobility in a van der Waals
  layered antiferromagnetic metal}},\ }\href@noop {} {\bibfield  {journal}
  {\bibinfo  {journal} {Science Advances}\ }\textbf {\bibinfo {volume} {6}},\
  \bibinfo {pages} {eaay6407} (\bibinfo {year} {2020})}\BibitemShut {NoStop}%
\bibitem [{\citenamefont {Che}\ \emph {et~al.}(2019)\citenamefont {Che},
  \citenamefont {Wang}, \citenamefont {Wu}, \citenamefont {Ma}, \citenamefont
  {Wen}, \citenamefont {Wu}, \citenamefont {Li}, \citenamefont {Zhao},
  \citenamefont {Wang}, \citenamefont {Zhang}, \citenamefont {Huang},
  \citenamefont {Li},\ and\ \citenamefont {Huang}}]{Chen2019}%
  \BibitemOpen
  \bibfield  {author} {\bibinfo {author} {\bibfnamefont {Y.}~\bibnamefont
  {Che}}, \bibinfo {author} {\bibfnamefont {P.}~\bibnamefont {Wang}}, \bibinfo
  {author} {\bibfnamefont {M.}~\bibnamefont {Wu}}, \bibinfo {author}
  {\bibfnamefont {J.}~\bibnamefont {Ma}}, \bibinfo {author} {\bibfnamefont
  {S.}~\bibnamefont {Wen}}, \bibinfo {author} {\bibfnamefont {X.}~\bibnamefont
  {Wu}}, \bibinfo {author} {\bibfnamefont {G.}~\bibnamefont {Li}}, \bibinfo
  {author} {\bibfnamefont {Y.}~\bibnamefont {Zhao}}, \bibinfo {author}
  {\bibfnamefont {K.}~\bibnamefont {Wang}}, \bibinfo {author} {\bibfnamefont
  {L.}~\bibnamefont {Zhang}}, \bibinfo {author} {\bibfnamefont
  {L.}~\bibnamefont {Huang}}, \bibinfo {author} {\bibfnamefont
  {W.}~\bibnamefont {Li}},\ and\ \bibinfo {author} {\bibfnamefont
  {M.}~\bibnamefont {Huang}},\ }\bibfield  {title} {\bibinfo {title} {{Raman
  spectra and dimensional effect on the charge density wave transition in
  GdTe$_3$}},\ }\href@noop {} {\bibfield  {journal} {\bibinfo  {journal}
  {Applied Physics Letters}\ }\textbf {\bibinfo {volume} {115}},\ \bibinfo
  {pages} {151905} (\bibinfo {year} {2019})}\BibitemShut {NoStop}%
\bibitem [{\citenamefont {Malliakas}\ and\ \citenamefont
  {Kanatzidis}(2006)}]{Malliakas2006}%
  \BibitemOpen
  \bibfield  {author} {\bibinfo {author} {\bibfnamefont {C.}~\bibnamefont
  {Malliakas}}\ and\ \bibinfo {author} {\bibfnamefont {M.}~\bibnamefont
  {Kanatzidis}},\ }\bibfield  {title} {\bibinfo {title} {{Divergence in the
  Behavior of the Charge Density Wave in RETe$_3$ (RE= Rare-Earth element) with
  Temperature and RE Element}},\ }\href@noop {} {\bibfield  {journal} {\bibinfo
   {journal} {Journal of the American Chemical Society}\ }\textbf {\bibinfo
  {volume} {128}},\ \bibinfo {pages} {12612} (\bibinfo {year}
  {2006})}\BibitemShut {NoStop}%
\bibitem [{\citenamefont {Laverock}\ \emph {et~al.}(2005)\citenamefont
  {Laverock}, \citenamefont {Dugdale}, \citenamefont {Major}, \citenamefont
  {Alam}, \citenamefont {Ru}, \citenamefont {Fisher}, \citenamefont {Santi},\
  and\ \citenamefont {Bruno}}]{Laverock2005}%
  \BibitemOpen
  \bibfield  {author} {\bibinfo {author} {\bibfnamefont {J.}~\bibnamefont
  {Laverock}}, \bibinfo {author} {\bibfnamefont {S.~B.}\ \bibnamefont
  {Dugdale}}, \bibinfo {author} {\bibfnamefont {Z.}~\bibnamefont {Major}},
  \bibinfo {author} {\bibfnamefont {M.~A.}\ \bibnamefont {Alam}}, \bibinfo
  {author} {\bibfnamefont {N.}~\bibnamefont {Ru}}, \bibinfo {author}
  {\bibfnamefont {I.~R.}\ \bibnamefont {Fisher}}, \bibinfo {author}
  {\bibfnamefont {G.}~\bibnamefont {Santi}},\ and\ \bibinfo {author}
  {\bibfnamefont {E.}~\bibnamefont {Bruno}},\ }\bibfield  {title} {\bibinfo
  {title} {{Fermi surface nesting and charge-density wave formation in
  rare-earth tritellurides}},\ }\href@noop {} {\bibfield  {journal} {\bibinfo
  {journal} {Physical Review B}\ }\textbf {\bibinfo {volume} {71}},\ \bibinfo
  {pages} {085114} (\bibinfo {year} {2005})}\BibitemShut {NoStop}%
\bibitem [{\citenamefont {Brouet}\ \emph {et~al.}(2008)\citenamefont {Brouet},
  \citenamefont {Yang}, \citenamefont {Zhou}, \citenamefont {Hussain},
  \citenamefont {Moore}, \citenamefont {He}, \citenamefont {Lu}, \citenamefont
  {Shen}, \citenamefont {Laverock}, \citenamefont {Dugdale}, \citenamefont
  {Ru},\ and\ \citenamefont {Fisher}}]{Brouet2008}%
  \BibitemOpen
  \bibfield  {author} {\bibinfo {author} {\bibfnamefont {V.}~\bibnamefont
  {Brouet}}, \bibinfo {author} {\bibfnamefont {W.~L.}\ \bibnamefont {Yang}},
  \bibinfo {author} {\bibfnamefont {X.~J.}\ \bibnamefont {Zhou}}, \bibinfo
  {author} {\bibfnamefont {Z.}~\bibnamefont {Hussain}}, \bibinfo {author}
  {\bibfnamefont {R.~G.}\ \bibnamefont {Moore}}, \bibinfo {author}
  {\bibfnamefont {R.}~\bibnamefont {He}}, \bibinfo {author} {\bibfnamefont
  {D.~H.}\ \bibnamefont {Lu}}, \bibinfo {author} {\bibfnamefont {Z.~X.}\
  \bibnamefont {Shen}}, \bibinfo {author} {\bibfnamefont {J.}~\bibnamefont
  {Laverock}}, \bibinfo {author} {\bibfnamefont {S.~B.}\ \bibnamefont
  {Dugdale}}, \bibinfo {author} {\bibfnamefont {N.}~\bibnamefont {Ru}},\ and\
  \bibinfo {author} {\bibfnamefont {I.~R.}\ \bibnamefont {Fisher}},\ }\bibfield
   {title} {\bibinfo {title} {{Angle-resolved photoemission study of the
  evolution of band structure and charge density wave properties in
  $R{\text{Te}}_{3}$ ($R=\text{Y}$, La, Ce, Sm, Gd, Tb, and Dy)}},\ }\href@noop
  {} {\bibfield  {journal} {\bibinfo  {journal} {Physical Review B}\ }\textbf
  {\bibinfo {volume} {77}},\ \bibinfo {pages} {235104} (\bibinfo {year}
  {2008})}\BibitemShut {NoStop}%
\bibitem [{\citenamefont {Chikina}\ \emph {et~al.}(2022)\citenamefont
  {Chikina}, \citenamefont {Lund}, \citenamefont {Bianchia}, \citenamefont
  {Curcio}, \citenamefont {Dalgaard}, \citenamefont {Bremholm}, \citenamefont
  {Lei}, \citenamefont {Singh}, \citenamefont {Schoop},\ and\ \citenamefont
  {Hofmann}}]{Chikina2022}%
  \BibitemOpen
  \bibfield  {author} {\bibinfo {author} {\bibfnamefont {A.}~\bibnamefont
  {Chikina}}, \bibinfo {author} {\bibfnamefont {H.}~\bibnamefont {Lund}},
  \bibinfo {author} {\bibfnamefont {M.}~\bibnamefont {Bianchia}}, \bibinfo
  {author} {\bibfnamefont {D.}~\bibnamefont {Curcio}}, \bibinfo {author}
  {\bibfnamefont {K.}~\bibnamefont {Dalgaard}}, \bibinfo {author}
  {\bibfnamefont {M.}~\bibnamefont {Bremholm}}, \bibinfo {author}
  {\bibfnamefont {S.}~\bibnamefont {Lei}}, \bibinfo {author} {\bibfnamefont
  {R.}~\bibnamefont {Singh}}, \bibinfo {author} {\bibfnamefont
  {L.}~\bibnamefont {Schoop}},\ and\ \bibinfo {author} {\bibfnamefont
  {P.}~\bibnamefont {Hofmann}},\ }\bibfield  {title} {\bibinfo {title} {{Charge
  density wave-generated Fermi surfaces in NdTe$_3$}},\ }\href@noop {} {\
  (\bibinfo {year} {2022})}\BibitemShut {NoStop}%
\bibitem [{\citenamefont {Ru}\ \emph {et~al.}(2008)\citenamefont {Ru},
  \citenamefont {Condron}, \citenamefont {Margulis}, \citenamefont {Shin},
  \citenamefont {Laverock}, \citenamefont {Dugdale}, \citenamefont {Toney},\
  and\ \citenamefont {Fisher}}]{Ru2008}%
  \BibitemOpen
  \bibfield  {author} {\bibinfo {author} {\bibfnamefont {N.}~\bibnamefont
  {Ru}}, \bibinfo {author} {\bibfnamefont {C.}~\bibnamefont {Condron}},
  \bibinfo {author} {\bibfnamefont {G.}~\bibnamefont {Margulis}}, \bibinfo
  {author} {\bibfnamefont {K.}~\bibnamefont {Shin}}, \bibinfo {author}
  {\bibfnamefont {J.}~\bibnamefont {Laverock}}, \bibinfo {author}
  {\bibfnamefont {S.}~\bibnamefont {Dugdale}}, \bibinfo {author} {\bibfnamefont
  {M.}~\bibnamefont {Toney}},\ and\ \bibinfo {author} {\bibfnamefont
  {I.}~\bibnamefont {Fisher}},\ }\bibfield  {title} {\bibinfo {title} {{Effect
  of chemical pressure on the charge density wave transition in rare-earth
  tritellurides {RTe$_3$}}},\ }\href@noop {} {\bibfield  {journal} {\bibinfo
  {journal} {Physical Review B}\ }\textbf {\bibinfo {volume} {77}},\ \bibinfo
  {pages} {035114} (\bibinfo {year} {2008})}\BibitemShut {NoStop}%
\bibitem [{\citenamefont {Zocco}(2011)}]{Zocco2011}%
  \BibitemOpen
  \bibfield  {author} {\bibinfo {author} {\bibfnamefont {D.~A.}\ \bibnamefont
  {Zocco}},\ }\emph {\bibinfo {title} {{Interplay of Superconductivity,
  Magnetism, and Density Waves in Rare-Earth Tritellurides and Iron-Based
  Superconducting Materials}}},\ \href@noop {} {Ph.D. thesis},\ \bibinfo
  {school} {University of California, San Diego} (\bibinfo {year}
  {2011})\BibitemShut {NoStop}%
\bibitem [{\citenamefont {Zocco}\ \emph {et~al.}(2015)\citenamefont {Zocco},
  \citenamefont {Hamlin}, \citenamefont {Grube}, \citenamefont {Chu},
  \citenamefont {Kuo}, \citenamefont {Fisher},\ and\ \citenamefont
  {Maple}}]{Zocco2015}%
  \BibitemOpen
  \bibfield  {author} {\bibinfo {author} {\bibfnamefont {D.~A.}\ \bibnamefont
  {Zocco}}, \bibinfo {author} {\bibfnamefont {J.~J.}\ \bibnamefont {Hamlin}},
  \bibinfo {author} {\bibfnamefont {K.}~\bibnamefont {Grube}}, \bibinfo
  {author} {\bibfnamefont {J.-H.}\ \bibnamefont {Chu}}, \bibinfo {author}
  {\bibfnamefont {H.-H.}\ \bibnamefont {Kuo}}, \bibinfo {author} {\bibfnamefont
  {I.~R.}\ \bibnamefont {Fisher}},\ and\ \bibinfo {author} {\bibfnamefont
  {M.~B.}\ \bibnamefont {Maple}},\ }\bibfield  {title} {\bibinfo {title}
  {{Pressure dependence of the charge-density-wave and superconducting states
  in GdTe$_3$, TbTe$_3$, and DyTe$_3$}},\ }\href@noop {} {\bibfield  {journal}
  {\bibinfo  {journal} {Physical Review B}\ }\textbf {\bibinfo {volume} {91}},\
  \bibinfo {pages} {205114} (\bibinfo {year} {2015})}\BibitemShut {NoStop}%
\bibitem [{\citenamefont {Sarkar}\ \emph {et~al.}(2023)\citenamefont {Sarkar},
  \citenamefont {Bhattacharya}, \citenamefont {Sadhukhan}, \citenamefont
  {Curcio}, \citenamefont {Dutt}, \citenamefont {Singh}, \citenamefont
  {Bianchi}, \citenamefont {Pariari}, \citenamefont {Roy}, \citenamefont
  {Mandal}, \citenamefont {Das}, \citenamefont {Hofmann}, \citenamefont
  {Chakrabarti},\ and\ \citenamefont {Barman}}]{Sarkar2022}%
  \BibitemOpen
  \bibfield  {author} {\bibinfo {author} {\bibfnamefont {S.}~\bibnamefont
  {Sarkar}}, \bibinfo {author} {\bibfnamefont {J.}~\bibnamefont
  {Bhattacharya}}, \bibinfo {author} {\bibfnamefont {P.}~\bibnamefont
  {Sadhukhan}}, \bibinfo {author} {\bibfnamefont {D.}~\bibnamefont {Curcio}},
  \bibinfo {author} {\bibfnamefont {R.}~\bibnamefont {Dutt}}, \bibinfo {author}
  {\bibfnamefont {V.}~\bibnamefont {Singh}}, \bibinfo {author} {\bibfnamefont
  {M.}~\bibnamefont {Bianchi}}, \bibinfo {author} {\bibfnamefont
  {A.}~\bibnamefont {Pariari}}, \bibinfo {author} {\bibfnamefont
  {S.}~\bibnamefont {Roy}}, \bibinfo {author} {\bibfnamefont {P.}~\bibnamefont
  {Mandal}}, \bibinfo {author} {\bibfnamefont {T.}~\bibnamefont {Das}},
  \bibinfo {author} {\bibfnamefont {P.}~\bibnamefont {Hofmann}}, \bibinfo
  {author} {\bibfnamefont {A.}~\bibnamefont {Chakrabarti}},\ and\ \bibinfo
  {author} {\bibfnamefont {S.}~\bibnamefont {Barman}},\ }\bibfield  {title}
  {\bibinfo {title} {{Charge density wave induced nodal lines in LaTe$_3$}},\
  }\href@noop {} {\bibfield  {journal} {\bibinfo  {journal} {Nature
  Communications}\ }\textbf {\bibinfo {volume} {14}},\ \bibinfo {pages} {3628}
  (\bibinfo {year} {2023})}\BibitemShut {NoStop}%
\bibitem [{\citenamefont {Iyeiri}\ \emph {et~al.}(2003)\citenamefont {Iyeiri},
  \citenamefont {Okumura}, \citenamefont {Michioka},\ and\ \citenamefont
  {Suzuki}}]{Iyeiri2003}%
  \BibitemOpen
  \bibfield  {author} {\bibinfo {author} {\bibfnamefont {Y.}~\bibnamefont
  {Iyeiri}}, \bibinfo {author} {\bibfnamefont {T.}~\bibnamefont {Okumura}},
  \bibinfo {author} {\bibfnamefont {C.}~\bibnamefont {Michioka}},\ and\
  \bibinfo {author} {\bibfnamefont {K.}~\bibnamefont {Suzuki}},\ }\bibfield
  {title} {\bibinfo {title} {{Magnetic properties of rare-earth metal
  tritellurides $R$Te$_3$ ($R$ = Ce, Pr, Nd, Gd, Dy)}},\ }\href@noop {}
  {\bibfield  {journal} {\bibinfo  {journal} {Physical Review B}\ }\textbf
  {\bibinfo {volume} {67}},\ \bibinfo {pages} {144417} (\bibinfo {year}
  {2003})}\BibitemShut {NoStop}%
\bibitem [{\citenamefont {Pfuner}\ \emph {et~al.}(2011)\citenamefont {Pfuner},
  \citenamefont {Gvasaliya}, \citenamefont {Zaharko}, \citenamefont {Keller},
  \citenamefont {Mesot}, \citenamefont {Pomjakushin}, \citenamefont {Chu},
  \citenamefont {Fisher},\ and\ \citenamefont {Degiorgi}}]{Pfuner2011}%
  \BibitemOpen
  \bibfield  {author} {\bibinfo {author} {\bibfnamefont {F.}~\bibnamefont
  {Pfuner}}, \bibinfo {author} {\bibfnamefont {S.}~\bibnamefont {Gvasaliya}},
  \bibinfo {author} {\bibfnamefont {O.}~\bibnamefont {Zaharko}}, \bibinfo
  {author} {\bibfnamefont {L.}~\bibnamefont {Keller}}, \bibinfo {author}
  {\bibfnamefont {J.}~\bibnamefont {Mesot}}, \bibinfo {author} {\bibfnamefont
  {V.}~\bibnamefont {Pomjakushin}}, \bibinfo {author} {\bibfnamefont {J.-H.}\
  \bibnamefont {Chu}}, \bibinfo {author} {\bibfnamefont {I.}~\bibnamefont
  {Fisher}},\ and\ \bibinfo {author} {\bibfnamefont {L.}~\bibnamefont
  {Degiorgi}},\ }\bibfield  {title} {\bibinfo {title} {{Incommensurate magnetic
  order in TbTe$_3$}},\ }\href@noop {} {\bibfield  {journal} {\bibinfo
  {journal} {Journal of Physics: Condensed Matter}\ }\textbf {\bibinfo {volume}
  {24}},\ \bibinfo {pages} {036001} (\bibinfo {year} {2011})}\BibitemShut
  {NoStop}%
\bibitem [{\citenamefont {Yang}\ \emph {et~al.}(2020)\citenamefont {Yang},
  \citenamefont {Drew}, \citenamefont {van Smaalen}, \citenamefont {van Well},
  \citenamefont {Pratt}, \citenamefont {Stenning}, \citenamefont {Karim},\ and\
  \citenamefont {Rabia}}]{Yang2020}%
  \BibitemOpen
  \bibfield  {author} {\bibinfo {author} {\bibfnamefont {Z.}~\bibnamefont
  {Yang}}, \bibinfo {author} {\bibfnamefont {A.}~\bibnamefont {Drew}}, \bibinfo
  {author} {\bibfnamefont {S.}~\bibnamefont {van Smaalen}}, \bibinfo {author}
  {\bibfnamefont {N.}~\bibnamefont {van Well}}, \bibinfo {author}
  {\bibfnamefont {F.}~\bibnamefont {Pratt}}, \bibinfo {author} {\bibfnamefont
  {G.}~\bibnamefont {Stenning}}, \bibinfo {author} {\bibfnamefont
  {A.}~\bibnamefont {Karim}},\ and\ \bibinfo {author} {\bibfnamefont
  {K.}~\bibnamefont {Rabia}},\ }\bibfield  {title} {\bibinfo {title} {{Multiple
  magnetic-phase transitions and critical behavior of charge-density wave
  compound TbTe$_3$}},\ }\href@noop {} {\bibfield  {journal} {\bibinfo
  {journal} {Journal of Physics: Condensed Matter}\ }\textbf {\bibinfo {volume}
  {32}},\ \bibinfo {pages} {305801} (\bibinfo {year} {2020})}\BibitemShut
  {NoStop}%
\bibitem [{\citenamefont {Guo}\ \emph {et~al.}(2021)\citenamefont {Guo},
  \citenamefont {Bao}, \citenamefont {Zhao},\ and\ \citenamefont
  {Ebisu}}]{Guo2021}%
  \BibitemOpen
  \bibfield  {author} {\bibinfo {author} {\bibfnamefont {Q.}~\bibnamefont
  {Guo}}, \bibinfo {author} {\bibfnamefont {D.}~\bibnamefont {Bao}}, \bibinfo
  {author} {\bibfnamefont {L.}~\bibnamefont {Zhao}},\ and\ \bibinfo {author}
  {\bibfnamefont {S.}~\bibnamefont {Ebisu}},\ }\bibfield  {title} {\bibinfo
  {title} {{Novel magnetic behavior of antiferromagnetic GdTe$_3$ induced by
  magnetic field}},\ }\href@noop {} {\bibfield  {journal} {\bibinfo  {journal}
  {Physica B: Condensed Matter}\ }\textbf {\bibinfo {volume} {617}},\ \bibinfo
  {pages} {413153} (\bibinfo {year} {2021})}\BibitemShut {NoStop}%
\bibitem [{\citenamefont {Volkova}\ \emph {et~al.}(2022)\citenamefont
  {Volkova}, \citenamefont {Hadj-Azzem}, \citenamefont {Remenyi}, \citenamefont
  {Lorenzo}, \citenamefont {Monceau}, \citenamefont {Sinchenko},\ and\
  \citenamefont {Vasiliev}}]{Volkova2022}%
  \BibitemOpen
  \bibfield  {author} {\bibinfo {author} {\bibfnamefont {O.}~\bibnamefont
  {Volkova}}, \bibinfo {author} {\bibfnamefont {A.}~\bibnamefont {Hadj-Azzem}},
  \bibinfo {author} {\bibfnamefont {G.}~\bibnamefont {Remenyi}}, \bibinfo
  {author} {\bibfnamefont {J.}~\bibnamefont {Lorenzo}}, \bibinfo {author}
  {\bibfnamefont {P.}~\bibnamefont {Monceau}}, \bibinfo {author} {\bibfnamefont
  {A.}~\bibnamefont {Sinchenko}},\ and\ \bibinfo {author} {\bibfnamefont
  {A.}~\bibnamefont {Vasiliev}},\ }\bibfield  {title} {\bibinfo {title}
  {{Magnetic Phase Diagram of van der Waals Antiferromagnet TbTe$_3$}},\
  }\href@noop {} {\bibfield  {journal} {\bibinfo  {journal} {Materials}\
  }\textbf {\bibinfo {volume} {15}},\ \bibinfo {pages} {8772} (\bibinfo {year}
  {2022})}\BibitemShut {NoStop}%
\bibitem [{\citenamefont {Chillal}\ \emph {et~al.}(2020)\citenamefont
  {Chillal}, \citenamefont {Schierle}, \citenamefont {Weschke}, \citenamefont
  {Yokaichiya}, \citenamefont {Hoffmann}, \citenamefont {Volkova},
  \citenamefont {Vasiliev}, \citenamefont {Sinchenko}, \citenamefont {Lejay},
  \citenamefont {Hadj-Azzem}, \citenamefont {Monceau},\ and\ \citenamefont
  {Lake}}]{Chillal2020}%
  \BibitemOpen
  \bibfield  {author} {\bibinfo {author} {\bibfnamefont {S.}~\bibnamefont
  {Chillal}}, \bibinfo {author} {\bibfnamefont {E.}~\bibnamefont {Schierle}},
  \bibinfo {author} {\bibfnamefont {E.}~\bibnamefont {Weschke}}, \bibinfo
  {author} {\bibfnamefont {F.}~\bibnamefont {Yokaichiya}}, \bibinfo {author}
  {\bibfnamefont {J.-U.}\ \bibnamefont {Hoffmann}}, \bibinfo {author}
  {\bibfnamefont {O.~S.}\ \bibnamefont {Volkova}}, \bibinfo {author}
  {\bibfnamefont {A.~N.}\ \bibnamefont {Vasiliev}}, \bibinfo {author}
  {\bibfnamefont {A.}~\bibnamefont {Sinchenko}}, \bibinfo {author}
  {\bibfnamefont {P.}~\bibnamefont {Lejay}}, \bibinfo {author} {\bibfnamefont
  {A.}~\bibnamefont {Hadj-Azzem}}, \bibinfo {author} {\bibfnamefont
  {P.}~\bibnamefont {Monceau}},\ and\ \bibinfo {author} {\bibfnamefont
  {B.}~\bibnamefont {Lake}},\ }\bibfield  {title} {\bibinfo {title} {{Strongly
  coupled charge, orbital, and spin order in TbTe$_3$}},\ }\href@noop {}
  {\bibfield  {journal} {\bibinfo  {journal} {Physical Review B}\ }\textbf
  {\bibinfo {volume} {102}},\ \bibinfo {pages} {241110(R)} (\bibinfo {year}
  {2020})}\BibitemShut {NoStop}%
\bibitem [{\citenamefont {Shin}\ \emph {et~al.}(2005)\citenamefont {Shin},
  \citenamefont {Brouet}, \citenamefont {Ru}, \citenamefont {Shen},\ and\
  \citenamefont {Fisher}}]{Shin2005}%
  \BibitemOpen
  \bibfield  {author} {\bibinfo {author} {\bibfnamefont {K.~Y.}\ \bibnamefont
  {Shin}}, \bibinfo {author} {\bibfnamefont {V.}~\bibnamefont {Brouet}},
  \bibinfo {author} {\bibfnamefont {N.}~\bibnamefont {Ru}}, \bibinfo {author}
  {\bibfnamefont {Z.~X.}\ \bibnamefont {Shen}},\ and\ \bibinfo {author}
  {\bibfnamefont {I.~R.}\ \bibnamefont {Fisher}},\ }\bibfield  {title}
  {\bibinfo {title} {{Electronic structure and charge-density wave formation in
  $\mathrm{La}{\mathrm{Te}}_{1.95}$ and $\mathrm{Ce}{\mathrm{Te}}_{2.00}$}},\
  }\href@noop {} {\bibfield  {journal} {\bibinfo  {journal} {Physical Review
  B}\ }\textbf {\bibinfo {volume} {72}},\ \bibinfo {pages} {085132} (\bibinfo
  {year} {2005})}\BibitemShut {NoStop}%
\bibitem [{\citenamefont {Straquadine}\ \emph {et~al.}(2022)\citenamefont
  {Straquadine}, \citenamefont {Ikeda},\ and\ \citenamefont
  {Fisher}}]{Straquadine2022}%
  \BibitemOpen
  \bibfield  {author} {\bibinfo {author} {\bibfnamefont {J.}~\bibnamefont
  {Straquadine}}, \bibinfo {author} {\bibfnamefont {M.~S.}\ \bibnamefont
  {Ikeda}},\ and\ \bibinfo {author} {\bibfnamefont {I.}~\bibnamefont
  {Fisher}},\ }\bibfield  {title} {\bibinfo {title} {{Evidence for Realignment
  of the Charge Density Wave State in ErTe$_3$ and TmTe$_3$ under Uniaxial
  Stress via Elastocaloric and Elastoresistivity Measurements}},\ }\href@noop
  {} {\bibfield  {journal} {\bibinfo  {journal} {Physical Review X}\ }\textbf
  {\bibinfo {volume} {12}},\ \bibinfo {pages} {021046} (\bibinfo {year}
  {2022})}\BibitemShut {NoStop}%
\bibitem [{\citenamefont {Dickinson}\ and\ \citenamefont
  {Pauling}(1923)}]{Dickinson1923}%
  \BibitemOpen
  \bibfield  {author} {\bibinfo {author} {\bibfnamefont {R.}~\bibnamefont
  {Dickinson}}\ and\ \bibinfo {author} {\bibfnamefont {L.}~\bibnamefont
  {Pauling}},\ }\bibfield  {title} {\bibinfo {title} {{The crystal structure of
  molybdenite}},\ }\href@noop {} {\bibfield  {journal} {\bibinfo  {journal}
  {Journal of the American Chemical Society}\ }\textbf {\bibinfo {volume}
  {45}},\ \bibinfo {pages} {1466} (\bibinfo {year} {1923})}\BibitemShut
  {NoStop}%
\bibitem [{\citenamefont {Manzeli}\ \emph {et~al.}(2017)\citenamefont
  {Manzeli}, \citenamefont {Ovchinnikov}, \citenamefont {Pasquier},
  \citenamefont {Yazyev},\ and\ \citenamefont {Kis}}]{Manzeli2017}%
  \BibitemOpen
  \bibfield  {author} {\bibinfo {author} {\bibfnamefont {S.}~\bibnamefont
  {Manzeli}}, \bibinfo {author} {\bibfnamefont {D.}~\bibnamefont
  {Ovchinnikov}}, \bibinfo {author} {\bibfnamefont {D.}~\bibnamefont
  {Pasquier}}, \bibinfo {author} {\bibfnamefont {O.}~\bibnamefont {Yazyev}},\
  and\ \bibinfo {author} {\bibfnamefont {A.}~\bibnamefont {Kis}},\ }\bibfield
  {title} {\bibinfo {title} {{2D transition metal dichalcogenides}},\
  }\href@noop {} {\bibfield  {journal} {\bibinfo  {journal} {Nature Reviews
  Materials}\ }\textbf {\bibinfo {volume} {2}},\ \bibinfo {pages} {17033}
  (\bibinfo {year} {2017})}\BibitemShut {NoStop}%
\bibitem [{\citenamefont {St{\"a}rk}\ \emph {et~al.}(2011)\citenamefont
  {St{\"a}rk}, \citenamefont {Kr{\"u}ger},\ and\ \citenamefont
  {Pollmann}}]{Pollmann2011}%
  \BibitemOpen
  \bibfield  {author} {\bibinfo {author} {\bibfnamefont {B.}~\bibnamefont
  {St{\"a}rk}}, \bibinfo {author} {\bibfnamefont {P.}~\bibnamefont
  {Kr{\"u}ger}},\ and\ \bibinfo {author} {\bibfnamefont {J.}~\bibnamefont
  {Pollmann}},\ }\bibfield  {title} {\bibinfo {title} {{Magnetic anisotropy of
  thin Co and Ni films on diamond surfaces}},\ }\href@noop {} {\bibfield
  {journal} {\bibinfo  {journal} {Physical Review B}\ }\textbf {\bibinfo
  {volume} {84}},\ \bibinfo {pages} {195316} (\bibinfo {year}
  {2011})}\BibitemShut {NoStop}%
\bibitem [{\citenamefont {Torun}\ \emph {et~al.}(2015)\citenamefont {Torun},
  \citenamefont {Sahin}, \citenamefont {Bacaksiz}, \citenamefont {Senger},\
  and\ \citenamefont {Peeters}}]{Torun2015}%
  \BibitemOpen
  \bibfield  {author} {\bibinfo {author} {\bibfnamefont {E.}~\bibnamefont
  {Torun}}, \bibinfo {author} {\bibfnamefont {H.}~\bibnamefont {Sahin}},
  \bibinfo {author} {\bibfnamefont {C.}~\bibnamefont {Bacaksiz}}, \bibinfo
  {author} {\bibfnamefont {R.~T.}\ \bibnamefont {Senger}},\ and\ \bibinfo
  {author} {\bibfnamefont {F.~M.}\ \bibnamefont {Peeters}},\ }\bibfield
  {title} {\bibinfo {title} {{Tuning the magnetic anisotropy in single-layer
  crystal structures}},\ }\href {https://doi.org/10.1103/PhysRevB.92.104407}
  {\bibfield  {journal} {\bibinfo  {journal} {Physical Review B}\ }\textbf
  {\bibinfo {volume} {92}},\ \bibinfo {pages} {104407} (\bibinfo {year}
  {2015})}\BibitemShut {NoStop}%
\bibitem [{\citenamefont {Gambardella}\ \emph {et~al.}(2009)\citenamefont
  {Gambardella}, \citenamefont {Stepanow}, \citenamefont {Dmitriev},
  \citenamefont {Honolka}, \citenamefont {de~Groot}, \citenamefont
  {Lingenfelder}, \citenamefont {Gupta}, \citenamefont {Sarma}, \citenamefont
  {Bencok}, \citenamefont {Stanescu}, \citenamefont {Clair}, \citenamefont
  {Pons}, \citenamefont {Lin}, \citenamefont {Seitsonen}, \citenamefont
  {Brune}, \citenamefont {Barth},\ and\ \citenamefont
  {Kern}}]{Gambardella2009}%
  \BibitemOpen
  \bibfield  {author} {\bibinfo {author} {\bibfnamefont {P.}~\bibnamefont
  {Gambardella}}, \bibinfo {author} {\bibfnamefont {S.}~\bibnamefont
  {Stepanow}}, \bibinfo {author} {\bibfnamefont {A.}~\bibnamefont {Dmitriev}},
  \bibinfo {author} {\bibfnamefont {J.}~\bibnamefont {Honolka}}, \bibinfo
  {author} {\bibfnamefont {F.}~\bibnamefont {de~Groot}}, \bibinfo {author}
  {\bibfnamefont {M.}~\bibnamefont {Lingenfelder}}, \bibinfo {author}
  {\bibfnamefont {S.}~\bibnamefont {Gupta}}, \bibinfo {author} {\bibfnamefont
  {D.}~\bibnamefont {Sarma}}, \bibinfo {author} {\bibfnamefont
  {P.}~\bibnamefont {Bencok}}, \bibinfo {author} {\bibfnamefont
  {S.}~\bibnamefont {Stanescu}}, \bibinfo {author} {\bibfnamefont
  {S.}~\bibnamefont {Clair}}, \bibinfo {author} {\bibfnamefont
  {S.}~\bibnamefont {Pons}}, \bibinfo {author} {\bibfnamefont {N.}~\bibnamefont
  {Lin}}, \bibinfo {author} {\bibfnamefont {A.}~\bibnamefont {Seitsonen}},
  \bibinfo {author} {\bibfnamefont {H.}~\bibnamefont {Brune}}, \bibinfo
  {author} {\bibfnamefont {J.}~\bibnamefont {Barth}},\ and\ \bibinfo {author}
  {\bibfnamefont {K.}~\bibnamefont {Kern}},\ }\bibfield  {title} {\bibinfo
  {title} {{Supramolecular control of the magnetic anisotropy in
  two-dimensional high-spin Fe arrays at a metal interface}},\ }\href@noop {}
  {\bibfield  {journal} {\bibinfo  {journal} {Nature Materials}\ }\textbf
  {\bibinfo {volume} {8}},\ \bibinfo {pages} {189} (\bibinfo {year}
  {2009})}\BibitemShut {NoStop}%
\bibitem [{\citenamefont {Honda}\ \emph {et~al.}(2017)\citenamefont {Honda},
  \citenamefont {White}, \citenamefont {Harris}, \citenamefont {Chapon},
  \citenamefont {Fennell}, \citenamefont {Roessli}, \citenamefont {Zaharko},
  \citenamefont {Murakami}, \citenamefont {Kenzelmann},\ and\ \citenamefont
  {Kimura}}]{Honda2017}%
  \BibitemOpen
  \bibfield  {author} {\bibinfo {author} {\bibfnamefont {T.}~\bibnamefont
  {Honda}}, \bibinfo {author} {\bibfnamefont {J.~S.}\ \bibnamefont {White}},
  \bibinfo {author} {\bibfnamefont {A.~B.}\ \bibnamefont {Harris}}, \bibinfo
  {author} {\bibfnamefont {L.~C.}\ \bibnamefont {Chapon}}, \bibinfo {author}
  {\bibfnamefont {A.}~\bibnamefont {Fennell}}, \bibinfo {author} {\bibfnamefont
  {B.}~\bibnamefont {Roessli}}, \bibinfo {author} {\bibfnamefont
  {O.}~\bibnamefont {Zaharko}}, \bibinfo {author} {\bibfnamefont
  {Y.}~\bibnamefont {Murakami}}, \bibinfo {author} {\bibfnamefont
  {M.}~\bibnamefont {Kenzelmann}},\ and\ \bibinfo {author} {\bibfnamefont
  {T.}~\bibnamefont {Kimura}},\ }\bibfield  {title} {\bibinfo {title} {{Coupled
  multiferroic domain switching in the canted conical spin spiral system
  Mn$_2$GeO$_4$}},\ }\href@noop {} {\bibfield  {journal} {\bibinfo  {journal}
  {Nature Communications}\ }\textbf {\bibinfo {volume} {8}},\ \bibinfo {pages}
  {15457} (\bibinfo {year} {2017})}\BibitemShut {NoStop}%
\bibitem [{\citenamefont {Riberolles}\ \emph {et~al.}(2021)\citenamefont
  {Riberolles}, \citenamefont {Trevisan}, \citenamefont {Kuthanazhi},
  \citenamefont {Heitmann}, \citenamefont {Ye}, \citenamefont {Johnston},
  \citenamefont {Bud'ko}, \citenamefont {Ryan}, \citenamefont {Canfield},
  \citenamefont {Kreyssig}, \citenamefont {Vishwanath}, \citenamefont
  {McQueeney}, \citenamefont {Wang}, \citenamefont {Orth},\ and\ \citenamefont
  {Ueland}}]{Riberolles2021}%
  \BibitemOpen
  \bibfield  {author} {\bibinfo {author} {\bibfnamefont {S.}~\bibnamefont
  {Riberolles}}, \bibinfo {author} {\bibfnamefont {T.}~\bibnamefont
  {Trevisan}}, \bibinfo {author} {\bibfnamefont {B.}~\bibnamefont
  {Kuthanazhi}}, \bibinfo {author} {\bibfnamefont {T.}~\bibnamefont
  {Heitmann}}, \bibinfo {author} {\bibfnamefont {F.}~\bibnamefont {Ye}},
  \bibinfo {author} {\bibfnamefont {D.}~\bibnamefont {Johnston}}, \bibinfo
  {author} {\bibfnamefont {S.}~\bibnamefont {Bud'ko}}, \bibinfo {author}
  {\bibfnamefont {D.}~\bibnamefont {Ryan}}, \bibinfo {author} {\bibfnamefont
  {P.}~\bibnamefont {Canfield}}, \bibinfo {author} {\bibfnamefont
  {A.}~\bibnamefont {Kreyssig}}, \bibinfo {author} {\bibfnamefont
  {A.}~\bibnamefont {Vishwanath}}, \bibinfo {author} {\bibfnamefont
  {R.}~\bibnamefont {McQueeney}}, \bibinfo {author} {\bibfnamefont {L.-L.}\
  \bibnamefont {Wang}}, \bibinfo {author} {\bibfnamefont {P.}~\bibnamefont
  {Orth}},\ and\ \bibinfo {author} {\bibfnamefont {B.}~\bibnamefont {Ueland}},\
  }\bibfield  {title} {\bibinfo {title} {{Magnetic
  crystalline-symmetry-protected axion electrodynamics and field-tunable
  unpinned Dirac cones in EuIn$_2$As$_2$}},\ }\href@noop {} {\bibfield
  {journal} {\bibinfo  {journal} {Nature Communications}\ }\textbf {\bibinfo
  {volume} {12}},\ \bibinfo {pages} {999} (\bibinfo {year} {2021})}\BibitemShut
  {NoStop}%
\bibitem [{\citenamefont {Lei}\ \emph {et~al.}(2019)\citenamefont {Lei},
  \citenamefont {Duppel}, \citenamefont {Lippmann}, \citenamefont {Nuss},
  \citenamefont {Lotsch},\ and\ \citenamefont {Schoop}}]{Lei2019}%
  \BibitemOpen
  \bibfield  {author} {\bibinfo {author} {\bibfnamefont {S.}~\bibnamefont
  {Lei}}, \bibinfo {author} {\bibfnamefont {V.}~\bibnamefont {Duppel}},
  \bibinfo {author} {\bibfnamefont {J.}~\bibnamefont {Lippmann}}, \bibinfo
  {author} {\bibfnamefont {J.}~\bibnamefont {Nuss}}, \bibinfo {author}
  {\bibfnamefont {B.}~\bibnamefont {Lotsch}},\ and\ \bibinfo {author}
  {\bibfnamefont {L.}~\bibnamefont {Schoop}},\ }\bibfield  {title} {\bibinfo
  {title} {{Charge Density Waves and Magnetism in Topological Semimetal
  Candidates GdSb$_x$Te$_{2-x-\delta}$}},\ }\href@noop {} {\bibfield  {journal}
  {\bibinfo  {journal} {Advanced Quantum Technologies}\ }\textbf {\bibinfo
  {volume} {2}},\ \bibinfo {pages} {1900045} (\bibinfo {year}
  {2019})}\BibitemShut {NoStop}%
\bibitem [{\citenamefont {Lei}\ \emph {et~al.}(2021)\citenamefont {Lei},
  \citenamefont {Saltzman},\ and\ \citenamefont {Schoop}}]{Lei2021}%
  \BibitemOpen
  \bibfield  {author} {\bibinfo {author} {\bibfnamefont {S.}~\bibnamefont
  {Lei}}, \bibinfo {author} {\bibfnamefont {A.}~\bibnamefont {Saltzman}},\ and\
  \bibinfo {author} {\bibfnamefont {L.}~\bibnamefont {Schoop}},\ }\bibfield
  {title} {\bibinfo {title} {{Complex magnetic phases enriched by charge
  density waves in the topological semimetals
  ${\mathrm{GdSb}}_{x}{\mathrm{Te}}_{2\ensuremath{-}x\ensuremath{-}\ensuremath{\delta}}$}},\
  }\href@noop {} {\bibfield  {journal} {\bibinfo  {journal} {Physical Review
  B}\ }\textbf {\bibinfo {volume} {103}},\ \bibinfo {pages} {134418} (\bibinfo
  {year} {2021})}\BibitemShut {NoStop}%
\bibitem [{\citenamefont {Gr{\"u}ner}(1988)}]{Gruner1988}%
  \BibitemOpen
  \bibfield  {author} {\bibinfo {author} {\bibfnamefont {G.}~\bibnamefont
  {Gr{\"u}ner}},\ }\bibfield  {title} {\bibinfo {title} {{The dynamics of
  charge-density waves}},\ }\href {https://doi.org/10.1103/RevModPhys.60.1129}
  {\bibfield  {journal} {\bibinfo  {journal} {Reviews of Modern Physics}\
  }\textbf {\bibinfo {volume} {60}},\ \bibinfo {pages} {1129} (\bibinfo {year}
  {1988})}\BibitemShut {NoStop}%
\bibitem [{\citenamefont {Sinchenko}\ \emph {et~al.}(2012)\citenamefont
  {Sinchenko}, \citenamefont {Lejay},\ and\ \citenamefont
  {Monceau}}]{Sinchenko2012}%
  \BibitemOpen
  \bibfield  {author} {\bibinfo {author} {\bibfnamefont {A.~A.}\ \bibnamefont
  {Sinchenko}}, \bibinfo {author} {\bibfnamefont {P.}~\bibnamefont {Lejay}},\
  and\ \bibinfo {author} {\bibfnamefont {P.}~\bibnamefont {Monceau}},\
  }\bibfield  {title} {\bibinfo {title} {{Sliding charge-density wave in
  two-dimensional rare-earth tellurides}},\ }\href@noop {} {\bibfield
  {journal} {\bibinfo  {journal} {Physical Review B}\ }\textbf {\bibinfo
  {volume} {85}},\ \bibinfo {pages} {241104} (\bibinfo {year}
  {2012})}\BibitemShut {NoStop}%
\bibitem [{\citenamefont {Moriya}(1960)}]{Moriya1960}%
  \BibitemOpen
  \bibfield  {author} {\bibinfo {author} {\bibfnamefont {T.}~\bibnamefont
  {Moriya}},\ }\bibfield  {title} {\bibinfo {title} {{Anisotropic Superexchange
  Interaction and Weak Ferromagnetism}},\ }\href@noop {} {\bibfield  {journal}
  {\bibinfo  {journal} {Physical Review}\ }\textbf {\bibinfo {volume} {120}},\
  \bibinfo {pages} {91} (\bibinfo {year} {1960})}\BibitemShut {NoStop}%
\bibitem [{\citenamefont {Ghader}\ \emph {et~al.}(2022)\citenamefont {Ghader},
  \citenamefont {Jabakhanji},\ and\ \citenamefont {Stroppa}}]{Ghader2022}%
  \BibitemOpen
  \bibfield  {author} {\bibinfo {author} {\bibfnamefont {D.}~\bibnamefont
  {Ghader}}, \bibinfo {author} {\bibfnamefont {B.}~\bibnamefont {Jabakhanji}},\
  and\ \bibinfo {author} {\bibfnamefont {A.}~\bibnamefont {Stroppa}},\
  }\bibfield  {title} {\bibinfo {title} {{Whirling interlayer fields as a
  source of stable topological order in Moir{\'e} CrI$_3$}},\ }\href@noop {}
  {\bibfield  {journal} {\bibinfo  {journal} {Communications Physics}\ }\textbf
  {\bibinfo {volume} {5}},\ \bibinfo {pages} {192} (\bibinfo {year}
  {2022})}\BibitemShut {NoStop}%
\bibitem [{\citenamefont {Volovik}(1987)}]{Volovik1987}%
  \BibitemOpen
  \bibfield  {author} {\bibinfo {author} {\bibfnamefont {G.}~\bibnamefont
  {Volovik}},\ }\bibfield  {title} {\bibinfo {title} {{Linear momentum in
  ferromagnets}},\ }\href@noop {} {\bibfield  {journal} {\bibinfo  {journal}
  {Journal of Physics C: Solid State Physics}\ }\textbf {\bibinfo {volume}
  {20}},\ \bibinfo {pages} {L83} (\bibinfo {year} {1987})}\BibitemShut
  {NoStop}%
\bibitem [{\citenamefont {Tokura}\ and\ \citenamefont
  {Nagaosa}(2018)}]{Tokura2018}%
  \BibitemOpen
  \bibfield  {author} {\bibinfo {author} {\bibfnamefont {Y.}~\bibnamefont
  {Tokura}}\ and\ \bibinfo {author} {\bibfnamefont {N.}~\bibnamefont
  {Nagaosa}},\ }\bibfield  {title} {\bibinfo {title} {{Nonreciprocal responses
  from non-centrosymmetric quantum materials}},\ }\href@noop {} {\bibfield
  {journal} {\bibinfo  {journal} {Nature Communications}\ }\textbf {\bibinfo
  {volume} {9}},\ \bibinfo {pages} {3740} (\bibinfo {year} {2018})}\BibitemShut
  {NoStop}%
\bibitem [{\citenamefont {Osborn}(1945)}]{Osborn1945}%
  \BibitemOpen
  \bibfield  {author} {\bibinfo {author} {\bibfnamefont {J.~A.}\ \bibnamefont
  {Osborn}},\ }\bibfield  {title} {\bibinfo {title} {{Demagnetizing Factors of
  the General Ellipsoid}},\ }\href@noop {} {\bibfield  {journal} {\bibinfo
  {journal} {Physical Review}\ }\textbf {\bibinfo {volume} {67}},\ \bibinfo
  {pages} {351} (\bibinfo {year} {1945})}\BibitemShut {NoStop}%
\bibitem [{\citenamefont {Scheie}(2021)}]{Scheie2021}%
  \BibitemOpen
  \bibfield  {author} {\bibinfo {author} {\bibfnamefont {A.}~\bibnamefont
  {Scheie}},\ }\bibfield  {title} {\bibinfo {title} {{PyCrystalField: Software
  for Calculation, Analysis and Fitting of Crystal Electric Field
  Hamiltonians}},\ }\href {https://doi.org/10.1107/S160057672001554X}
  {\bibfield  {journal} {\bibinfo  {journal} {Journal of Applied
  Crystallography}\ }\textbf {\bibinfo {volume} {54}},\ \bibinfo {pages} {356}
  (\bibinfo {year} {2021})}\BibitemShut {NoStop}%
\bibitem [{\citenamefont {Gao}\ \emph {et~al.}(2020)\citenamefont {Gao},
  \citenamefont {Yin}, \citenamefont {Wang}, \citenamefont {Li}, \citenamefont
  {Cai}, \citenamefont {Zhao}, \citenamefont {Lei}, \citenamefont {Wang},
  \citenamefont {Zhang},\ and\ \citenamefont {Shen}}]{Gao2020}%
  \BibitemOpen
  \bibfield  {author} {\bibinfo {author} {\bibfnamefont {Y.}~\bibnamefont
  {Gao}}, \bibinfo {author} {\bibfnamefont {Q.}~\bibnamefont {Yin}}, \bibinfo
  {author} {\bibfnamefont {Q.}~\bibnamefont {Wang}}, \bibinfo {author}
  {\bibfnamefont {Z.}~\bibnamefont {Li}}, \bibinfo {author} {\bibfnamefont
  {J.}~\bibnamefont {Cai}}, \bibinfo {author} {\bibfnamefont {T.}~\bibnamefont
  {Zhao}}, \bibinfo {author} {\bibfnamefont {H.}~\bibnamefont {Lei}}, \bibinfo
  {author} {\bibfnamefont {S.}~\bibnamefont {Wang}}, \bibinfo {author}
  {\bibfnamefont {Y.}~\bibnamefont {Zhang}},\ and\ \bibinfo {author}
  {\bibfnamefont {B.}~\bibnamefont {Shen}},\ }\bibfield  {title} {\bibinfo
  {title} {{Spontaneous (Anti)meron Chains in the Domain Walls of van der Waals
  Ferromagnetic Fe$_{5\ensuremath{-}x}$GeTe$_2$}},\ }\href@noop {} {\bibfield
  {journal} {\bibinfo  {journal} {Advanced Materials}\ }\textbf {\bibinfo
  {volume} {32}},\ \bibinfo {pages} {2005228} (\bibinfo {year}
  {2020})}\BibitemShut {NoStop}%
\bibitem [{\citenamefont {Ly}\ \emph {et~al.}(2021)\citenamefont {Ly},
  \citenamefont {Park}, \citenamefont {Kim}, \citenamefont {Ahn}, \citenamefont
  {Lee}, \citenamefont {Kim}, \citenamefont {Park}, \citenamefont {Duvjir},
  \citenamefont {Lam}, \citenamefont {Jang}, \citenamefont {You}, \citenamefont
  {Jo}, \citenamefont {Kim}, \citenamefont {Lee}, \citenamefont {Kim},\ and\
  \citenamefont {Kim}}]{Ly2021}%
  \BibitemOpen
  \bibfield  {author} {\bibinfo {author} {\bibfnamefont {T.~T.}\ \bibnamefont
  {Ly}}, \bibinfo {author} {\bibfnamefont {J.}~\bibnamefont {Park}}, \bibinfo
  {author} {\bibfnamefont {K.}~\bibnamefont {Kim}}, \bibinfo {author}
  {\bibfnamefont {H.-B.}\ \bibnamefont {Ahn}}, \bibinfo {author} {\bibfnamefont
  {N.~J.}\ \bibnamefont {Lee}}, \bibinfo {author} {\bibfnamefont
  {K.}~\bibnamefont {Kim}}, \bibinfo {author} {\bibfnamefont {T.-E.}\
  \bibnamefont {Park}}, \bibinfo {author} {\bibfnamefont {G.}~\bibnamefont
  {Duvjir}}, \bibinfo {author} {\bibfnamefont {N.~H.}\ \bibnamefont {Lam}},
  \bibinfo {author} {\bibfnamefont {K.}~\bibnamefont {Jang}}, \bibinfo {author}
  {\bibfnamefont {C.-Y.}\ \bibnamefont {You}}, \bibinfo {author} {\bibfnamefont
  {Y.}~\bibnamefont {Jo}}, \bibinfo {author} {\bibfnamefont {S.~K.}\
  \bibnamefont {Kim}}, \bibinfo {author} {\bibfnamefont {C.}~\bibnamefont
  {Lee}}, \bibinfo {author} {\bibfnamefont {S.}~\bibnamefont {Kim}},\ and\
  \bibinfo {author} {\bibfnamefont {J.}~\bibnamefont {Kim}},\ }\bibfield
  {title} {\bibinfo {title} {{Direct Observation of Fe-Ge Ordering in
  Fe$_{5\ensuremath{-}x}$GeTe$_2$ Crystals and Resultant Helimagnetism}},\
  }\href {https://doi.org/https://doi.org/10.1002/adfm.202009758} {\bibfield
  {journal} {\bibinfo  {journal} {Advanced Functional Materials}\ }\textbf
  {\bibinfo {volume} {31}},\ \bibinfo {pages} {2009758} (\bibinfo {year}
  {2021})}\BibitemShut {NoStop}%
\bibitem [{\citenamefont {May}\ \emph {et~al.}(2019)\citenamefont {May},
  \citenamefont {Bridges},\ and\ \citenamefont {McGuire}}]{May2019}%
  \BibitemOpen
  \bibfield  {author} {\bibinfo {author} {\bibfnamefont {A.~F.}\ \bibnamefont
  {May}}, \bibinfo {author} {\bibfnamefont {C.~A.}\ \bibnamefont {Bridges}},\
  and\ \bibinfo {author} {\bibfnamefont {M.~A.}\ \bibnamefont {McGuire}},\
  }\bibfield  {title} {\bibinfo {title} {{Physical properties and thermal
  stability of Fe$_{5\ensuremath{-}x}$GeTe$_2$ single crystals}},\ }\href
  {https://doi.org/10.1103/PhysRevMaterials.3.104401} {\bibfield  {journal}
  {\bibinfo  {journal} {Physical Review Materials}\ }\textbf {\bibinfo {volume}
  {3}},\ \bibinfo {pages} {104401} (\bibinfo {year} {2019})}\BibitemShut
  {NoStop}%
\bibitem [{\citenamefont {Baenitz}\ \emph {et~al.}(2021)\citenamefont
  {Baenitz}, \citenamefont {Piva}, \citenamefont {Luther}, \citenamefont
  {Sichelschmidt}, \citenamefont {Ranjith}, \citenamefont
  {Dawczak-D\ifmmode~\mbox{\c{e}}\else \c{e}\fi{}bicki}, \citenamefont
  {Ajeesh}, \citenamefont {Kim}, \citenamefont {Siemann}, \citenamefont {Bigi},
  \citenamefont {Manuel}, \citenamefont {Khalyavin}, \citenamefont {Sokolov},
  \citenamefont {Mokhtari}, \citenamefont {Zhang}, \citenamefont {Yasuoka},
  \citenamefont {King}, \citenamefont {Vinai}, \citenamefont {Polewczyk},
  \citenamefont {Torelli}, \citenamefont {Wosnitza}, \citenamefont {Burkhardt},
  \citenamefont {Schmidt}, \citenamefont {Rosner}, \citenamefont {Wirth},
  \citenamefont {K\"uhne}, \citenamefont {Nicklas},\ and\ \citenamefont
  {Schmidt}}]{Baenitz2021}%
  \BibitemOpen
  \bibfield  {author} {\bibinfo {author} {\bibfnamefont {M.}~\bibnamefont
  {Baenitz}}, \bibinfo {author} {\bibfnamefont {M.~M.}\ \bibnamefont {Piva}},
  \bibinfo {author} {\bibfnamefont {S.}~\bibnamefont {Luther}}, \bibinfo
  {author} {\bibfnamefont {J.}~\bibnamefont {Sichelschmidt}}, \bibinfo {author}
  {\bibfnamefont {K.~M.}\ \bibnamefont {Ranjith}}, \bibinfo {author}
  {\bibfnamefont {H.}~\bibnamefont {Dawczak-D\ifmmode~\mbox{\c{e}}\else
  \c{e}\fi{}bicki}}, \bibinfo {author} {\bibfnamefont {M.~O.}\ \bibnamefont
  {Ajeesh}}, \bibinfo {author} {\bibfnamefont {S.-J.}\ \bibnamefont {Kim}},
  \bibinfo {author} {\bibfnamefont {G.}~\bibnamefont {Siemann}}, \bibinfo
  {author} {\bibfnamefont {C.}~\bibnamefont {Bigi}}, \bibinfo {author}
  {\bibfnamefont {P.}~\bibnamefont {Manuel}}, \bibinfo {author} {\bibfnamefont
  {D.}~\bibnamefont {Khalyavin}}, \bibinfo {author} {\bibfnamefont {D.~A.}\
  \bibnamefont {Sokolov}}, \bibinfo {author} {\bibfnamefont {P.}~\bibnamefont
  {Mokhtari}}, \bibinfo {author} {\bibfnamefont {H.}~\bibnamefont {Zhang}},
  \bibinfo {author} {\bibfnamefont {H.}~\bibnamefont {Yasuoka}}, \bibinfo
  {author} {\bibfnamefont {P.~D.~C.}\ \bibnamefont {King}}, \bibinfo {author}
  {\bibfnamefont {G.}~\bibnamefont {Vinai}}, \bibinfo {author} {\bibfnamefont
  {V.}~\bibnamefont {Polewczyk}}, \bibinfo {author} {\bibfnamefont
  {P.}~\bibnamefont {Torelli}}, \bibinfo {author} {\bibfnamefont
  {J.}~\bibnamefont {Wosnitza}}, \bibinfo {author} {\bibfnamefont
  {U.}~\bibnamefont {Burkhardt}}, \bibinfo {author} {\bibfnamefont
  {B.}~\bibnamefont {Schmidt}}, \bibinfo {author} {\bibfnamefont
  {H.}~\bibnamefont {Rosner}}, \bibinfo {author} {\bibfnamefont
  {S.}~\bibnamefont {Wirth}}, \bibinfo {author} {\bibfnamefont
  {H.}~\bibnamefont {K\"uhne}}, \bibinfo {author} {\bibfnamefont
  {M.}~\bibnamefont {Nicklas}},\ and\ \bibinfo {author} {\bibfnamefont
  {M.}~\bibnamefont {Schmidt}},\ }\bibfield  {title} {\bibinfo {title} {{Planar
  triangular $S=3/2$ magnet AgCrSe$_2$: Magnetic frustration, short range
  correlations, and field-tuned anisotropic cycloidal magnetic order}},\ }\href
  {https://doi.org/10.1103/PhysRevB.104.134410} {\bibfield  {journal} {\bibinfo
   {journal} {Physical Review B}\ }\textbf {\bibinfo {volume} {104}},\ \bibinfo
  {pages} {134410} (\bibinfo {year} {2021})}\BibitemShut {NoStop}%
\bibitem [{\citenamefont {Gautam}\ \emph {et~al.}(2002)\citenamefont {Gautam},
  \citenamefont {Seshadri}, \citenamefont {Vasudevan},\ and\ \citenamefont
  {Maignan}}]{Gautam2002}%
  \BibitemOpen
  \bibfield  {author} {\bibinfo {author} {\bibfnamefont {U.~K.}\ \bibnamefont
  {Gautam}}, \bibinfo {author} {\bibfnamefont {R.}~\bibnamefont {Seshadri}},
  \bibinfo {author} {\bibfnamefont {S.}~\bibnamefont {Vasudevan}},\ and\
  \bibinfo {author} {\bibfnamefont {A.}~\bibnamefont {Maignan}},\ }\bibfield
  {title} {\bibinfo {title} {{Magnetic and transport properties, and electronic
  structure of the layered chalcogenide AgCrSe$_2$}},\ }\href@noop {}
  {\bibfield  {journal} {\bibinfo  {journal} {Solid state communications}\
  }\textbf {\bibinfo {volume} {122}},\ \bibinfo {pages} {607} (\bibinfo {year}
  {2002})}\BibitemShut {NoStop}%
\bibitem [{\citenamefont {Kurumaji}\ \emph {et~al.}(2013)\citenamefont
  {Kurumaji}, \citenamefont {Seki}, \citenamefont {Ishiwata}, \citenamefont
  {Murakawa}, \citenamefont {Kaneko},\ and\ \citenamefont
  {Tokura}}]{Kurumaji2013}%
  \BibitemOpen
  \bibfield  {author} {\bibinfo {author} {\bibfnamefont {T.}~\bibnamefont
  {Kurumaji}}, \bibinfo {author} {\bibfnamefont {S.}~\bibnamefont {Seki}},
  \bibinfo {author} {\bibfnamefont {S.}~\bibnamefont {Ishiwata}}, \bibinfo
  {author} {\bibfnamefont {H.}~\bibnamefont {Murakawa}}, \bibinfo {author}
  {\bibfnamefont {Y.}~\bibnamefont {Kaneko}},\ and\ \bibinfo {author}
  {\bibfnamefont {Y.}~\bibnamefont {Tokura}},\ }\bibfield  {title} {\bibinfo
  {title} {{Magnetoelectric responses induced by domain rearrangement and spin
  structural change in triangular-lattice helimagnets NiI$_2$ and CoI$_2$}},\
  }\href@noop {} {\bibfield  {journal} {\bibinfo  {journal} {Physical Review
  B}\ }\textbf {\bibinfo {volume} {87}},\ \bibinfo {pages} {014429} (\bibinfo
  {year} {2013})}\BibitemShut {NoStop}%
\bibitem [{\citenamefont {Lebedev}\ \emph {et~al.}(2023)\citenamefont
  {Lebedev}, \citenamefont {Gish}, \citenamefont {Garvey}, \citenamefont
  {Stanev}, \citenamefont {Choi}, \citenamefont {Georgopoulos}, \citenamefont
  {Song}, \citenamefont {Park}, \citenamefont {Watanabe}, \citenamefont
  {Taniguchi}, \citenamefont {Stern}, \citenamefont {Sangwan},\ and\
  \citenamefont {Hersam}}]{Lebedev2023}%
  \BibitemOpen
  \bibfield  {author} {\bibinfo {author} {\bibfnamefont {D.}~\bibnamefont
  {Lebedev}}, \bibinfo {author} {\bibfnamefont {J.~T.}\ \bibnamefont {Gish}},
  \bibinfo {author} {\bibfnamefont {E.~S.}\ \bibnamefont {Garvey}}, \bibinfo
  {author} {\bibfnamefont {T.~K.}\ \bibnamefont {Stanev}}, \bibinfo {author}
  {\bibfnamefont {J.}~\bibnamefont {Choi}}, \bibinfo {author} {\bibfnamefont
  {L.}~\bibnamefont {Georgopoulos}}, \bibinfo {author} {\bibfnamefont {T.~W.}\
  \bibnamefont {Song}}, \bibinfo {author} {\bibfnamefont {H.~Y.}\ \bibnamefont
  {Park}}, \bibinfo {author} {\bibfnamefont {K.}~\bibnamefont {Watanabe}},
  \bibinfo {author} {\bibfnamefont {T.}~\bibnamefont {Taniguchi}}, \bibinfo
  {author} {\bibfnamefont {N.~P.}\ \bibnamefont {Stern}}, \bibinfo {author}
  {\bibfnamefont {V.~K.}\ \bibnamefont {Sangwan}},\ and\ \bibinfo {author}
  {\bibfnamefont {M.~C.}\ \bibnamefont {Hersam}},\ }\bibfield  {title}
  {\bibinfo {title} {{Electrical Interrogation of Thickness-Dependent
  Multiferroic Phase Transitions in the 2D Antiferromagnetic Semiconductor
  $\mathrm{NiI}_2$}},\ }\href
  {https://doi.org/https://doi.org/10.1002/adfm.202212568} {\bibfield
  {journal} {\bibinfo  {journal} {Advanced Functional Materials}\ }\textbf
  {\bibinfo {volume} {33}},\ \bibinfo {pages} {2212568} (\bibinfo {year}
  {2023})}\BibitemShut {NoStop}%
\bibitem [{\citenamefont {Adam}\ \emph {et~al.}(1980)\citenamefont {Adam},
  \citenamefont {Billerey}, \citenamefont {Terrier}, \citenamefont {Mainard},
  \citenamefont {Regnault}, \citenamefont {Rossat-Mignod},\ and\ \citenamefont
  {M{\'e}riel}}]{Adam1980}%
  \BibitemOpen
  \bibfield  {author} {\bibinfo {author} {\bibfnamefont {A.}~\bibnamefont
  {Adam}}, \bibinfo {author} {\bibfnamefont {D.}~\bibnamefont {Billerey}},
  \bibinfo {author} {\bibfnamefont {C.}~\bibnamefont {Terrier}}, \bibinfo
  {author} {\bibfnamefont {R.}~\bibnamefont {Mainard}}, \bibinfo {author}
  {\bibfnamefont {L.}~\bibnamefont {Regnault}}, \bibinfo {author}
  {\bibfnamefont {J.}~\bibnamefont {Rossat-Mignod}},\ and\ \bibinfo {author}
  {\bibfnamefont {P.}~\bibnamefont {M{\'e}riel}},\ }\bibfield  {title}
  {\bibinfo {title} {{Neutron diffraction study of the commensurate and
  incommensurate magnetic structures of NiBr$_2$}},\ }\href
  {https://doi.org/https://doi.org/10.1016/0038-1098(80)90757-7} {\bibfield
  {journal} {\bibinfo  {journal} {Solid State Communications}\ }\textbf
  {\bibinfo {volume} {35}},\ \bibinfo {pages} {1} (\bibinfo {year}
  {1980})}\BibitemShut {NoStop}%
\bibitem [{\citenamefont {Tokunaga}\ \emph {et~al.}(2011)\citenamefont
  {Tokunaga}, \citenamefont {Okuyama}, \citenamefont {Kurumaji}, \citenamefont
  {Arima}, \citenamefont {Nakao}, \citenamefont {Murakami}, \citenamefont
  {Taguchi},\ and\ \citenamefont {Tokura}}]{Tokunaga2011}%
  \BibitemOpen
  \bibfield  {author} {\bibinfo {author} {\bibfnamefont {Y.}~\bibnamefont
  {Tokunaga}}, \bibinfo {author} {\bibfnamefont {D.}~\bibnamefont {Okuyama}},
  \bibinfo {author} {\bibfnamefont {T.}~\bibnamefont {Kurumaji}}, \bibinfo
  {author} {\bibfnamefont {T.}~\bibnamefont {Arima}}, \bibinfo {author}
  {\bibfnamefont {H.}~\bibnamefont {Nakao}}, \bibinfo {author} {\bibfnamefont
  {Y.}~\bibnamefont {Murakami}}, \bibinfo {author} {\bibfnamefont
  {Y.}~\bibnamefont {Taguchi}},\ and\ \bibinfo {author} {\bibfnamefont
  {Y.}~\bibnamefont {Tokura}},\ }\bibfield  {title} {\bibinfo {title}
  {{Multiferroicity in NiBr$_2$ with long-wavelength cycloidal spin structure
  on a triangular lattice}},\ }\href
  {https://doi.org/10.1103/PhysRevB.84.060406} {\bibfield  {journal} {\bibinfo
  {journal} {Physical Review B}\ }\textbf {\bibinfo {volume} {84}},\ \bibinfo
  {pages} {060406} (\bibinfo {year} {2011})}\BibitemShut {NoStop}%
\bibitem [{\citenamefont {Ronda}\ \emph {et~al.}(1987)\citenamefont {Ronda},
  \citenamefont {Arends},\ and\ \citenamefont {Haas}}]{Ronda1987}%
  \BibitemOpen
  \bibfield  {author} {\bibinfo {author} {\bibfnamefont {C.~R.}\ \bibnamefont
  {Ronda}}, \bibinfo {author} {\bibfnamefont {G.~J.}\ \bibnamefont {Arends}},\
  and\ \bibinfo {author} {\bibfnamefont {C.}~\bibnamefont {Haas}},\ }\bibfield
  {title} {\bibinfo {title} {{Photoconductivity of the nickel dihalides and the
  nature of the energy gap}},\ }\href@noop {} {\bibfield  {journal} {\bibinfo
  {journal} {Physical Review B}\ }\textbf {\bibinfo {volume} {35}},\ \bibinfo
  {pages} {4038} (\bibinfo {year} {1987})}\BibitemShut {NoStop}%
\bibitem [{\citenamefont {Kurumaji}\ \emph {et~al.}(2011)\citenamefont
  {Kurumaji}, \citenamefont {Seki}, \citenamefont {Ishiwata}, \citenamefont
  {Murakawa}, \citenamefont {Tokunaga}, \citenamefont {Kaneko},\ and\
  \citenamefont {Tokura}}]{Kurumaji2011}%
  \BibitemOpen
  \bibfield  {author} {\bibinfo {author} {\bibfnamefont {T.}~\bibnamefont
  {Kurumaji}}, \bibinfo {author} {\bibfnamefont {S.}~\bibnamefont {Seki}},
  \bibinfo {author} {\bibfnamefont {S.}~\bibnamefont {Ishiwata}}, \bibinfo
  {author} {\bibfnamefont {H.}~\bibnamefont {Murakawa}}, \bibinfo {author}
  {\bibfnamefont {Y.}~\bibnamefont {Tokunaga}}, \bibinfo {author}
  {\bibfnamefont {Y.}~\bibnamefont {Kaneko}},\ and\ \bibinfo {author}
  {\bibfnamefont {Y.}~\bibnamefont {Tokura}},\ }\bibfield  {title} {\bibinfo
  {title} {{Magnetic-Field Induced Competition of Two Multiferroic Orders in a
  Triangular-Lattice Helimagnet MnI$_2$}},\ }\href@noop {} {\bibfield
  {journal} {\bibinfo  {journal} {Physical Review Letters}\ }\textbf {\bibinfo
  {volume} {106}},\ \bibinfo {pages} {167206} (\bibinfo {year}
  {2011})}\BibitemShut {NoStop}%
\bibitem [{\citenamefont {Ghimire}\ \emph {et~al.}(2013)\citenamefont
  {Ghimire}, \citenamefont {McGuire}, \citenamefont {Parker}, \citenamefont
  {Sipos}, \citenamefont {Tang}, \citenamefont {Yan}, \citenamefont {Sales},\
  and\ \citenamefont {Mandrus}}]{Ghimire2013}%
  \BibitemOpen
  \bibfield  {author} {\bibinfo {author} {\bibfnamefont {N.~J.}\ \bibnamefont
  {Ghimire}}, \bibinfo {author} {\bibfnamefont {M.~A.}\ \bibnamefont
  {McGuire}}, \bibinfo {author} {\bibfnamefont {D.~S.}\ \bibnamefont {Parker}},
  \bibinfo {author} {\bibfnamefont {B.}~\bibnamefont {Sipos}}, \bibinfo
  {author} {\bibfnamefont {S.}~\bibnamefont {Tang}}, \bibinfo {author}
  {\bibfnamefont {J.-Q.}\ \bibnamefont {Yan}}, \bibinfo {author} {\bibfnamefont
  {B.~C.}\ \bibnamefont {Sales}},\ and\ \bibinfo {author} {\bibfnamefont
  {D.}~\bibnamefont {Mandrus}},\ }\bibfield  {title} {\bibinfo {title}
  {{Magnetic phase transition in single crystals of the chiral helimagnet
  Cr$_{1/3}$NbS$_2$}},\ }\href {https://doi.org/10.1103/PhysRevB.87.104403}
  {\bibfield  {journal} {\bibinfo  {journal} {Physical Review B}\ }\textbf
  {\bibinfo {volume} {87}},\ \bibinfo {pages} {104403} (\bibinfo {year}
  {2013})}\BibitemShut {NoStop}%
\bibitem [{\citenamefont {Lu}\ \emph {et~al.}(2022)\citenamefont {Lu},
  \citenamefont {Murzabekova}, \citenamefont {Shim}, \citenamefont {Park},
  \citenamefont {Kim}, \citenamefont {Kish}, \citenamefont {Wu}, \citenamefont
  {DeBeer-Schmitt}, \citenamefont {Aczel}, \citenamefont {Schleife},
  \citenamefont {Mason}, \citenamefont {Mahmood},\ and\ \citenamefont
  {MacDougall}}]{Lu2022}%
  \BibitemOpen
  \bibfield  {author} {\bibinfo {author} {\bibfnamefont {K.}~\bibnamefont
  {Lu}}, \bibinfo {author} {\bibfnamefont {A.}~\bibnamefont {Murzabekova}},
  \bibinfo {author} {\bibfnamefont {S.}~\bibnamefont {Shim}}, \bibinfo {author}
  {\bibfnamefont {J.}~\bibnamefont {Park}}, \bibinfo {author} {\bibfnamefont
  {S.}~\bibnamefont {Kim}}, \bibinfo {author} {\bibfnamefont {L.}~\bibnamefont
  {Kish}}, \bibinfo {author} {\bibfnamefont {Y.}~\bibnamefont {Wu}}, \bibinfo
  {author} {\bibfnamefont {L.}~\bibnamefont {DeBeer-Schmitt}}, \bibinfo
  {author} {\bibfnamefont {A.~A.}\ \bibnamefont {Aczel}}, \bibinfo {author}
  {\bibfnamefont {A.}~\bibnamefont {Schleife}}, \bibinfo {author}
  {\bibfnamefont {N.}~\bibnamefont {Mason}}, \bibinfo {author} {\bibfnamefont
  {F.}~\bibnamefont {Mahmood}},\ and\ \bibinfo {author} {\bibfnamefont {G.~J.}\
  \bibnamefont {MacDougall}},\ }\href@noop {} {\bibinfo {title} {{Understanding
  the Anomalous Hall effect in Co$_{1/3}$NbS$_2$ from crystal and magnetic
  structures}}} (\bibinfo {year} {2022}),\ \Eprint
  {https://arxiv.org/abs/2212.14762} {arXiv:2212.14762 [cond-mat.mtrl-sci]}
  \BibitemShut {NoStop}%
\bibitem [{\citenamefont {Tenasini}\ \emph {et~al.}(2020)\citenamefont
  {Tenasini}, \citenamefont {Martino}, \citenamefont {Ubrig}, \citenamefont
  {Ghimire}, \citenamefont {Berger}, \citenamefont {Zaharko}, \citenamefont
  {Wu}, \citenamefont {Mitchell}, \citenamefont {Martin}, \citenamefont
  {Forr\'o},\ and\ \citenamefont {Morpurgo}}]{Tenasini2020}%
  \BibitemOpen
  \bibfield  {author} {\bibinfo {author} {\bibfnamefont {G.}~\bibnamefont
  {Tenasini}}, \bibinfo {author} {\bibfnamefont {E.}~\bibnamefont {Martino}},
  \bibinfo {author} {\bibfnamefont {N.}~\bibnamefont {Ubrig}}, \bibinfo
  {author} {\bibfnamefont {N.~J.}\ \bibnamefont {Ghimire}}, \bibinfo {author}
  {\bibfnamefont {H.}~\bibnamefont {Berger}}, \bibinfo {author} {\bibfnamefont
  {O.}~\bibnamefont {Zaharko}}, \bibinfo {author} {\bibfnamefont
  {F.}~\bibnamefont {Wu}}, \bibinfo {author} {\bibfnamefont {J.~F.}\
  \bibnamefont {Mitchell}}, \bibinfo {author} {\bibfnamefont {I.}~\bibnamefont
  {Martin}}, \bibinfo {author} {\bibfnamefont {L.}~\bibnamefont {Forr\'o}},\
  and\ \bibinfo {author} {\bibfnamefont {A.~F.}\ \bibnamefont {Morpurgo}},\
  }\bibfield  {title} {\bibinfo {title} {{Giant anomalous Hall effect in
  quasi-two-dimensional layered antiferromagnet Co$_{1/3}$NbS$_2$}},\ }\href
  {https://doi.org/10.1103/PhysRevResearch.2.023051} {\bibfield  {journal}
  {\bibinfo  {journal} {Physical Review Research}\ }\textbf {\bibinfo {volume}
  {2}},\ \bibinfo {pages} {023051} (\bibinfo {year} {2020})}\BibitemShut
  {NoStop}%
\bibitem [{\citenamefont {Takagi}\ \emph {et~al.}(2023)\citenamefont {Takagi},
  \citenamefont {Takagi}, \citenamefont {Minami}, \citenamefont {Nomoto},
  \citenamefont {Ohishi}, \citenamefont {Suzuki}, \citenamefont {Yanagi},
  \citenamefont {Hirayama}, \citenamefont {Khanh}, \citenamefont {Karube},
  \citenamefont {Saito}, \citenamefont {Hashizume}, \citenamefont {Kiyanagi},
  \citenamefont {Tokura}, \citenamefont {Arita}, \citenamefont {Nakajima},\
  and\ \citenamefont {Seki}}]{Takagi2023}%
  \BibitemOpen
  \bibfield  {author} {\bibinfo {author} {\bibfnamefont {H.}~\bibnamefont
  {Takagi}}, \bibinfo {author} {\bibfnamefont {R.}~\bibnamefont {Takagi}},
  \bibinfo {author} {\bibfnamefont {S.}~\bibnamefont {Minami}}, \bibinfo
  {author} {\bibfnamefont {T.}~\bibnamefont {Nomoto}}, \bibinfo {author}
  {\bibfnamefont {K.}~\bibnamefont {Ohishi}}, \bibinfo {author} {\bibfnamefont
  {M.-T.}\ \bibnamefont {Suzuki}}, \bibinfo {author} {\bibfnamefont
  {Y.}~\bibnamefont {Yanagi}}, \bibinfo {author} {\bibfnamefont
  {M.}~\bibnamefont {Hirayama}}, \bibinfo {author} {\bibfnamefont
  {N.}~\bibnamefont {Khanh}}, \bibinfo {author} {\bibfnamefont
  {K.}~\bibnamefont {Karube}}, \bibinfo {author} {\bibfnamefont
  {H.}~\bibnamefont {Saito}}, \bibinfo {author} {\bibfnamefont
  {D.}~\bibnamefont {Hashizume}}, \bibinfo {author} {\bibfnamefont
  {R.}~\bibnamefont {Kiyanagi}}, \bibinfo {author} {\bibfnamefont
  {Y.}~\bibnamefont {Tokura}}, \bibinfo {author} {\bibfnamefont
  {R.}~\bibnamefont {Arita}}, \bibinfo {author} {\bibfnamefont
  {T.}~\bibnamefont {Nakajima}},\ and\ \bibinfo {author} {\bibfnamefont
  {S.}~\bibnamefont {Seki}},\ }\bibfield  {title} {\bibinfo {title}
  {{Spontaneous topological Hall effect induced by non-coplanar
  antiferromagnetic order in intercalated van der Waals materials}},\
  }\href@noop {} {\bibfield  {journal} {\bibinfo  {journal} {Nature Physics}\
  }\textbf {\bibinfo {volume} {19}},\ \bibinfo {pages} {961} (\bibinfo {year}
  {2023})}\BibitemShut {NoStop}%
\bibitem [{\citenamefont {Kousaka}\ \emph {et~al.}(2016)\citenamefont
  {Kousaka}, \citenamefont {Ogura}, \citenamefont {Zhang}, \citenamefont
  {Miao}, \citenamefont {Lee}, \citenamefont {Torii}, \citenamefont {Kamiyama},
  \citenamefont {Campo}, \citenamefont {Inoue},\ and\ \citenamefont
  {Akimitsu}}]{Kousaka2016}%
  \BibitemOpen
  \bibfield  {author} {\bibinfo {author} {\bibfnamefont {Y.}~\bibnamefont
  {Kousaka}}, \bibinfo {author} {\bibfnamefont {T.}~\bibnamefont {Ogura}},
  \bibinfo {author} {\bibfnamefont {J.}~\bibnamefont {Zhang}}, \bibinfo
  {author} {\bibfnamefont {P.}~\bibnamefont {Miao}}, \bibinfo {author}
  {\bibfnamefont {S.}~\bibnamefont {Lee}}, \bibinfo {author} {\bibfnamefont
  {S.}~\bibnamefont {Torii}}, \bibinfo {author} {\bibfnamefont
  {T.}~\bibnamefont {Kamiyama}}, \bibinfo {author} {\bibfnamefont
  {J.}~\bibnamefont {Campo}}, \bibinfo {author} {\bibfnamefont
  {K.}~\bibnamefont {Inoue}},\ and\ \bibinfo {author} {\bibfnamefont
  {J.}~\bibnamefont {Akimitsu}},\ }\bibfield  {title} {\bibinfo {title} {{Long
  Periodic Helimagnetic Ordering in CrM$_3$S$_6$ ($M =$ Nb and Ta)}},\ }\href
  {https://doi.org/10.1088/1742-6596/746/1/012061} {\bibfield  {journal}
  {\bibinfo  {journal} {Journal of Physics: Conference Series}\ }\textbf
  {\bibinfo {volume} {746}},\ \bibinfo {pages} {012061} (\bibinfo {year}
  {2016})}\BibitemShut {NoStop}%
\bibitem [{\citenamefont {Miyadai}\ \emph {et~al.}(1983)\citenamefont
  {Miyadai}, \citenamefont {Kikuchi}, \citenamefont {Kondo}, \citenamefont
  {Sakka}, \citenamefont {Arai},\ and\ \citenamefont {Ishikawa}}]{Miyadai1983}%
  \BibitemOpen
  \bibfield  {author} {\bibinfo {author} {\bibfnamefont {T.}~\bibnamefont
  {Miyadai}}, \bibinfo {author} {\bibfnamefont {K.}~\bibnamefont {Kikuchi}},
  \bibinfo {author} {\bibfnamefont {H.}~\bibnamefont {Kondo}}, \bibinfo
  {author} {\bibfnamefont {S.}~\bibnamefont {Sakka}}, \bibinfo {author}
  {\bibfnamefont {M.}~\bibnamefont {Arai}},\ and\ \bibinfo {author}
  {\bibfnamefont {Y.}~\bibnamefont {Ishikawa}},\ }\bibfield  {title} {\bibinfo
  {title} {{Magnetic Properties of Cr$_{1/3}$NbS$_2$}},\ }\href
  {https://doi.org/10.1143/JPSJ.52.1394} {\bibfield  {journal} {\bibinfo
  {journal} {Journal of the Physical Society of Japan}\ }\textbf {\bibinfo
  {volume} {52}},\ \bibinfo {pages} {1394} (\bibinfo {year}
  {1983})}\BibitemShut {NoStop}%
\bibitem [{\citenamefont {Wang}\ \emph {et~al.}(2017)\citenamefont {Wang},
  \citenamefont {Chepiga}, \citenamefont {Ki}, \citenamefont {Li},
  \citenamefont {Li}, \citenamefont {Zhu}, \citenamefont {Kato}, \citenamefont
  {Ovchinnikova}, \citenamefont {Mila}, \citenamefont {Martin}, \citenamefont
  {Mandrus},\ and\ \citenamefont {Morpurgo}}]{Wang2017}%
  \BibitemOpen
  \bibfield  {author} {\bibinfo {author} {\bibfnamefont {L.}~\bibnamefont
  {Wang}}, \bibinfo {author} {\bibfnamefont {N.}~\bibnamefont {Chepiga}},
  \bibinfo {author} {\bibfnamefont {D.-K.}\ \bibnamefont {Ki}}, \bibinfo
  {author} {\bibfnamefont {L.}~\bibnamefont {Li}}, \bibinfo {author}
  {\bibfnamefont {F.}~\bibnamefont {Li}}, \bibinfo {author} {\bibfnamefont
  {W.}~\bibnamefont {Zhu}}, \bibinfo {author} {\bibfnamefont {Y.}~\bibnamefont
  {Kato}}, \bibinfo {author} {\bibfnamefont {O.~S.}\ \bibnamefont
  {Ovchinnikova}}, \bibinfo {author} {\bibfnamefont {F.}~\bibnamefont {Mila}},
  \bibinfo {author} {\bibfnamefont {I.}~\bibnamefont {Martin}}, \bibinfo
  {author} {\bibfnamefont {D.}~\bibnamefont {Mandrus}},\ and\ \bibinfo {author}
  {\bibfnamefont {A.~F.}\ \bibnamefont {Morpurgo}},\ }\bibfield  {title}
  {\bibinfo {title} {{Controlling the Topological Sector of Magnetic Solitons
  in Exfoliated Cr$_{1/3}$NbS$_2$ Crystals}},\ }\href
  {https://doi.org/10.1103/PhysRevLett.118.257203} {\bibfield  {journal}
  {\bibinfo  {journal} {Physical Revie Letters}\ }\textbf {\bibinfo {volume}
  {118}},\ \bibinfo {pages} {257203} (\bibinfo {year} {2017})}\BibitemShut
  {NoStop}%
\bibitem [{\citenamefont {Obeysekera}\ \emph {et~al.}(2021)\citenamefont
  {Obeysekera}, \citenamefont {Gamage}, \citenamefont {Gao}, \citenamefont
  {Cheong},\ and\ \citenamefont {Yang}}]{Obeysekera2021}%
  \BibitemOpen
  \bibfield  {author} {\bibinfo {author} {\bibfnamefont {D.}~\bibnamefont
  {Obeysekera}}, \bibinfo {author} {\bibfnamefont {K.}~\bibnamefont {Gamage}},
  \bibinfo {author} {\bibfnamefont {Y.}~\bibnamefont {Gao}}, \bibinfo {author}
  {\bibfnamefont {S.-w.}\ \bibnamefont {Cheong}},\ and\ \bibinfo {author}
  {\bibfnamefont {J.}~\bibnamefont {Yang}},\ }\bibfield  {title} {\bibinfo
  {title} {{The Magneto-Transport Properties of Cr$_{1/3}$TaS$_2$ with Chiral
  Magnetic Solitons}},\ }\href
  {https://doi.org/https://doi.org/10.1002/aelm.202100424} {\bibfield
  {journal} {\bibinfo  {journal} {Advanced Electronic Materials}\ }\textbf
  {\bibinfo {volume} {7}},\ \bibinfo {pages} {2100424} (\bibinfo {year}
  {2021})}\BibitemShut {NoStop}%
\bibitem [{\citenamefont {Zhang}\ \emph {et~al.}(2021)\citenamefont {Zhang},
  \citenamefont {Zhang}, \citenamefont {Liu}, \citenamefont {Zhang},
  \citenamefont {Yuan}, \citenamefont {Li}, \citenamefont {Wen}, \citenamefont
  {Jiang}, \citenamefont {Zhou}, \citenamefont {Lei}, \citenamefont {Zheng},
  \citenamefont {Song}, \citenamefont {Hou}, \citenamefont {Mi}, \citenamefont
  {Schwingenschl{\"o}gl}, \citenamefont {Manchon}, \citenamefont {Qiu},
  \citenamefont {Alshareef}, \citenamefont {Peng},\ and\ \citenamefont
  {Zhang}}]{Zhang2021}%
  \BibitemOpen
  \bibfield  {author} {\bibinfo {author} {\bibfnamefont {C.}~\bibnamefont
  {Zhang}}, \bibinfo {author} {\bibfnamefont {J.}~\bibnamefont {Zhang}},
  \bibinfo {author} {\bibfnamefont {C.}~\bibnamefont {Liu}}, \bibinfo {author}
  {\bibfnamefont {S.}~\bibnamefont {Zhang}}, \bibinfo {author} {\bibfnamefont
  {Y.}~\bibnamefont {Yuan}}, \bibinfo {author} {\bibfnamefont {P.}~\bibnamefont
  {Li}}, \bibinfo {author} {\bibfnamefont {Y.}~\bibnamefont {Wen}}, \bibinfo
  {author} {\bibfnamefont {Z.}~\bibnamefont {Jiang}}, \bibinfo {author}
  {\bibfnamefont {B.}~\bibnamefont {Zhou}}, \bibinfo {author} {\bibfnamefont
  {Y.}~\bibnamefont {Lei}}, \bibinfo {author} {\bibfnamefont {D.}~\bibnamefont
  {Zheng}}, \bibinfo {author} {\bibfnamefont {C.}~\bibnamefont {Song}},
  \bibinfo {author} {\bibfnamefont {Z.}~\bibnamefont {Hou}}, \bibinfo {author}
  {\bibfnamefont {W.}~\bibnamefont {Mi}}, \bibinfo {author} {\bibfnamefont
  {U.}~\bibnamefont {Schwingenschl{\"o}gl}}, \bibinfo {author} {\bibfnamefont
  {A.}~\bibnamefont {Manchon}}, \bibinfo {author} {\bibfnamefont {Z.~Q.}\
  \bibnamefont {Qiu}}, \bibinfo {author} {\bibfnamefont {H.~N.}\ \bibnamefont
  {Alshareef}}, \bibinfo {author} {\bibfnamefont {Y.}~\bibnamefont {Peng}},\
  and\ \bibinfo {author} {\bibfnamefont {X.-X.}\ \bibnamefont {Zhang}},\
  }\bibfield  {title} {\bibinfo {title} {{Chiral Helimagnetism and
  One-Dimensional Magnetic Solitons in a Cr-Intercalated Transition Metal
  Dichalcogenide}},\ }\href
  {https://doi.org/https://doi.org/10.1002/adma.202101131} {\bibfield
  {journal} {\bibinfo  {journal} {Advanced Materials}\ }\textbf {\bibinfo
  {volume} {33}},\ \bibinfo {pages} {2101131} (\bibinfo {year}
  {2021})}\BibitemShut {NoStop}%
\bibitem [{\citenamefont {Zhang}\ \emph {et~al.}(2022)\citenamefont {Zhang},
  \citenamefont {Algaidi}, \citenamefont {Li}, \citenamefont {Yuan},\ and\
  \citenamefont {Zhang}}]{Zhang2022}%
  \BibitemOpen
  \bibfield  {author} {\bibinfo {author} {\bibfnamefont {C.-H.}\ \bibnamefont
  {Zhang}}, \bibinfo {author} {\bibfnamefont {H.}~\bibnamefont {Algaidi}},
  \bibinfo {author} {\bibfnamefont {P.}~\bibnamefont {Li}}, \bibinfo {author}
  {\bibfnamefont {Y.}~\bibnamefont {Yuan}},\ and\ \bibinfo {author}
  {\bibfnamefont {X.-X.}\ \bibnamefont {Zhang}},\ }\bibfield  {title} {\bibinfo
  {title} {{Magnetic soliton confinement and discretization effects in
  Cr$_{1/3}$TaS$_2$ nanoflakes}},\ }\href@noop {} {\bibfield  {journal}
  {\bibinfo  {journal} {Rare Metals}\ }\textbf {\bibinfo {volume} {41}},\
  \bibinfo {pages} {3005} (\bibinfo {year} {2022})}\BibitemShut {NoStop}%
\bibitem [{ILL()}]{ILLffacts2023}%
  \BibitemOpen
  \href@noop {} {\bibinfo {title} {{Tables of Form Factors - Institut Laue
  Langevin, Grenoble, France}}},\ \bibinfo {howpublished}
  {\url{https://www.ill.eu/sites/ccsl/ffacts/}},\ \bibinfo {note} {accessed:
  2023-03-31}\BibitemShut {NoStop}%
\bibitem [{\citenamefont {Squires}(2012)}]{Squires2012}%
  \BibitemOpen
  \bibfield  {author} {\bibinfo {author} {\bibfnamefont {G.~L.}\ \bibnamefont
  {Squires}},\ }\href {https://doi.org/10.1017/CBO9781139107808} {\emph
  {\bibinfo {title} {{Introduction to the Theory of Thermal Neutron
  Scattering}}}},\ \bibinfo {edition} {3rd}\ ed.\ (\bibinfo  {publisher}
  {Cambridge University Press},\ \bibinfo {year} {2012})\BibitemShut {NoStop}%
\bibitem [{\citenamefont {Slovyanskikh}\ \emph {et~al.}(1985)\citenamefont
  {Slovyanskikh}, \citenamefont {Kuznetsov},\ and\ \citenamefont
  {Gracheva}}]{Slovyanskikh1985}%
  \BibitemOpen
  \bibfield  {author} {\bibinfo {author} {\bibfnamefont {V.}~\bibnamefont
  {Slovyanskikh}}, \bibinfo {author} {\bibfnamefont {N.}~\bibnamefont
  {Kuznetsov}},\ and\ \bibinfo {author} {\bibfnamefont {N.}~\bibnamefont
  {Gracheva}},\ }\bibfield  {title} {\bibinfo {title} {{The Dy-U-Te system}},\
  }\href@noop {} {\bibfield  {journal} {\bibinfo  {journal} {Russian Journal of
  Inorganic Chemistry}\ }\textbf {\bibinfo {volume} {30}},\ \bibinfo {pages}
  {1666} (\bibinfo {year} {1985})}\BibitemShut {NoStop}%
\bibitem [{\citenamefont {Malliakas}\ \emph {et~al.}(2005)\citenamefont
  {Malliakas}, \citenamefont {Billinge}, \citenamefont {Kim},\ and\
  \citenamefont {Kanatzidis}}]{Malliakas2005}%
  \BibitemOpen
  \bibfield  {author} {\bibinfo {author} {\bibfnamefont {C.}~\bibnamefont
  {Malliakas}}, \bibinfo {author} {\bibfnamefont {S.~J.~L.}\ \bibnamefont
  {Billinge}}, \bibinfo {author} {\bibfnamefont {H.~J.}\ \bibnamefont {Kim}},\
  and\ \bibinfo {author} {\bibfnamefont {M.~G.}\ \bibnamefont {Kanatzidis}},\
  }\bibfield  {title} {\bibinfo {title} {{Square Nets of Tellurium: Rare-Earth
  Dependent Variation in the Charge-Density Wave of RETe$_3$ (RE = Rare-Earth
  Element)}},\ }\href@noop {} {\bibfield  {journal} {\bibinfo  {journal}
  {Journal of the American Chemical Society}\ }\textbf {\bibinfo {volume}
  {127}},\ \bibinfo {pages} {6510} (\bibinfo {year} {2005})}\BibitemShut
  {NoStop}%
\bibitem [{\citenamefont {Aroyo}\ \emph {et~al.}(2006)\citenamefont {Aroyo},
  \citenamefont {Kirov}, \citenamefont {Capillas}, \citenamefont {Perez-Mato},\
  and\ \citenamefont {Wondratschek}}]{Bilbao}%
  \BibitemOpen
  \bibfield  {author} {\bibinfo {author} {\bibfnamefont {M.~I.}\ \bibnamefont
  {Aroyo}}, \bibinfo {author} {\bibfnamefont {A.}~\bibnamefont {Kirov}},
  \bibinfo {author} {\bibfnamefont {C.}~\bibnamefont {Capillas}}, \bibinfo
  {author} {\bibfnamefont {J.~M.}\ \bibnamefont {Perez-Mato}},\ and\ \bibinfo
  {author} {\bibfnamefont {H.}~\bibnamefont {Wondratschek}},\ }\bibfield
  {title} {\bibinfo {title} {{Bilbao Crystallographic Server II:
  Representations of crystallographic point groups and space groups}},\
  }\href@noop {} {\bibfield  {journal} {\bibinfo  {journal} {Acta
  Crystallographica}\ }\textbf {\bibinfo {volume} {A62}},\ \bibinfo {pages}
  {115} (\bibinfo {year} {2006})}\BibitemShut {NoStop}%
\bibitem [{\citenamefont {Kenzelmann}\ \emph {et~al.}(2006)\citenamefont
  {Kenzelmann}, \citenamefont {Harris}, \citenamefont {Aharony}, \citenamefont
  {Entin-Wohlman}, \citenamefont {Yildirim}, \citenamefont {Huang},
  \citenamefont {Park}, \citenamefont {Lawes}, \citenamefont {Broholm},
  \citenamefont {Rogado}, \citenamefont {Cava}, \citenamefont {Kim},
  \citenamefont {Jorge},\ and\ \citenamefont {Ramirez}}]{Kenzelmann2006}%
  \BibitemOpen
  \bibfield  {author} {\bibinfo {author} {\bibfnamefont {M.}~\bibnamefont
  {Kenzelmann}}, \bibinfo {author} {\bibfnamefont {A.~B.}\ \bibnamefont
  {Harris}}, \bibinfo {author} {\bibfnamefont {A.}~\bibnamefont {Aharony}},
  \bibinfo {author} {\bibfnamefont {O.}~\bibnamefont {Entin-Wohlman}}, \bibinfo
  {author} {\bibfnamefont {T.}~\bibnamefont {Yildirim}}, \bibinfo {author}
  {\bibfnamefont {Q.}~\bibnamefont {Huang}}, \bibinfo {author} {\bibfnamefont
  {S.}~\bibnamefont {Park}}, \bibinfo {author} {\bibfnamefont {G.}~\bibnamefont
  {Lawes}}, \bibinfo {author} {\bibfnamefont {C.}~\bibnamefont {Broholm}},
  \bibinfo {author} {\bibfnamefont {N.}~\bibnamefont {Rogado}}, \bibinfo
  {author} {\bibfnamefont {R.~J.}\ \bibnamefont {Cava}}, \bibinfo {author}
  {\bibfnamefont {K.~H.}\ \bibnamefont {Kim}}, \bibinfo {author} {\bibfnamefont
  {G.}~\bibnamefont {Jorge}},\ and\ \bibinfo {author} {\bibfnamefont {A.~P.}\
  \bibnamefont {Ramirez}},\ }\bibfield  {title} {\bibinfo {title} {{Field
  dependence of magnetic ordering in Kagom\'e-staircase compound
  ${\mathrm{Ni}}_{3}{\mathrm{V}}_{2}{\mathrm{O}}_{8}$}},\ }\href@noop {}
  {\bibfield  {journal} {\bibinfo  {journal} {Phys. Rev. B}\ }\textbf {\bibinfo
  {volume} {74}},\ \bibinfo {pages} {014429} (\bibinfo {year}
  {2006})}\BibitemShut {NoStop}%
\bibitem [{\citenamefont {Brown}\ \emph {et~al.}(1991)\citenamefont {Brown},
  \citenamefont {Chattopadhyay}, \citenamefont {Forsyth},\ and\ \citenamefont
  {Nunez}}]{Brown1991}%
  \BibitemOpen
  \bibfield  {author} {\bibinfo {author} {\bibfnamefont {P.}~\bibnamefont
  {Brown}}, \bibinfo {author} {\bibfnamefont {T.}~\bibnamefont
  {Chattopadhyay}}, \bibinfo {author} {\bibfnamefont {J.}~\bibnamefont
  {Forsyth}},\ and\ \bibinfo {author} {\bibfnamefont {V.}~\bibnamefont
  {Nunez}},\ }\bibfield  {title} {\bibinfo {title} {{Magnetic phase transitions
  of ${\mathrm{MnWO}}_{4}$ studied by the use of neutron diffraction}},\
  }\href@noop {} {\bibfield  {journal} {\bibinfo  {journal} {J. Phys.: Condens.
  Matter}\ }\textbf {\bibinfo {volume} {48}},\ \bibinfo {pages} {4281}
  (\bibinfo {year} {1991})}\BibitemShut {NoStop}%
\bibitem [{\citenamefont {Lautenschl{\"a}ger}\ \emph
  {et~al.}(1993)\citenamefont {Lautenschl{\"a}ger}, \citenamefont {Weitzel},
  \citenamefont {Vogt}, \citenamefont {Hock}, \citenamefont {B{\"o}hm},
  \citenamefont {Bonnet},\ and\ \citenamefont {Fuess}}]{Lautenschlaeger1993}%
  \BibitemOpen
  \bibfield  {author} {\bibinfo {author} {\bibfnamefont {G.}~\bibnamefont
  {Lautenschl{\"a}ger}}, \bibinfo {author} {\bibfnamefont {H.}~\bibnamefont
  {Weitzel}}, \bibinfo {author} {\bibfnamefont {T.}~\bibnamefont {Vogt}},
  \bibinfo {author} {\bibfnamefont {R.}~\bibnamefont {Hock}}, \bibinfo {author}
  {\bibfnamefont {A.}~\bibnamefont {B{\"o}hm}}, \bibinfo {author}
  {\bibfnamefont {M.}~\bibnamefont {Bonnet}},\ and\ \bibinfo {author}
  {\bibfnamefont {H.}~\bibnamefont {Fuess}},\ }\bibfield  {title} {\bibinfo
  {title} {{Magnetic phase transitions of ${\mathrm{MnWO}}_{4}$ studied by the
  use of neutron diffraction}},\ }\href@noop {} {\bibfield  {journal} {\bibinfo
   {journal} {Phys. Rev. B}\ }\textbf {\bibinfo {volume} {48}},\ \bibinfo
  {pages} {6087} (\bibinfo {year} {1993})}\BibitemShut {NoStop}%
\bibitem [{\citenamefont {Biffin}\ \emph {et~al.}(2014)\citenamefont {Biffin},
  \citenamefont {Johnson}, \citenamefont {Kimchi}, \citenamefont {Morris},
  \citenamefont {Bombardi}, \citenamefont {Analytis}, \citenamefont
  {Vishwanath},\ and\ \citenamefont {Coldea}}]{Biffin2014}%
  \BibitemOpen
  \bibfield  {author} {\bibinfo {author} {\bibfnamefont {A.}~\bibnamefont
  {Biffin}}, \bibinfo {author} {\bibfnamefont {R.~D.}\ \bibnamefont {Johnson}},
  \bibinfo {author} {\bibfnamefont {I.}~\bibnamefont {Kimchi}}, \bibinfo
  {author} {\bibfnamefont {R.}~\bibnamefont {Morris}}, \bibinfo {author}
  {\bibfnamefont {A.}~\bibnamefont {Bombardi}}, \bibinfo {author}
  {\bibfnamefont {J.~G.}\ \bibnamefont {Analytis}}, \bibinfo {author}
  {\bibfnamefont {A.}~\bibnamefont {Vishwanath}},\ and\ \bibinfo {author}
  {\bibfnamefont {R.}~\bibnamefont {Coldea}},\ }\bibfield  {title} {\bibinfo
  {title} {{Noncoplanar and Counterrotating Incommensurate Magnetic Order
  Stabilized by Kitaev Interactions in
  $\ensuremath{\gamma}\text{\ensuremath{-}}{\mathrm{Li}}_{2}{\mathrm{IrO}}_{3}$}},\
  }\href@noop {} {\bibfield  {journal} {\bibinfo  {journal} {Phys. Rev. Lett.}\
  }\textbf {\bibinfo {volume} {113}},\ \bibinfo {pages} {197201} (\bibinfo
  {year} {2014})}\BibitemShut {NoStop}%
\bibitem [{\citenamefont {Pardo}\ and\ \citenamefont
  {Flahaut}(1967)}]{Pardo1967}%
  \BibitemOpen
  \bibfield  {author} {\bibinfo {author} {\bibfnamefont {M.~P.}\ \bibnamefont
  {Pardo}}\ and\ \bibinfo {author} {\bibfnamefont {J.}~\bibnamefont
  {Flahaut}},\ }\bibfield  {title} {\bibinfo {title} {{Les tellurures
  superieurs des elements des terres rares, de formules L$_2$Te$_5$ et
  LTe$_3$}},\ }\href@noop {} {\bibfield  {journal} {\bibinfo  {journal}
  {Bulletin de la Soci\'et\'e Chimique de France}\ ,\ \bibinfo {pages} {3658}}
  (\bibinfo {year} {1967})}\BibitemShut {NoStop}%
\bibitem [{\citenamefont {Kenzelmann}\ \emph {et~al.}(2005)\citenamefont
  {Kenzelmann}, \citenamefont {Harris}, \citenamefont {Jonas}, \citenamefont
  {Broholm}, \citenamefont {Schefer}, \citenamefont {Kim}, \citenamefont
  {Zhang}, \citenamefont {Cheong}, \citenamefont {Vajk},\ and\ \citenamefont
  {Lynn}}]{Kenzelmann2005}%
  \BibitemOpen
  \bibfield  {author} {\bibinfo {author} {\bibfnamefont {M.}~\bibnamefont
  {Kenzelmann}}, \bibinfo {author} {\bibfnamefont {A.~B.}\ \bibnamefont
  {Harris}}, \bibinfo {author} {\bibfnamefont {S.}~\bibnamefont {Jonas}},
  \bibinfo {author} {\bibfnamefont {C.}~\bibnamefont {Broholm}}, \bibinfo
  {author} {\bibfnamefont {J.}~\bibnamefont {Schefer}}, \bibinfo {author}
  {\bibfnamefont {S.~B.}\ \bibnamefont {Kim}}, \bibinfo {author} {\bibfnamefont
  {C.~L.}\ \bibnamefont {Zhang}}, \bibinfo {author} {\bibfnamefont {S.-W.}\
  \bibnamefont {Cheong}}, \bibinfo {author} {\bibfnamefont {O.~P.}\
  \bibnamefont {Vajk}},\ and\ \bibinfo {author} {\bibfnamefont {J.~W.}\
  \bibnamefont {Lynn}},\ }\bibfield  {title} {\bibinfo {title} {{Magnetic
  Inversion Symmetry Breaking and Ferroelectricity in
  ${\mathrm{TbMnO}}_{3}$}},\ }\href@noop {} {\bibfield  {journal} {\bibinfo
  {journal} {Phys. Rev. Lett.}\ }\textbf {\bibinfo {volume} {95}},\ \bibinfo
  {pages} {087206} (\bibinfo {year} {2005})}\BibitemShut {NoStop}%
\bibitem [{\citenamefont {Dos~Santos}\ \emph {et~al.}(2011)\citenamefont
  {Dos~Santos}, \citenamefont {De~Campos}, \citenamefont {Da~Luz},
  \citenamefont {White}, \citenamefont {Neumeier}, \citenamefont {De~Lima},\
  and\ \citenamefont {Shigue}}]{DosSantos2011}%
  \BibitemOpen
  \bibfield  {author} {\bibinfo {author} {\bibfnamefont {C.}~\bibnamefont
  {Dos~Santos}}, \bibinfo {author} {\bibfnamefont {A.}~\bibnamefont
  {De~Campos}}, \bibinfo {author} {\bibfnamefont {M.}~\bibnamefont {Da~Luz}},
  \bibinfo {author} {\bibfnamefont {B.}~\bibnamefont {White}}, \bibinfo
  {author} {\bibfnamefont {J.}~\bibnamefont {Neumeier}}, \bibinfo {author}
  {\bibfnamefont {B.}~\bibnamefont {De~Lima}},\ and\ \bibinfo {author}
  {\bibfnamefont {C.}~\bibnamefont {Shigue}},\ }\bibfield  {title} {\bibinfo
  {title} {{Procedure for measuring electrical resistivity of anisotropic
  materials: A revision of the Montgomery method}},\ }\href@noop {} {\bibfield
  {journal} {\bibinfo  {journal} {Journal of Applied Physics}\ }\textbf
  {\bibinfo {volume} {110}} (\bibinfo {year} {2011})}\BibitemShut {NoStop}%
\end{thebibliography}
\end{document}